\documentclass[5p,twocolumn,final]{elsarticle}
\usepackage{amsmath,bm,epsfig,subfigure}

\usepackage{indentfirst}
\usepackage{amssymb}
\usepackage{graphicx}
\usepackage{epstopdf}
\usepackage{float}
\usepackage{romanbar}
\makeatletter
\renewcommand{\fnum@figure}{Fig. \thefigure}
\makeatother
\usepackage{color}
\usepackage[normalem]{ulem}
\usepackage{lineno,hyperref}
\modulolinenumbers[5]
\usepackage[toc,page]{appendix}
\usepackage{natbib}
\usepackage{ mathrsfs }
\usepackage{ifpdf}

\journal{Chaos, Solitons \& Fractals}

\begin{document}

\begin{frontmatter}

\title{Stability of compact breathers in translationally-invariant nonlinear chains with flat dispersion bands}


\author{Nathan Perchikov}

\author{O.V. Gendelman}
\address{Faculty of Mechanical Engineering, Technion, Haifa 32000, Israel}

\begin{abstract}
The paper addresses compact oscillatory states (compact breathers) in translationally-invariant lattices with flat dispersion bands. The compact breathers appear in such systems even in the linear approximation. If the interactions are nonlinear, but comply with the flat-band symmetry, the compact breather solutions exist, but can lose their stability for certain parameter values. As benchmark nonlinear potentials, we use the $\beta$-FPU (Fermi-Pasta-Ulam) and vibro-impact models. Loss of stability is numerically observed to occur through either pitchfork or Hopf bifurcations. The loss of stability can occur through two qualitatively different mechanisms -- through internal instability in the basic lattice elements, or through interaction of the compact breather with the linear passband of the lattice. The former scenario is more typical for high-amplitude breathers, and the latter -- for low amplitudes. For the high-amplitude case, insights into the nature of compact-mode loss-of-stability are obtained by resorting to the limit of a piecewise-linear system, where interactions are represented by conservative impacts. This issue calls for detailed introspection into integrability of piecewise-linear (impacting) systems and their relation to the smooth system. An idea for a sensor based on the studied mechanisms is suggested.
\end{abstract}


\end{frontmatter}


\section{Introduction}
\label{Sec1}

Following the considerable interest of the recent years in studying breathers (using the terminology and presuppositions introduced in \cite{Aubry}) and solitons in meta-materials, with special emphasis placed on the phenomena of resonance and spatial localization, this paper proposes a new model system with several advantageous features, and carries out theoretical analysis with possible engineering applications.

The main object of scientific investigation is a simplified model for a meta-material -- a complex spatially-extended dynamical system, in which a `large-scale' excitation applied to the boundary can have resonant interaction with specific degrees-of-freedom. By fine-tuning the `structural' parameters, enhanced engineering functionality can be obtained, such as acoustic attenuation or shock absorbance or, alternatively, high sensitivity and amplification of weak excitation.

Special interest is usually placed on optical meta-materials, such as the ones comprised of arrays of optical fibers, possibly with the nonlinear dielectric (Kerr) effect. In such systems, usually, identification of spatially localized nonlinear periodic solutions (modes), and analysis of their stability or resonant interaction is performed after averaging, using the so-called discrete nonlinear Shr\"odinger equation (DNLS), which is itself generally non-integrable (in contrast to some known integrable discrete non-consistent analogues of the continuous nonlinear Schr\"odinger equation). The focus on optical meta-materials is justified by a multitude of engineering applications and by the physical-model validity, associated with the fact that energetic losses in optical fibers are reasonably negligible.

Interesting work has also been done in the context of acoustic (rather than optical) meta-materials. Although acoustic systems, especially ones operating in the nonlinear regime, may suffer from relatively prominent energy dissipation effects, their study may be well justified. The first reason would be, obviously, related to the sought application -- acoustic noise attenuation, and in some cases also acoustic sensing, can only be based on \emph{acoustic} meta-materials. The second reason for one's interest in acoustic meta-materials is mainly theoretical. It appears that for mechanical systems, at least in some simplified geometries, one can exactly solve the equations of motion, obtaining, for example, spatially-localized temporally-periodic solutions. Moreover, for certain types of physically justifiable interaction potentials, the stability problem for the aforementioned solutions appears to be tractable, at least partially (and not only for the averaged system).

Addressing, in specific, the interesting question of spatial localization, it should be mentioned that of particular interest would be perfectly-localized periodic solutions. Such solutions may be traveling wave envelopes, as in optical applications (the so-called compact solitons, or compactons), or `standing' perfectly-compact discrete breathers, as observed in acoustic settings. In both cases, perfect-compactness (in contrast to super-exponential but non-perfect compactness) was shown to be possible to obtain when lattices with special local configurational symmetry are used, ones for which linear spectral analysis reveals the emergence of flat bands (or curves for one-dimensional lattices).

The terminology used for the dynamic regime central for the study reported in this paper is that of \emph{compact breathers} (CBs). The label `compact breather' is used here to make a distinction from a soliton (self-similar moving nonlinear wave with a localized profile) with perfect compactness (the tail-free case), such as the one considered by Rosenau in the continuum-limit and named `compacton' \cite{Rosenau}. In the same time, the relation to a breather owes to the fact that a discrete medium is considered here and the obtained spatially-localized and temporally-periodic solution is not moving with respect to the medium. Finally, the observed nonlinear normal mode (NNM) appears strictly spatially-compact (tail-free), and hence the term \emph{compact} breather (CB).

The study of systems with flat bands (FB) originated in purely-linear solid-state quantum settings, where flat-band generating systems were studied in many different geometries, and the effect of symmetry-breaking fields was examined \cite{Flach1}. In subsequent publications, it was found that classical systems, whether optical or mechanical, can be constructed, using the quantum analogy, such that in the linear setting the system would have a flat band (or several flat bands), but then, when augmenting the interaction potential by anharmonic additions, new noteworthy effects would be observed. 

One of the observed effects is the fact that the spatially-detached, perfectly local, periodic mode of the system will still exist even for anharmonic addition to the potential, given that this addition respects the original symmetries of the system. Moreover, one can study the stability of such compact localized states (CLSs, see \cite{Flach2}). Numerical stability-results for compact periodic solutions in physically realizable optical models (with cubic `Kerr' nonlinearity) were given in \cite{Johansson2015b}. A later work showed analytic stability-analysis results for a specific set of parameters for a similar optical setting \cite{ZegadloMalomed}. Asymptotic stability-analysis results for such optical systems were subsequently given for a family of solutions in \cite{DanieliFlach}. 

Analytic stability-analysis results for s specific set of parameters and asymptotic results for a family of compact solutions for the acoustic setting (with exact description of the degrees of freedom) were given by the authors in \cite{NPOVG2017}. The limitation of the latter work is the fact that the nonliniarity assumed therein was of the impact type, provided by fixed, translationally non-invariant constraints. Similarly, in the studies dedicated to the optical setting, the model equations (DNLS) are formulated in a moving framework. However, with respect to \emph{that} framework, no translational invariance holds, and thus strictly mathematically, the nonlinear dynamic difference equations in which a compact periodic solution is sought, are not translationally invariant.

Owing to all of the aforementioned and to the possible applicability of translationally-invariant systems with compact, marginally-stable, periodic modes in acoustic sensor engineering (to give but one example), the present work opts to address the problem directly, by exploiting the piecewise-linear interaction-potential limit, with the analysis-advantages associated with it.

Among the recent works dedicated to the study of translationally-invariant nonlinear mechanical lattices with compact solutions, it is worth mentioning \cite{Sergyeyev}, where a prestressed granular one-dimensional chain is examined, and a continuous Nesterenko type equation of motion is constructed consistently from the discrete lattice limit (for large wavelengths). Then, compact moving periodic solutions (compact solitons -- compactons) are obtained and shown to be stable for certain parameters, both for the bright and the dark soliton cases. It should be noted that compactness is obtained not owing to local complexity and symmetry, but due to taking the continuous limit.

In \cite{Maraver}, non-integrable DNLS models related to the Ablowitz-Ladik equation and Bose-Hubbard systems with power-law interaction nonlinearitis are examined, with super-exponentially localized (descrete) soliton solutions observed formally and their stability examined variationally.

In \cite{Aravena}, perfectly localized modes in photonic lattices described by the discrete linear Schr\"odinger equation are studied for the case of complex arrangements of one-dimensional lattices with one or multiple flat bands with non-trivial edge effects. Noteworthy is the possibility that the model provides for the study of transport in such systems.

A very interesting recent study in a mechanical setting is presented in \cite{James}, where an array of pendula with beads having (generalized) Hertzian interaction (Newton's cradle) is assumed. The slowly modulated amplitude of traveling solutions is examined through the associated discrete $p$-Schr\"odinger equation. Traveling waves with super-exponentially decaying tails are observed. The considered system is translationally invariant, and moreover, the vibro-impact limit is addressed, albeit with no explicit results presented regarding analytical stability investigation.

In \cite{Real}, a photonic lattice with two-dimensional Lieb structure and Kerr interaction nonlinearity is studied by averaging, using the associated DNLS equation. Flatband-related compact discrete (mobile) solitons are observed (numerically) at the zero-power limit.

In a very recent work, \cite{Qin}, a two-component Bose-Einstein condensate with cubic short-range interactions and long-range magnetic coupling is studied, and perfectly compact (tail-free) accelerating solitons are observed analytically and numerically.

The issue of localization in translationally-invariant nonlinear chains was considered in a noteworthy earlier work, \cite{Kim}, where the Klein-Gordon equation was studied and a rich resonance structure involving nontrivial transmission properties was observed (associated with emergence of discrete breathers).

A very recent paper, \cite{Maimaiti}, addresses the problem of obtaining a two-dimensional lattice with configurational symmetries and translational invariance, characterized by an arbitrary number of flat bands, generated from compact localized states (CLSs) occupying an arbitrary number of lattice site. The generating algorithm is derived by solving an inverse eigenvalue problem numerically employing chiral symmetry.

Interesting results are presented in \cite{Maraver2}, where the authors explore the spectral stability of Kuznetsov-Ma breathers, which are generalized Peregrine solitons. The stability is analyzed using Floquet theory, where the period is successively increased until the limiting structure of Perigrine solitons is reached. In the examined systems, flat dispersion modes are identified and observed to be related to compact nonlinear-case solutions with a parametrically increasing period. Loss of stability through pitchfork bifurcation is demonstrated.

Another work, \cite{Maraver3}, performs comparative stability analysis of solitary traveling waves and discrete breathers in Fermi-Pasta-Ulam (FPU) and Toda lattices, using energy-based criteria for the Hamiltonian case. Although perfect compactness is not addressed there, the correspondence between stationary discrete breathers and traveling continuous solitons, based on energy considerations, is studied within the context of stability analysis, which is relevant for the present work.

A noteworthy study is reported on in \cite{Doi}, where a special one-dimensional lattice is constructed, which supports solutions in the form of (smoothly propagating) `tail-free' traveling discrete breathers. The spatial compactness of the traveling discrete breathers in the specially-designed lattice arises due to lack of resonance of the localized nonlinear mode of the discrete breather with phonon modes, contrary to the standard case reported on in $\beta$-FPU lattices. It should be noted that the spatial compactness observed for the special lattice is not perfect. Rather, instead of a constant-amplitude weak tail, which is absent for continuous compactons (in waveguides with strongly nonlinear dispersion) but does emerge due to lattice discreteness in, say, $\beta$-FPU lattices, one finds that in the case of the aforementioned special lattice, the tail is super-exponentially decaying in the co-traveling frame.

In \cite{Suchkov}, a system of two linearly-coupled photonic chains with cubic on-site nonlinearity is examined. Stationary (immobile) solitons are obtained in the discrete (DNLS) case and are found to be related to Peierls-Nabarro potential pinning. In the continuous limit, the immobile modes transform to solitons, stable for subcritical amplitudes.

The central goal of the present work is to give (analytic) insight on the stability of perfectly-compact periodic modes in conservative translationally-invariant nonlinear mechanical (quasi-)one-dimensional lattices.

The structure of the paper is as follows. Section \ref{Sec2} presents a one-dimensional smoothly-nonlinear translationally-invariant mechanical system admitting a flat band in the linear regime and compact solutions in its nonlinear extension; Section \ref{Sec3} presents the linear analysis; Section \ref{Sec4} studies the (smoothly) nonlinear regime; Section \ref{Sec5} furthers the analysis of the nonlinear regime by examining a nonsmooth interaction-potential analogue for the large-amplitude limit; Section \ref{Sec6} examines a possible application of the theoretical insights gained, and Section \ref{Sec7} concludes.

\section{The model system}
\label{Sec2}

As a model system, we employ a one-dimensional lattice with internal symmetry. The system is sketched in Fig. \ref{Fig1} below, with linear properties specified by parameters and nonlinear augmentation denoted by the crossed-out spring symbols.

The representative element (unit cell) of the system comprises a particle (of mass $M$), to which two generally anharmonic identical oscillators (with mass $m$ each) are attached in parallel. The coupling between the two aforementioned oscillators is realized through common boundary conditions. For the symmetric mode, corresponding to synchronous motion of the two oscillators, the system degenerates to a standard dimeric chain of alternating particles of two types, with anharmonic potential. For the case $M=m$, a trivial uniform generally nonlinear chain with identical particles is recovered. The anti-synchronous mode corresponds to the chain decomposing to a collection of non-interacting unit-cells. The local stability of the modes and their interaction, in view of possible traveling-wave (acoustic) perturbations, suggest a rich dynamic picture. To start, linear analysis is performed next.

\begin{figure*}[h]
\begin{center}
{\includegraphics[scale = 0.4,trim={0 4cm 0 4cm},clip]{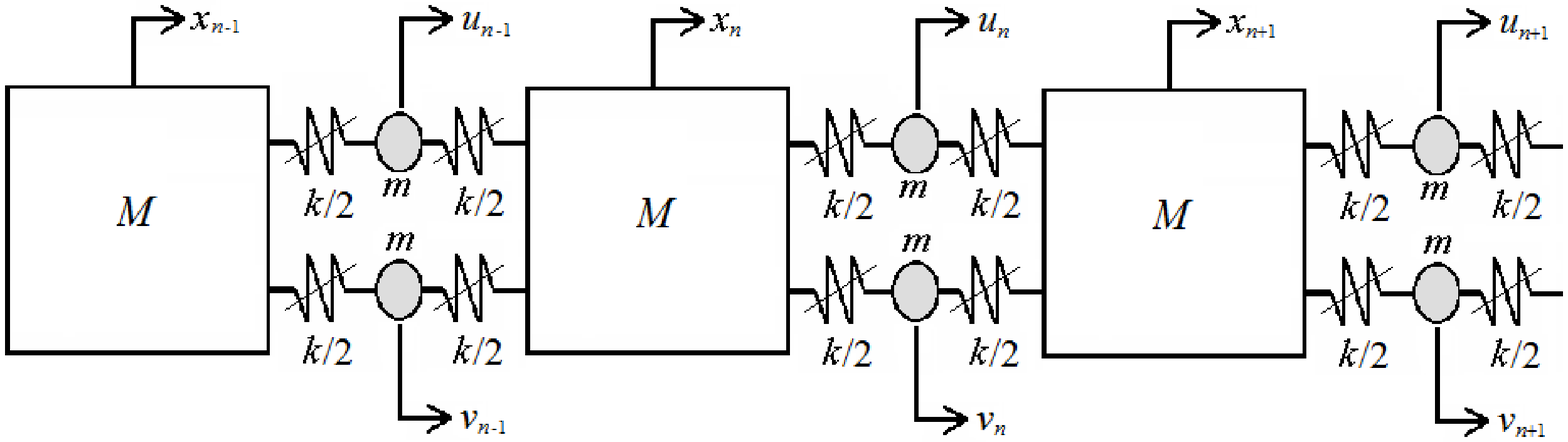}}
\end{center}
\caption{\small Sketch of the (quasi-)one-dimensional (horizontal) model chain}
\label{Fig1}
\end{figure*}

\section{Linear analysis -- dispersion bands}
\label{Sec3}
In the linear regime, neglecting boundary conditions, the system is assumed to satisfy the following equations of motion:
\begin{equation}
\label{eq1}
\begin{split}
M\ddot{x}_n-\frac{k}{2}(u_n-x_n)-\frac{k}{2}(v_n-x_n)+\\+\frac{k}{2}(x_n-u_{n-1})+\frac{k}{2}(x_n-v_{n-1})=0, \\
m\ddot{u}_n-\frac{k}{2}(x_{n+1}-u_n)+\frac{k}{2}(u_n-x_n)=0, \\ m\ddot{v}_n-\frac{k}{2}(x_{n+1}-v_n)+\frac{k}{2}(v_n-x_n)=0
\end{split}
\end{equation}

Dispersion analysis, that is, assumption of (generally) different-amplitude equal-frequency planar waves for all three vector displacements $\textbf{x},\textbf{u},\textbf{v}$, yields three dispersion relations. The first one corresponds to different amplitudes of $\textbf{u}$ and $\textbf{v}$, and reads (in units of $\sqrt{k/m}$):
\begin{equation}
\label{eq2}
\hat{\omega}_1(q)=1
\end{equation}
where $q$ is the dimensionless wavenumber. This is a \emph{flat band}. Two additional dispersion relations are obtained under the assumption of equal amplitudes of $\textbf{u}$ and $\textbf{v}$, and they are as follows:
\begin{equation}
\label{eq3}
\hat{\omega}_{2,3}(q)=\frac{\sqrt{1+\mu\pm \sqrt{(1+\mu)^2-4\mu\sin^2\frac{q}{2}}}}{\sqrt{2\mu}}
\end{equation}

The following normalization is used:
\begin{equation}
\label{eq4}
\hat{\omega}_i(q)= \frac{\omega_i(q)}{\sqrt{k/m}} \ , \ \ \mu \triangleq \frac{M/2}{m}
\end{equation}

For the special case of $\mu=1$, one has:
\begin{equation}
\label{eq5}
\hat{\omega}^{\mu=1}_{2,3}(q)=\sqrt{1\pm \cos{(q/2)}}
\end{equation}

For this special case, the three bands intersect at $q=\pi$.

For $\mu >1$, there is a frequency \emph{gap} between the acoustic and optical bands. The dispersion relations are
\begin{equation}
\label{eq6}
\begin{split}
\hat{\omega}_2(q)\le\sqrt{\frac{1+\mu- |1-\mu|}{2\mu}}=\begin{cases} \mu^{-1/2}<1, \ \mu>1 \\ 1, \ \mu\le 1\end{cases}\\
\hat{\omega}_3(q)\ge\sqrt{\frac{1+\mu+|1-\mu|}{2\mu}}=\begin{cases} 1, \ \mu>1 \\ \mu^{-1/2}>1, \ \mu\le 1\end{cases}
\end{split}
\end{equation}
which corresponds to a frequency gap of $|\Delta{\hat{\omega}}|\ge|\mu^{-1/2}-1|$.

For $\mu<1$, there is intersection between the acoustic and flat bands. This intersection occurs for vanishing group velocity. This means that when energy is not transported, it may be transferred to a local mode.

\begin{figure}[H]
\begin{center}
{\includegraphics[scale = 0.45,trim={0.28cm 0 1cm 0},clip]{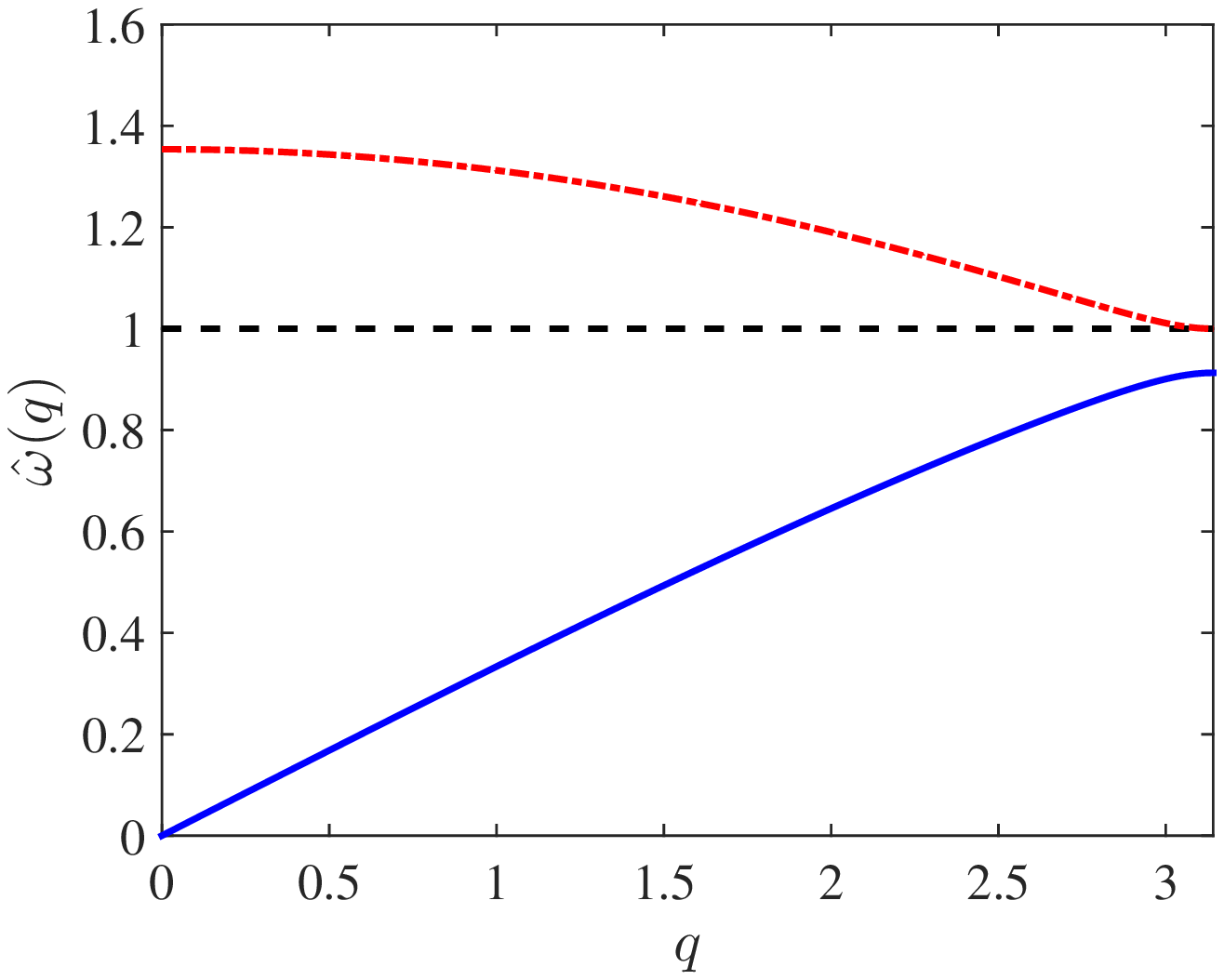}}
{\includegraphics[scale = 0.45,trim={0.28cm 0 1cm 0},clip]{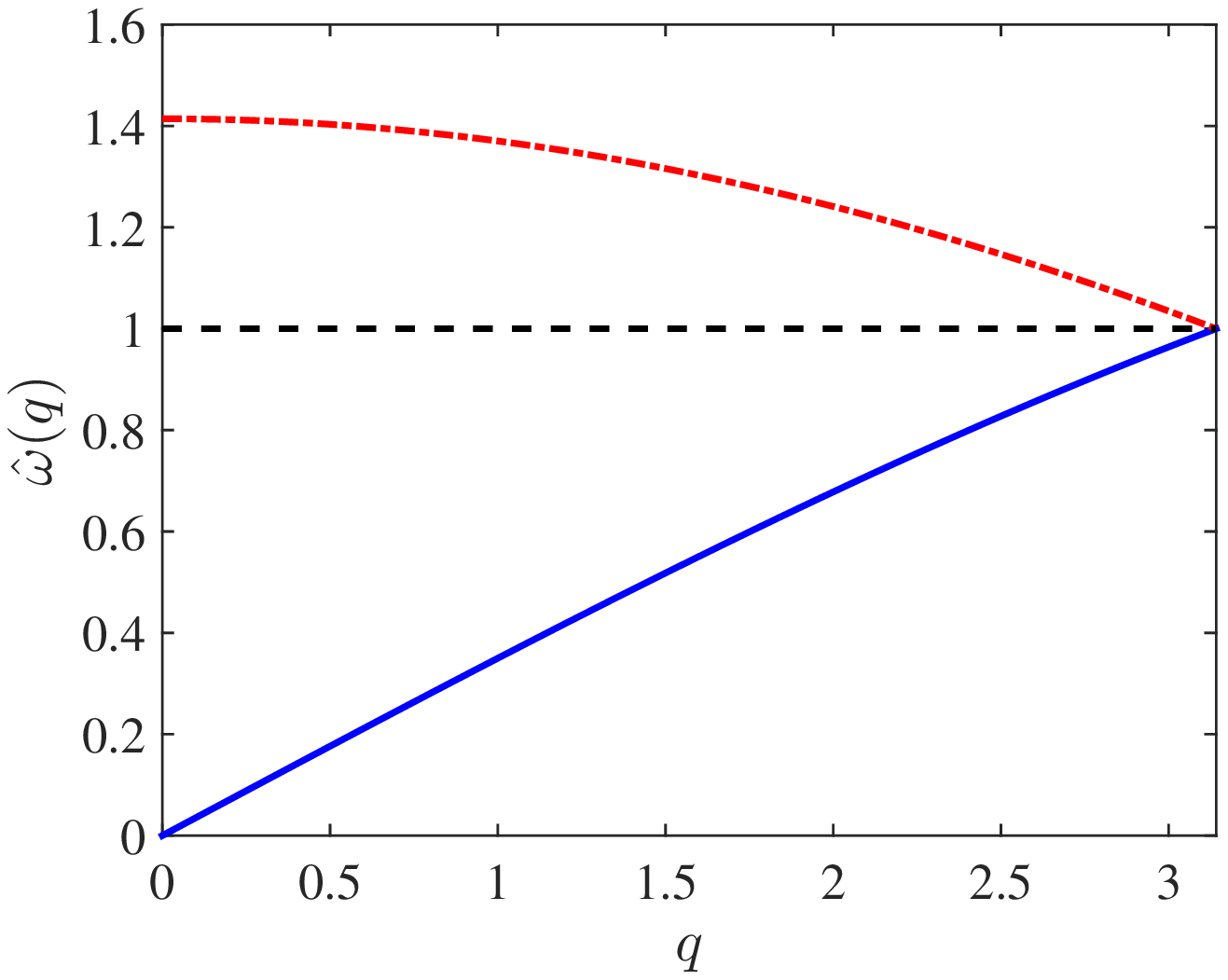}}
{\includegraphics[scale = 0.45,trim={0.28cm 0 1cm 0},clip]{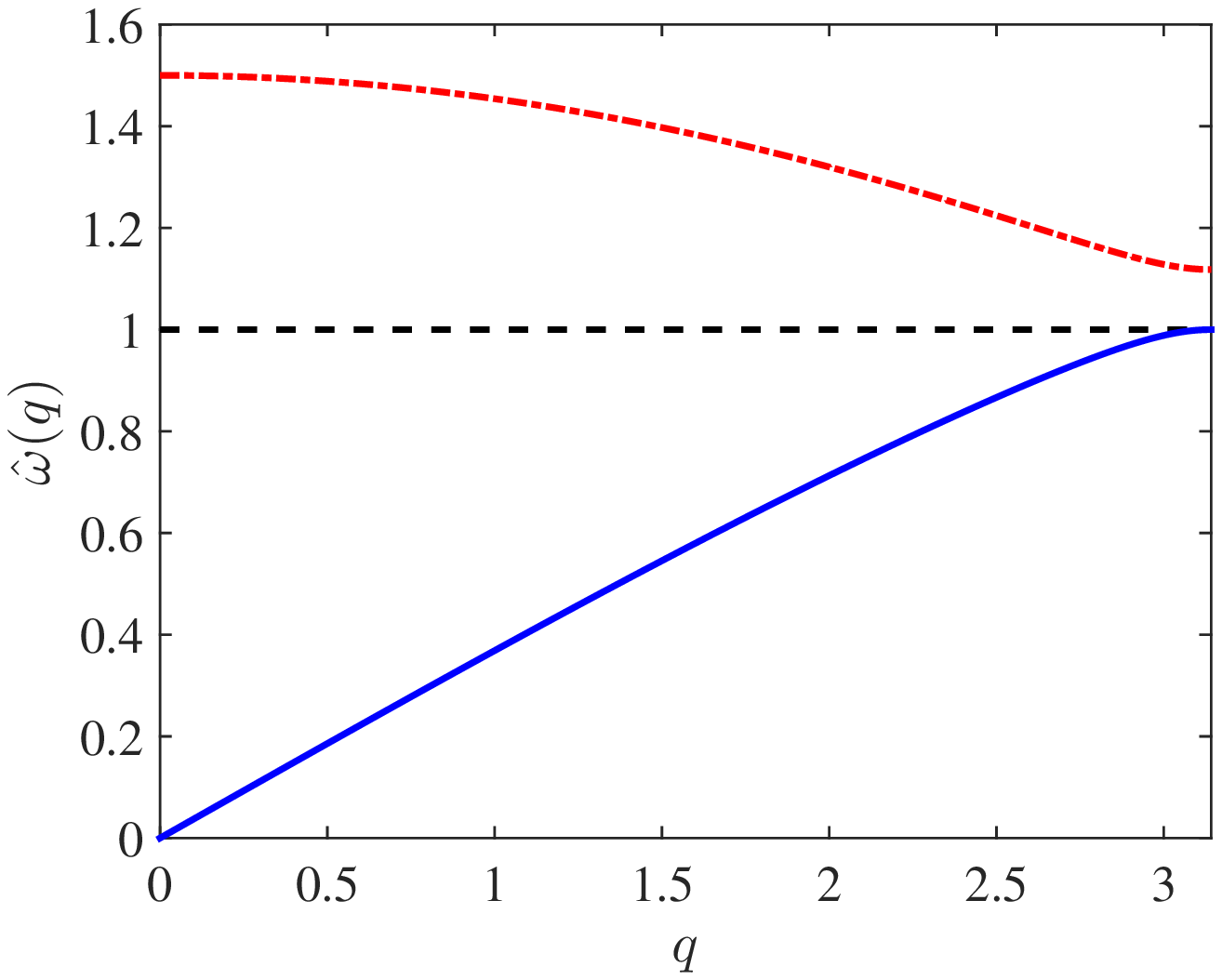}}
\end{center}
\caption{\small Dispersion bands for the model chain for $\mu=1.2$ (top), $\mu=1$ (center) and $\mu=0.8$ (bottom).}
\label{Fig2}
\end{figure}

Figure \ref{Fig2} shows the dispersion bands for $\mu>1,\mu=1$ and $\mu<1$.

For the special case of $\mu=1$, the acoustic branch is nearly dispersionless, with the group velocity ($c_g$) decreasing from $1/\sqrt{8}$ to 1/4 as $q$ increases from 0 to $\pi$. Generally one has
\begin{equation}
\label{eq7}
\begin{split}
\left.c_{g}^{(2)}\right|_{q=0}=\frac{1}{2\sqrt{1+\mu}}, \ \left.c_{g}^{(3)}\right|_{q=0}=0, \\  \left.c_{g}^{(2,3)}\right|_{q=\pi}=\begin{cases} \pm\frac{1}{4}, \ \mu=1 \\ 0, \ \ \ \ \mu\neq1 \end{cases}
\end{split}
\end{equation}

\section{The (smoothly) nonlinear regime}
\label{Sec4}
Here we assume potential anharmonicity by adding cubic force terms of the Fermi-Pasta-Ulam type. Neglecting boundary conditions, the system is assumed to satisfy the following equations of motion:
\begin{equation}
\label{eq8}
\begin{split}
M\ddot{x}_n-\frac{k}{2}(u_n-x_n)-p(u_n-x_n)^3-\frac{k}{2}(v_n-x_n)\\-p(v_n-x_n)^3+\frac{k}{2}(x_n-u_{n-1})+p(x_n-u_{n-1})^3+\\+\frac{k}{2}(x_n-v_{n-1})+p(x_n-v_{n-1})^3=0, \\
m\ddot{u}_n-\frac{k}{2}(x_{n+1}-u_n)-p(x_{n+1}-u_n)^3+\frac{k}{2}(u_n-x_n)\\+p(u_n-x_n)^3=0,  m\ddot{v}_n-\frac{k}{2}(x_{n+1}-v_n)\\-p(x_{n+1}-v_n)^3+\frac{k}{2}(v_n-x_n)+p(v_n-x_n)^3=0
\end{split}
\end{equation}
(where we added a cubic force coefficient $p>0$ for each link between masses $M$ and $m$).

\subsection{Single-element analysis}
\begin{figure}[H]
\begin{center}
{\includegraphics[scale = 0.55,trim={6cm 4cm 6cm 4cm},clip]{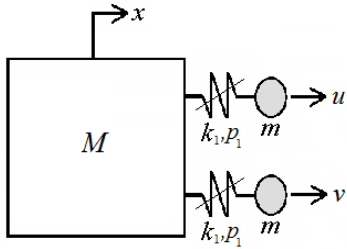}}
\end{center}
\caption{\small Sketch of a representative element of the model chain}
\label{Fig3}
\end{figure}

In order to gain insight on local stability, a single element is analyzed first. An unconnected element as shown in Fig. \ref{Fig3} is assumed, for the problem to be solvable analytically.

The element is assumed to be disconnected, `floating', and thus there is conservation of linear momentum (and position). In addition, there is energy conservation. Therefore, the solution lies on a three-dimensional manifold and can be described entirely on Poincar\'e sections. Moreover, nonlinear normal modes (NNMs) can be identified. For each NNM, the remaining integral can be obtained by direct time integration. After one NNM is obtained, its linear stability can be established by analysis of a single Hill equation using Hill's determinants method.
The equations of motion are as follows:
\begin{equation}
\label{eq9}
\begin{split}
x=-\frac{1}{2\mu}(u+v),m\ddot{u}+k_1(u-x)+p_1(u-x)^3=0, \\ m\ddot{v}+k_1(v-x)+p_1(v-x)^3=0
\end{split}
\end{equation}
which reduces to a system of two differential equations:
\begin{equation}
\label{eq10}
\begin{split}
m\ddot{u}+\frac{k_1}{2\mu}[(2\mu+1)u+v]+\frac{p_1}{8\mu^3}[(2\mu+1)u+v]^3=0 \\
m\ddot{v}+\frac{k_1}{2\mu}[u+(2\mu+1)v]+\frac{p_1}{8\mu^3}[u+(2\mu+1)v]^3=0
\end{split}
\end{equation}

Introducing the sum of the displacements and their difference, as follows:
\begin{equation}
\label{eq11}
y\triangleq u+v,z\triangleq u-v
\end{equation}
enables rewriting Eqs. \ref{eq10} as evolution equations for the symmetric and antisymmetric modes.

The equation for the antisymmetric mode is
\begin{equation}
\label{eq12}
m\ddot{z}+\left[{k_1+\frac{3}{4}\left({\frac{1+\mu}{\mu}}\right)^2p_1 y^2}\right]z+\frac{1}{4}p_1z^3=0
\end{equation}

The equation for the symmetric mode is
\begin{equation}
\label{eq13}
m\ddot{y}+\frac{1+\mu}{\mu}\left({k_1+\frac{3}{4}p_1 z^2}\right)y+\frac{1}{4}\left({\frac{1+\mu}{\mu}}\right)^3p_1y^3=0
\end{equation}

The motivation of the present work is, in part, to investigate (possibly resonant) interaction between CBs and phonons. Therefore, at this point, we are interested in finding the parameters corresponding to the loss of stability of a CB due to resonance with a propagating wave, which is represented by the symmetric mode. Hence, we need to obtain the compact antisymmetric mode itself and its stability equation. In order to obtain the antisymmetric mode, we take Eq. \ref{eq12} to the $y\to0$ limit, which yields:
\begin{equation}
\label{eq14}
m\ddot{\bar z}+k_1\bar z+\frac{1}{4}p_1\bar z^3=0
\end{equation}
where the bar denotes that it is the NNM solution.

In order to be able to interpret the single-element analysis results in the context of a chain, the coupling parameters should be identified correctly. In the case of the antisymmetric mode, which is the case we are interested in within the context of stability analysis, we have $y=x=0$. This means that, within the single element, each of the two masses is attached to a stationary point by a \emph{single} linear-cubic force, with parameters $k_1$ and $p_1$ for each. However, in the chain depicted in Fig. \ref{Fig1}, for stationary masses $M$, each mass $m$ is attached to stationary points by \emph{two} linear-cubic forces with parameters $k_1$ and $p_1$ for each. Ordinarily, forces acting in a queue are not additive. However, since they attach the masses $m$ to two stationary points, they are equivalent to forces acting in parallel, and thus they are additive. Consequently, for a stable stationary compact mode, the following relations between the potential parameters can be identified:
\begin{equation}
\label{eq15}
k_1=2(k/2)=k,p_1=2p
\end{equation}

Therefore, the antisymmetric-mode equation of motion would be:
\begin{equation}
\label{eq16}
m\ddot{\bar z}+k\bar z+\frac{1}{2}p\bar z^3=0
\end{equation}

Introducing the following dimensionless quantities,
\begin{equation}
\label{eq18}
\begin{split}
\omega_0\triangleq \sqrt{\frac{k}{m}}, \ \tau \triangleq \omega_0 (t-t|_{\bar{z}=0}), \  ()'\triangleq \frac{d}{d\tau}, \\ A \triangleq \underset{t}{\text{max}}\lbrace\bar{z}(t)\rbrace, \ \alpha \triangleq \frac{pA^2}{k}, \ \hat{z} \triangleq \frac{\bar{z}}{A}
\end{split}
\end{equation}
we obtain the dimensionless form of the first integral of motion (in Lagrangian description), similar to what is derived in \cite{Perchikov2016}:
\begin{equation}
\label{eq19}
(\hat{z}')^2+\hat{z}^2+\frac{1}{4}\alpha(\hat{z}^4-1)=1
\end{equation}

This equation has a solution that can be expressed using the fifth Jacobi elliptic function \cite{Perchikov2016}:
\begin{equation}
\label{eq20}
\hat{z}(\tau)=\sqrt{\frac{{1+\alpha/4}}{{1+\alpha/2}}}\text{sd}\left({\sqrt{1+\alpha/2} \ \tau}\left|\frac{1}{2}\frac{\alpha/2}{1+\alpha/2}\right.\right)
\end{equation}

Moreover, $\hat{z}(\tau)$ has a Fourier series representation, as follows:
\begin{equation}
\label{eq21}
\hat{z}(\hat\tau)=\sum_{n=1}^{\infty}{Z_n\sin{(n\hat\tau)}}
\end{equation}
where
\begin{equation}
\begin{split}
\label{eq22}
\hat\tau=\Omega_z\tau, \hat{m}\triangleq \frac{1}{2}\frac{\alpha}{\alpha+2},\Omega_z\triangleq \frac{\pi}{2}\frac{\sqrt{1+\alpha/2}}{K(\hat{m})}, \\ Z_n\triangleq \pi\sqrt{\frac{{1+\alpha/4}}{{1+\alpha/2}}}\frac{\sqrt{1+4/\alpha}}{K(\hat{m})}\frac{(-1)^{\frac{n-1}{2}}\delta_{n,2\mathbb{N}-1}}{\cosh{\left[{\frac{n\pi K(1-\hat{m})}{2K(\hat{m})}}\right]}}
\end{split}
\end{equation}
and $K(\hat{m})$ is the complete elliptic integral of the first kind (and $\delta_{a,b}$ is Kronecker's delta).

The linear stability of the derived antisymmetric mode can be examined by linearizing Eq. (\ref{eq13}), which employing Eqs. \ref{eq15} and \ref{eq18}, takes the following form (where linearization yields the exchange of $z$ by $\bar{z}$):
\begin{equation}
\label{eq23}
\begin{split}
\frac{d^2y(\hat\tau)}{d\hat\tau^2}+h(\hat{\tau})y(\hat\tau)=0, \\ h(\hat{\tau})\triangleq \frac{1+\mu}{\mu}\Omega_z^{-2}\left[{1+\frac{3}{2}\alpha \hat{z}^2(\hat\tau)}\right]
\end{split}
\end{equation}
(the use of Eq. \ref{eq15} is justified since for small $y$, which is assumed within the linearization framework, the box of mass $M$ is nearly stationary and the two springs are nearly equivalent to being connected in parallel).

The determination of the Hill function in Eq. \ref{eq23} requires the presentation of the square of the antisymmetric mode as a Fourier series, which becomes:
\begin{equation}
\label{eq24}
[\hat{z}(\hat\tau)]^2=\frac{Z_0^{(2)}}{2}+\sum_{k=1}^{\infty}{Z_k^{(2)}\cos{(2k\hat\tau)}}
\end{equation}
where
\begin{equation}
\label{eq25}
\begin{split}
Z_k^{(2)}\triangleq \left[\sum_{n=1,3,5}^{\infty}{Z_n Z_{2k+n}-\frac{1}{2}Z_k^2\delta_{k,2\mathbb{N}-1}}\right. \\ \left. -(1-\delta_{k,1})\sum_{n=1,3,5}^{\infty}{Z_{k-n}Z_{k+n}}\right]\delta_{k,\mathbb{N}-1}
\end{split}
\end{equation}

Thus, the corresponding Hill function has the following Fourier series representation:
\begin{equation}
\label{eq26}
\begin{split}
h(\hat{\tau})=c_0+\sum_{k=1}^{\infty}{c_k\cos{(2k\hat\tau)}}, \\ c_0 = \frac{1+\mu}{\mu}\Omega_z^{-2} +\frac{3}{4} \frac{1+\mu}{\mu}\alpha\Omega_z^{-2} Z_0^{(2)}, \\
c_k = \frac{3}{2} \frac{1+\mu}{\mu}\alpha\Omega_z^{-2} Z_k^{(2)} \\
\end{split}
\end{equation}

Consequently, we can plot a stability map in the amplitude--mass-ratio plane using Hill's infinite determinants method.

\subsubsection{The small-amplitude limit}

For $\alpha\to0$, Eq. (\ref{eq20}) yields a sine and thus:
\begin{equation}
\label{eq27}
\begin{split}
\hat{m}\to\frac{\alpha}{4},Z_n\to\delta_{n,1},\Omega_z\to1+\frac{3\alpha}{16},\hat{z}(\hat\tau)\to\sin{\hat\tau}\\ \Rightarrow h(\hat{\tau})\to {\frac{1+\mu}{\mu}\left(1+ \frac{3}{8}\alpha\right)-\frac{3}{4}\frac{1+\mu}{\mu}\alpha\cos{\left(2\hat{\tau}\right)}}
\end{split}
\end{equation}

This limit corresponds to the Mathieu equation, for which the first instability tongue is absent and only higher-order instability tongues exist, with the following zero-amplitude-limit critical mass ratios:
\begin{equation}
\label{eq28}
\mu_{\text{cr}}^{\alpha\to0}=\frac{1}{(1+N)^2-1}, N\in \mathbb{N}
\end{equation}
for which, as expected, $\mu_{\text{cr}}<1$, and the value closest to unity is 1/3.

\begin{figure}[H]
\begin{center}
\includegraphics[scale = 0.5,trim={0.28cm 0 1cm 0},clip]{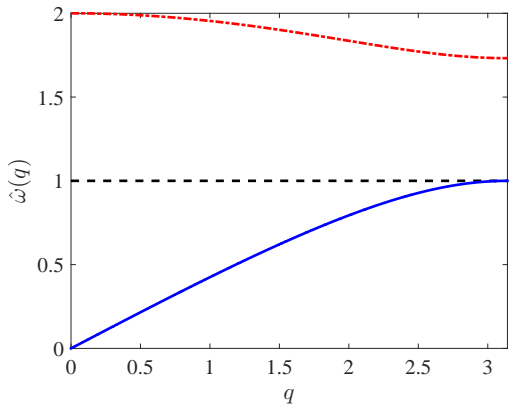}
\end{center}
\caption{\small Dispersion bands for $\mu=1/3$}
\label{Fig4}
\end{figure}

Plotting the dispersion bands for this value shows a finite region of possible intersection between the flat and acoustic bands due to weak nonlinearity, coinciding with the expectation of parametric resonance as derived above. The dispersion bands plot for the aforementioned mass-ratio is shown in Fig. \ref{Fig4}.

\begin{figure}[H]
\begin{center}
\includegraphics[scale =0.5,trim={0.28cm 0 1cm 0},clip]{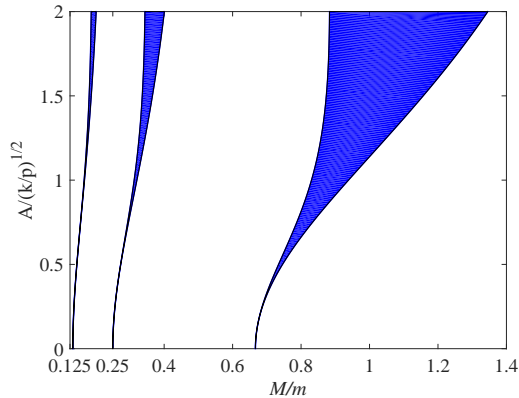}
\end{center}
\caption{\small Instability tongues (the first three) for the antisymmetric mode for the representative element of the model chain}
\label{Fig5}
\end{figure}

The stability map for finite amplitudes is shown in Fig. \ref{Fig5}.

Additional instability tongues, with decreasing widths, forming an infinite sequence, emerge to the left of the presented range, for $M/m<1/8$, at least for $\sqrt{\alpha}<2$. The details of the analysis used to produce Fig. \ref{Fig5} can be found in \cite{Perchikov2016}. The right boundary of the principal instability tongue has a vertical asymptote with $\mu_{+}^{\text{cr}}\underset{\alpha\to\infty}\to 1$. The left boundary has a vertical asymptote with $\mu_{-}^{\text{cr}}\underset{\alpha\to\infty}\to3/7$. In addition, the left boundary (of the principal tongue) has a local maximum with $\mu'(\alpha)=0$ at $\alpha\approx 4.33,\mu\approx 0.44$. For $\mu>1$, when there is no intersection between the acoustic and flat bands, the antisymmetric CB is linearly stable for all amplitudes and could only be destroyed by finite, large-enough effects.

The single-element model is valid whenever the antisymmetric mode is stable, thus the white areas in Fig. \ref{Fig5} imply (linear) stability of CBs in the chain. The shaded areas correspond to instability in the single-element model, and they are thus suspected to correspond to instability of CBs in the chain. However, since once instability occurs, the single-element model is no longer representative of the chain, one should observe the behavior of finite perturbations to CBs in a chain, not unlike when validating the implications of linear stability in a general case. The following section examines the dynamics of small perturbations to a single CB solution, assuming, consecutively, local (white) noise, and propagating (phonon or small-amplitude soliton) perturbations.

\subsection{Numerical integration for a chain}
\label{Sec4b}

\subsubsection{Full equations of motion}

The equations of motion for a chain in dimensionless form (where all displacements are relative to the CB amplitude, $A$), in modal coordinates, in terms of previously defined quantities, constitute the following dynamical system:
\begin{equation}
\label{eq28}
\begin{split}
\hat{x}_n''=\frac{\hat{y}_n+\hat{y}_{n-1}-4\hat{x}_n}{4\mu}\\-\frac{\alpha}{8\mu}\left\lbrace(2\hat{x}_n-\hat{y}_n)\left[(2\hat{x}_n-\hat{y}_n)^2+3\hat{z}_n^2\right]+\right.\\ \left.+(2\hat{x}_n-\hat{y}_{n-1})\left[(2\hat{x}_n-\hat{y}_{n-1})^2+3\hat{z}_{n-1}^2\right]\right\rbrace, \\
\hat{y}_n''= -\hat{y}_n+\hat{x}_n+\hat{x}_{n+1}+\\+\frac{\alpha}{4}\left\lbrace(2\hat{x}_n-\hat{y}_n)\left[(2\hat{x}_n-\hat{y}_n)^2+3\hat{z}_n^2\right]+\right.\\+\left. (2\hat{x}_{n+1}-\hat{y}_n)\left[(2\hat{x}_{n+1}-\hat{y}_n)^2+3\hat{z}_n^2\right]\right\rbrace, \\
\hat{z}_n''= -\hat{z}_n\\-\alpha\frac{{3\left[(2\hat{x}_n-\hat{y}_n)^2+(2\hat{x}_{n+1}-\hat{y}_n)^2\right]\hat{z}_n+2\hat{z}_n^3}}{4}, \\ \ \forall \   n\in \mathbb N, \  2 \le n \le N-1
\end{split}
\end{equation}

In addition, there are the equations at the boundaries. Assuming free boundary conditions (to coincide with the single element analysis), one has
\begin{equation}
\label{eq29}
\begin{split}
\hat{y}_N''= \frac{2\hat{x}_N-\hat{y}_N}{2}+\\+\alpha\frac{{(2\hat{x}_N-\hat{y}_N)[(2\hat{x}_N-\hat{y}_N)^2+3\hat{z}_N^2]}}{4} , \\
\hat{z}_N''= -\frac{\hat{z}_N}{2}-\alpha\frac{{3(2\hat{x}_N-\hat{y}_N)^2\hat{z}_N+\hat{z}_N^3}}{4}
\end{split}
\end{equation}
and the expression for $\hat{x}''_N$ can also be obtained from Eq. \ref{eq28}. The equations for $\hat{x}''_1$ can be replaced by a conservation law giving an expression for $\hat{x}_1$, which for initial conditions of a CB would be $\hat{x}_1=-\sum\limits_{n=2}^{N}\hat{x}_n-\frac{1}{2\mu}\sum\limits_{n=1}^{N}\hat{y}_n$, yielding the following two remaining equations for $N=1$:
\begin{equation}
\label{eq29b}
\begin{split}
\hat{y}_1''= -\frac{1+2\mu}{2\mu}\hat{y}_1-\sum\limits_{n=3}^{N}\hat{x}_n\\-\frac{1}{2\mu}\sum\limits_{n=2}^{N}\hat{y}_n+\frac{\alpha}{4}(2\hat{x}_2-\hat{y}_1)[(2\hat{x}_2-\hat{y}_1)^2+3\hat{z}_1^2]\\-\frac{\alpha}{4}\left[\sum\limits_{n=2}^{N}\left(2\hat{x}_n+\frac{1}{\mu}\hat{y}_n\right)+\frac{1+\mu}{\mu}\hat{y}_1\right]\times \\ \times\left\lbrace\left[\sum\limits_{n=2}^{N}\left(2\hat{x}_n+\frac{1}{\mu}\hat{y}_n\right)+\frac{1+\mu}{\mu}\hat{y}_1\right]^2+3\hat{z}_1^2\right\rbrace , \\
\hat{z}_1''= -\hat{z}_1\\-\frac{\alpha}{4}\left(3\left\lbrace\left[\sum\limits_{n=2}^{N}\left(2\hat{x}_n+\frac{1}{\mu}\hat{y}_n\right)+\frac{1+\mu}{\mu}\hat{y}_1\right]^2+\right. \right.\\  \left.  \left. + \vphantom{\left[\sum\limits_{n=2}^{N}\left(2\hat{x}_n+\frac{1}{\mu}\hat{y}_n\right)+\frac{1+\mu}{\mu}\hat{y}_1\right]^2}(2\hat{x}_2-\hat{y}_1)^2\right\rbrace\hat{z}_1+2\hat{z}_1^3\right)
\end{split}
\end{equation}

\subsubsection{Floquet-theory analysis of the stability of the compact breather solution}

Writing the state of the chain in vector form, as $\textbf{w}=(\hat{\textbf{x}}^\top,\hat{\textbf{y}}^\top,\hat{\textbf{z}}^\top,\hat{\textbf{x}}'^\top,\hat{\textbf{y}}'^\top,\hat{\textbf{z}}')^\top$, the evolution equation for the variation of the solution from that of a CB at a given location, under the assumption that the variation is small, can be written as:
\begin{equation}
\label{eq30}
\delta{\textbf{w}}'={\textbf{A}}(\tau)\delta\textbf{w}
\end{equation}
where the time-dependent matrix $\textbf{A}(\tau)$ can be expressed in a block-form as follows:
\begin{equation}
\label{eq31}
{\textbf{A}}(\tau)=\begin{bmatrix}
    \textbf{0}_{3N-1} & \textbf{I}_{3N-1}\\
    \textbf{B}(\tau) & \textbf{0}_{3N-1}
  \end{bmatrix}
\end{equation}
where the matrix $\textbf{0}_d$ is a square zeros-matrix of dimension $d$, $N$ is the number of representative-cells in the chain; the matrix $\textbf{I}_d$ is the identity matrix of dimension $d$, and the matrix ${\textbf{B}}(\tau)$ is the gradient of the right-hand side of the equations of motion in Eqs. \ref{eq28}-\ref{eq29b}, estimated at the CB solution. The full expression for this matrix (for both free-ends and periodic boundary conditions) is given in \ref{AppendixA}. 

\begin{figure}[H]
\begin{center}
\includegraphics[scale =0.5,trim={0.28cm 0 1cm 0},clip]{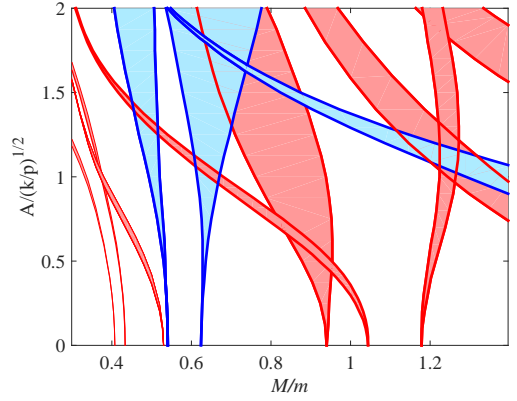}
\end{center}
\caption{\small Instability tongues for the (compact) antisymmetric mode for the model chain in a typical rectangular subspace situated around $\alpha=1,M/m=1$, for $N=3,N_0=2$, with pitchfork instability bounds in blue (online) and Neimark-Sacker instability bounds in red (online).}
\label{Fig6}
\end{figure}

The analysis for the smallest chain with a symmetric single CB shows the emergence of resonance overlay, as depicted in Fig. \ref{Fig6}.

\begin{figure}[H]
\begin{center}
\includegraphics[scale = 0.53]{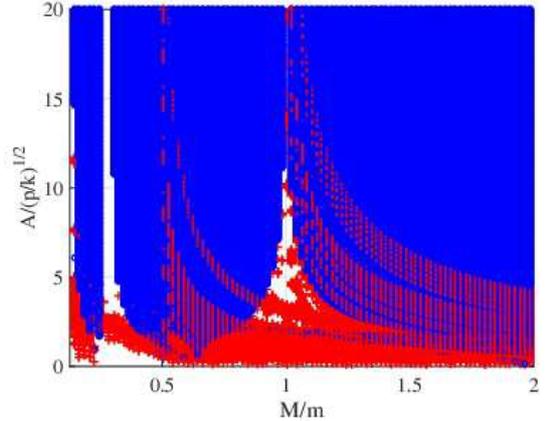}
\end{center}
\caption{\small Central-site CB stability map for $N=19$ -- pitchfork instability points denoted by (blue online) dots, Neimark-Sacker instability points denoted by (red online) `+' symbols.}
\label{Fig7}
\end{figure}

A CB stability map for a chain of $N=19$ sites with \emph{free} ends and excitation in the central site is given in Fig. \ref{Fig7}. Stability regions are scarce.

\subsubsection{Numerical integration for a large-amplitude compact breather in the center and a zero-momentum perturbation at the boundary}

\begin{figure}[H]
\begin{center}
{\includegraphics[scale = 0.5,trim={0.28cm 0 1cm 0},clip]  {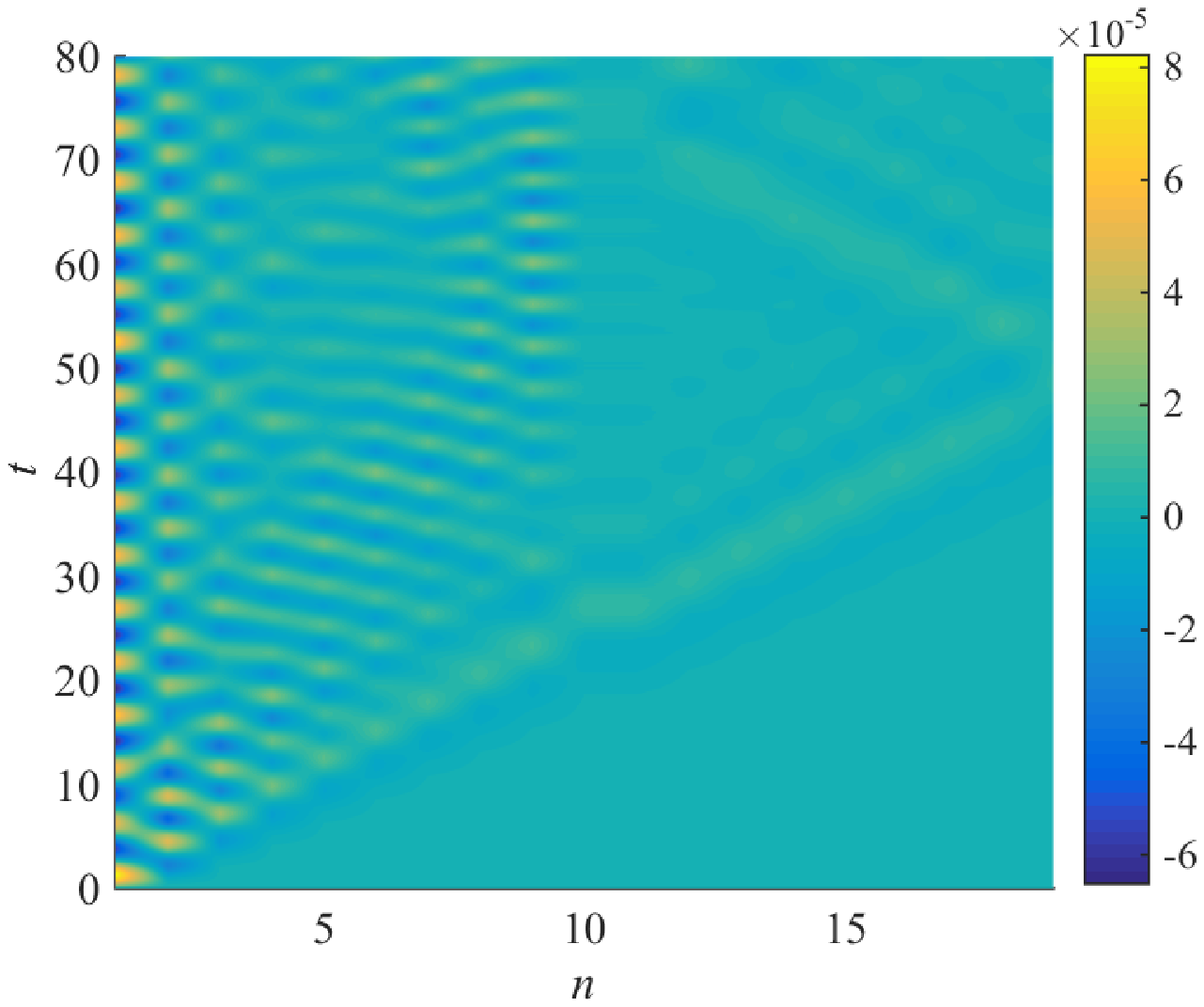}
\includegraphics[scale = 0.5,trim={0.28cm 0 1cm 0},clip] {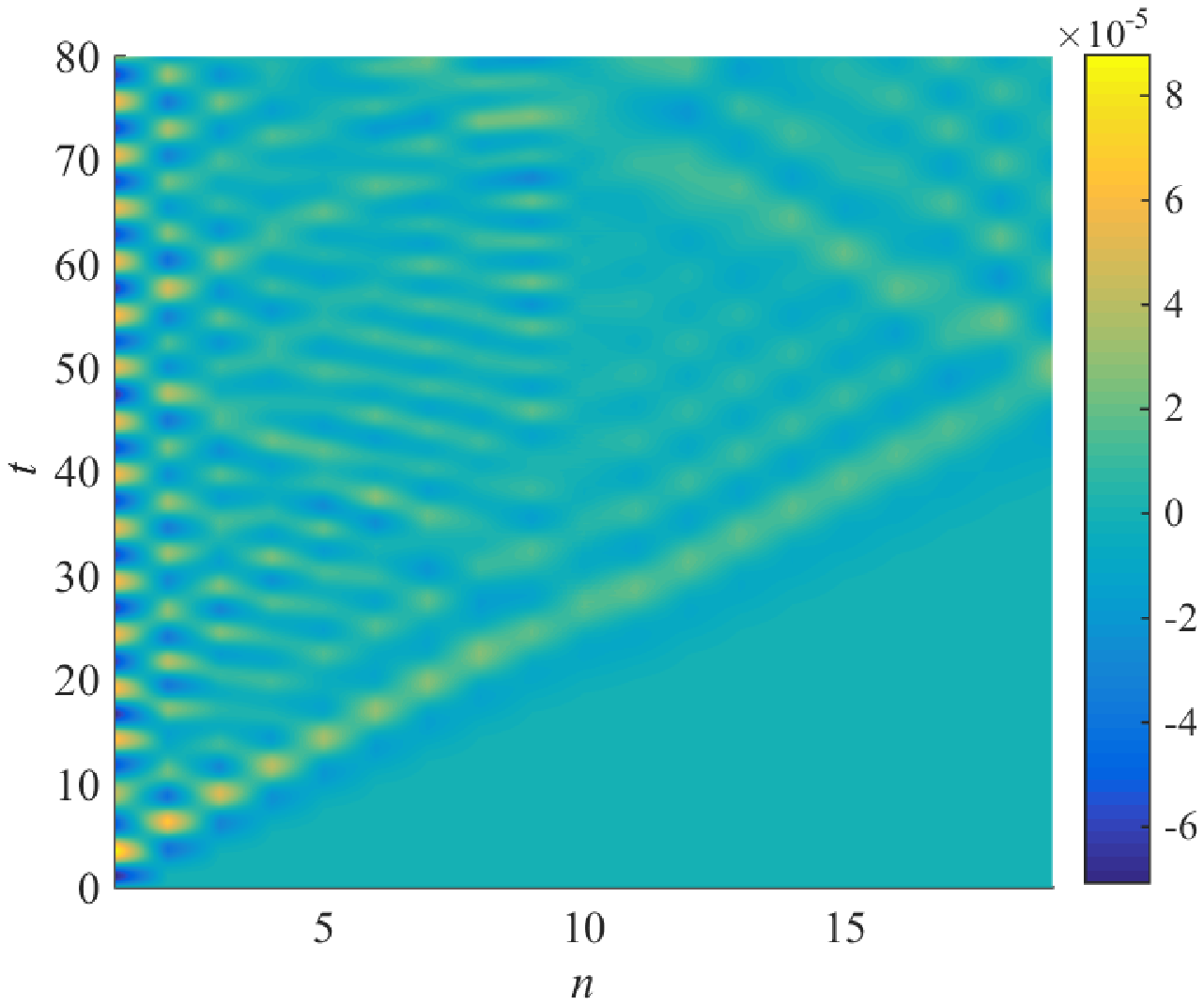}
\includegraphics[scale = 0.5,trim={0.28cm 0 1cm 0},clip]{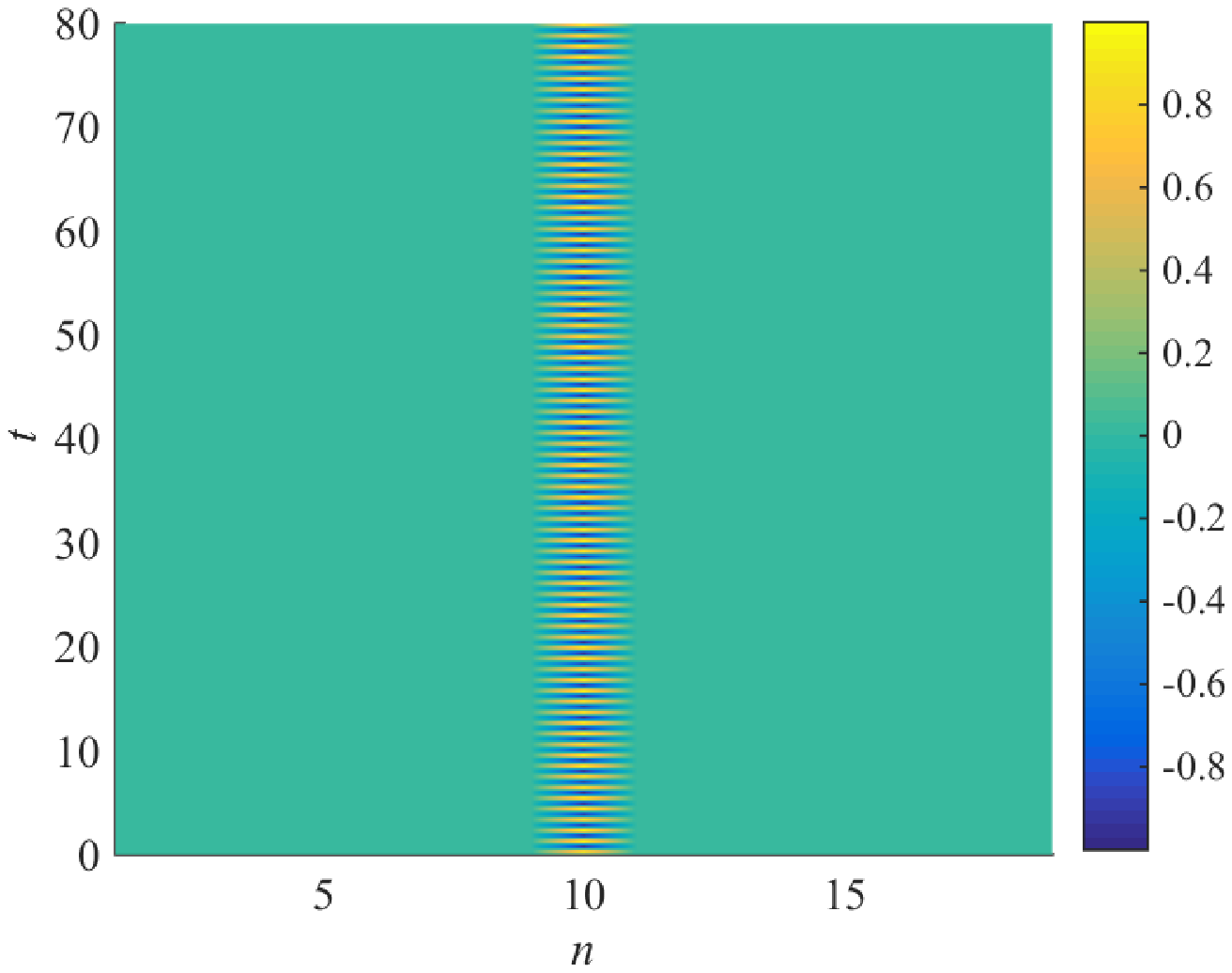}
\includegraphics[scale = 0.5,trim={0.28cm 0 1cm 0},clip]{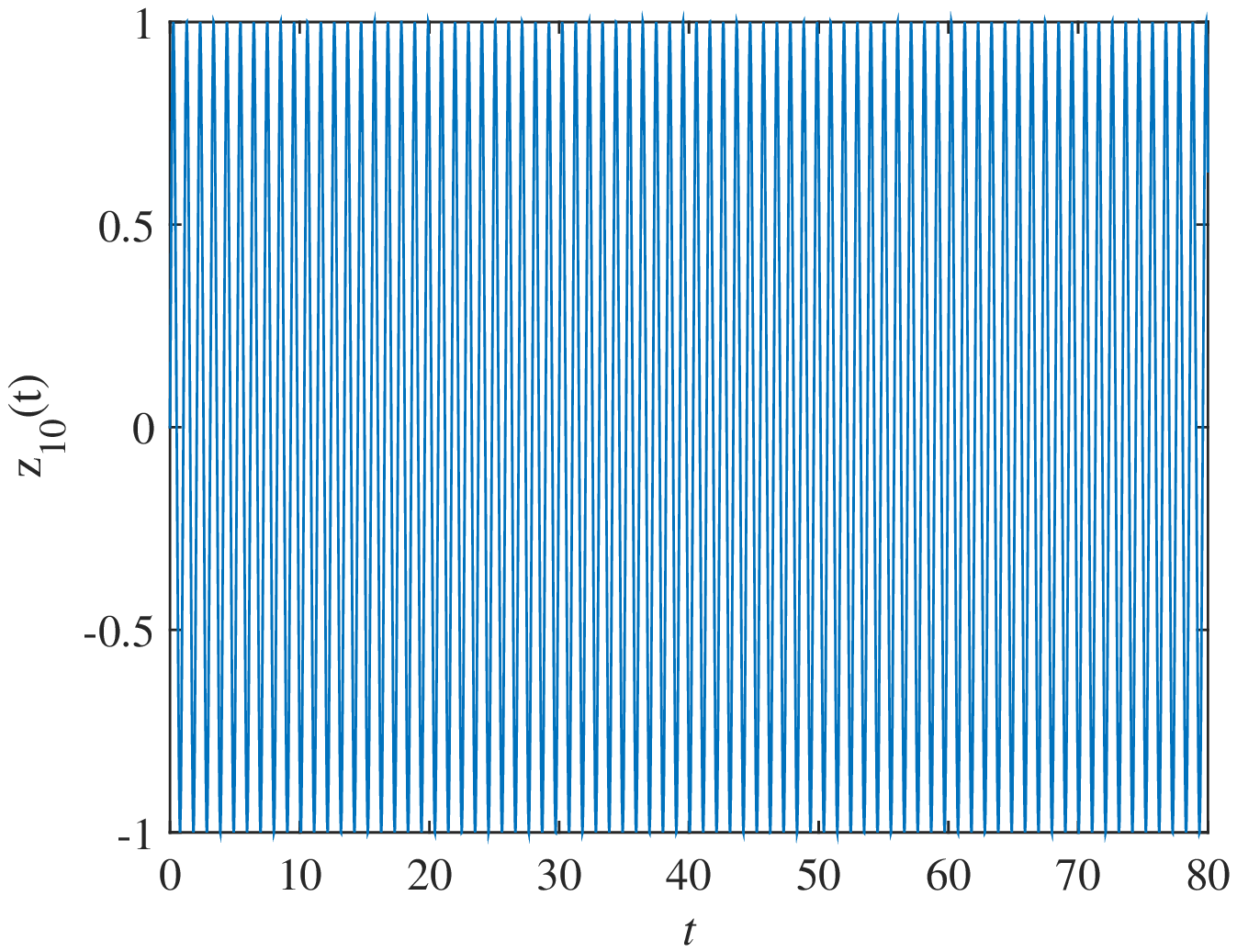}}
\end{center}
\caption{\small Central-site CB stability validation for $N=19,\mu=\frac{1}{2},\alpha=100,x'(1)|_{t=0}=-y'(1)|_{t=0}=10^{-4}$. The plots show propagation maps of (top to bottom) $x_n(t)$, $y_n(t)$, $z_n(t)$, and also $z_{10}(t)$}
\label{Fig8}
\end{figure}

Figure \ref{Fig8} illustrates the stability of a central CB of large amplitude ($\alpha=100$) with respect to a small zero-momentum perturbation introduced at the left boundary (by setting $x'(1)=-\frac{y'(1)}{2\mu}=10^{-4},\mu=\frac{1}{2}$).

\begin{figure}[H]
\begin{center}
{\includegraphics[scale = 0.5,trim={0.28cm 0 1cm 0},clip]  {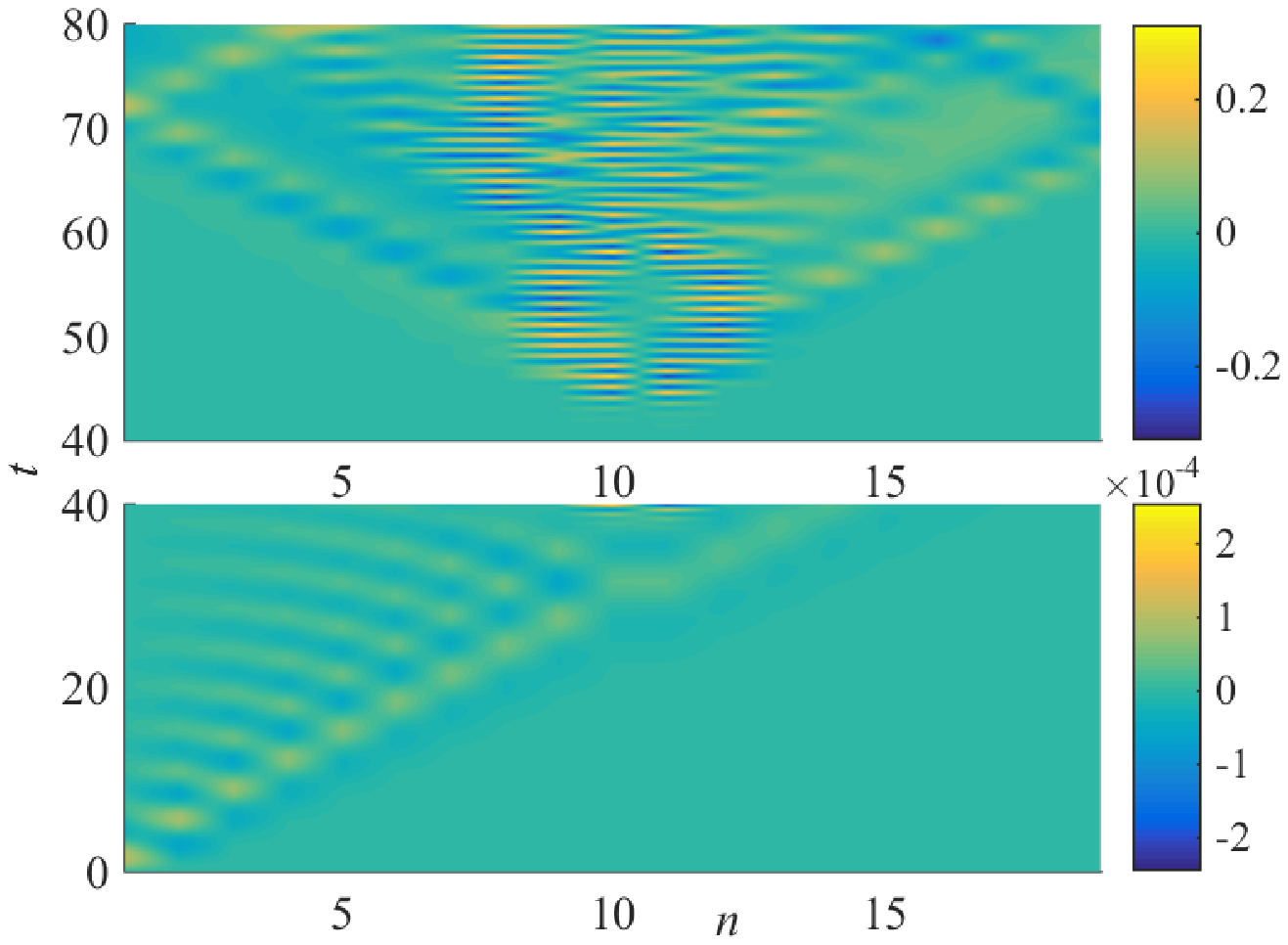}
\includegraphics[scale = 0.5,trim={0.28cm 0 1cm 0},clip]   {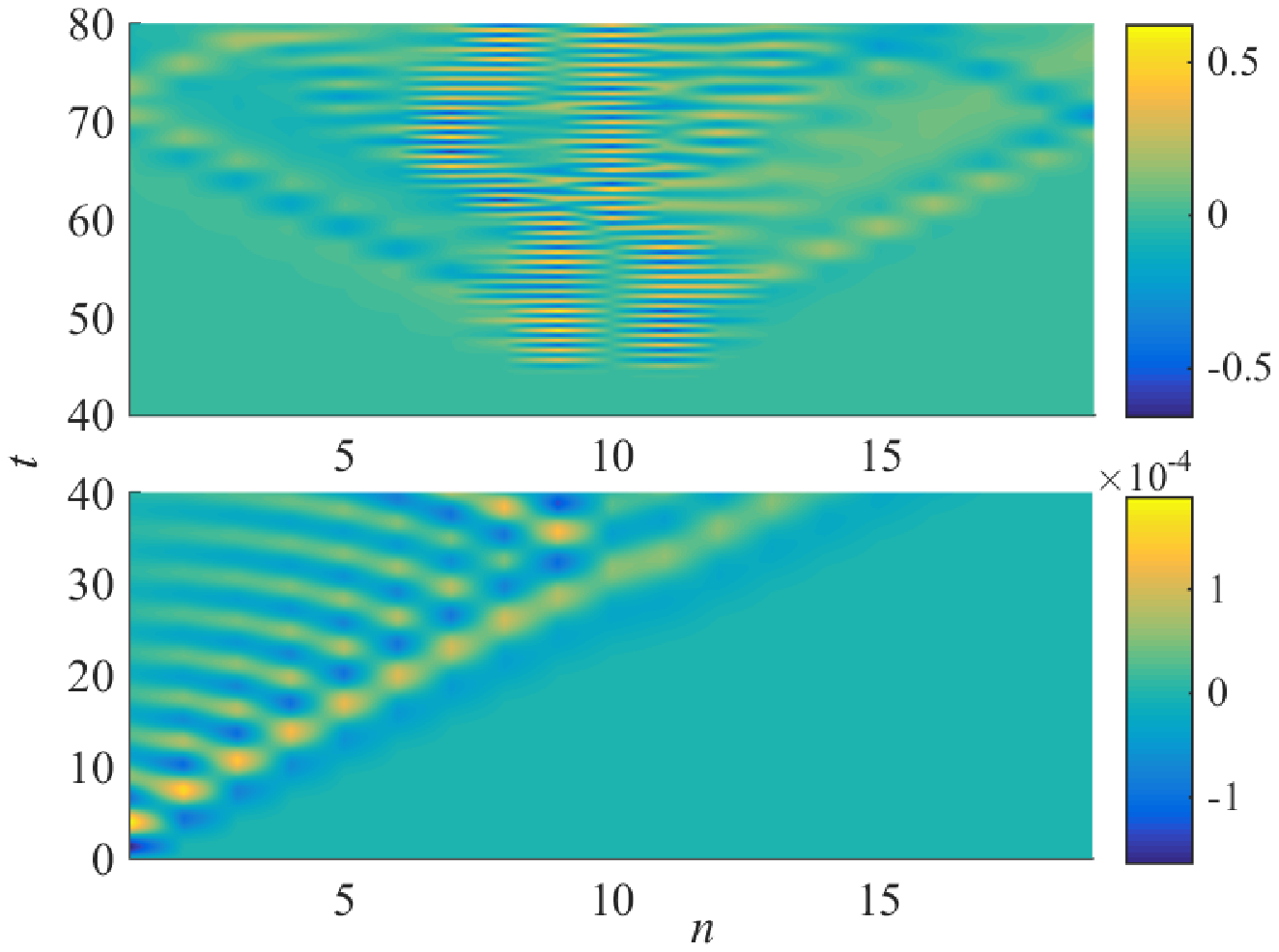}
\includegraphics[scale = 0.5,trim={0.28cm 0 1cm 0},clip]{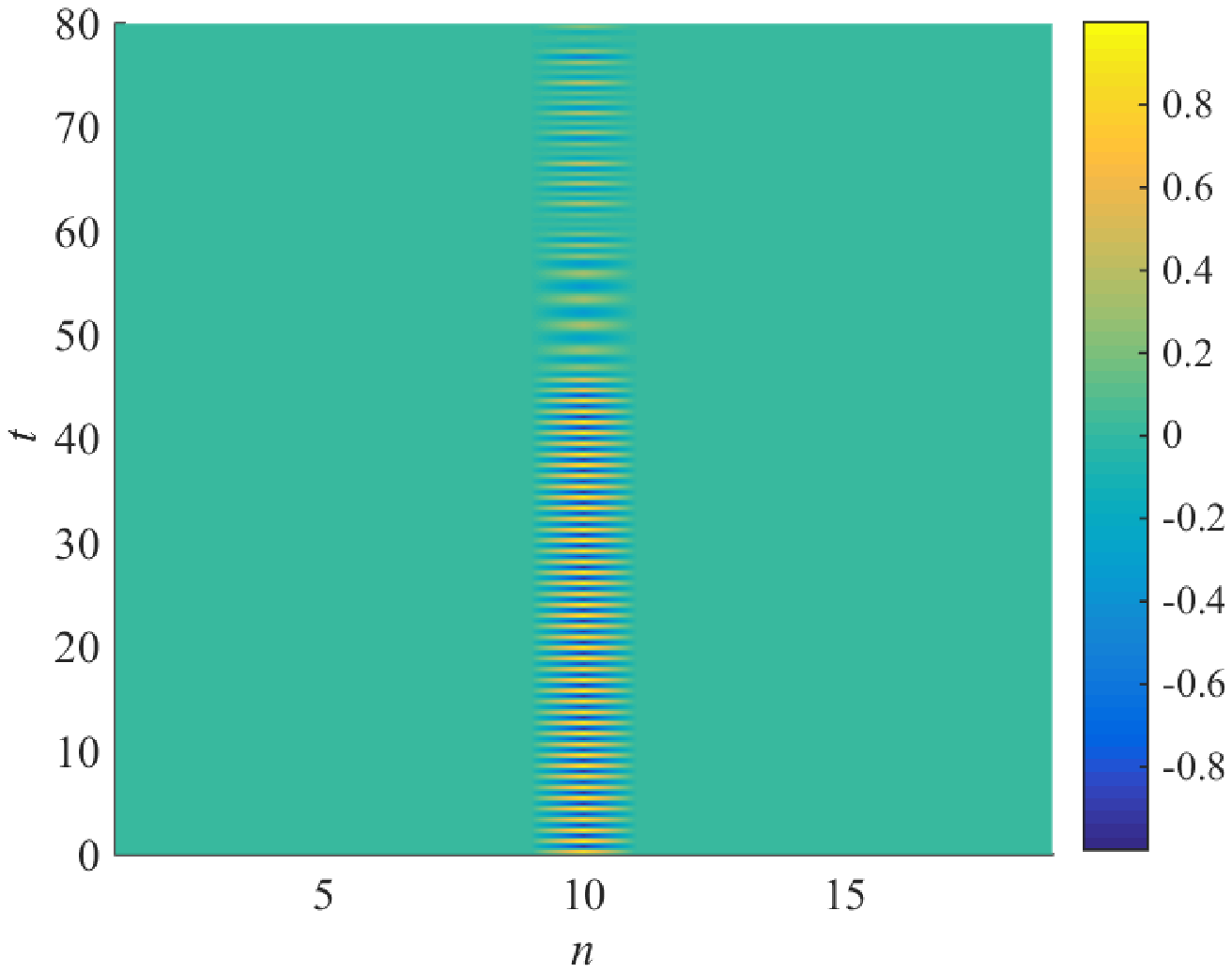}
\includegraphics[scale = 0.5,trim={0.28cm 0 1cm 0},clip]{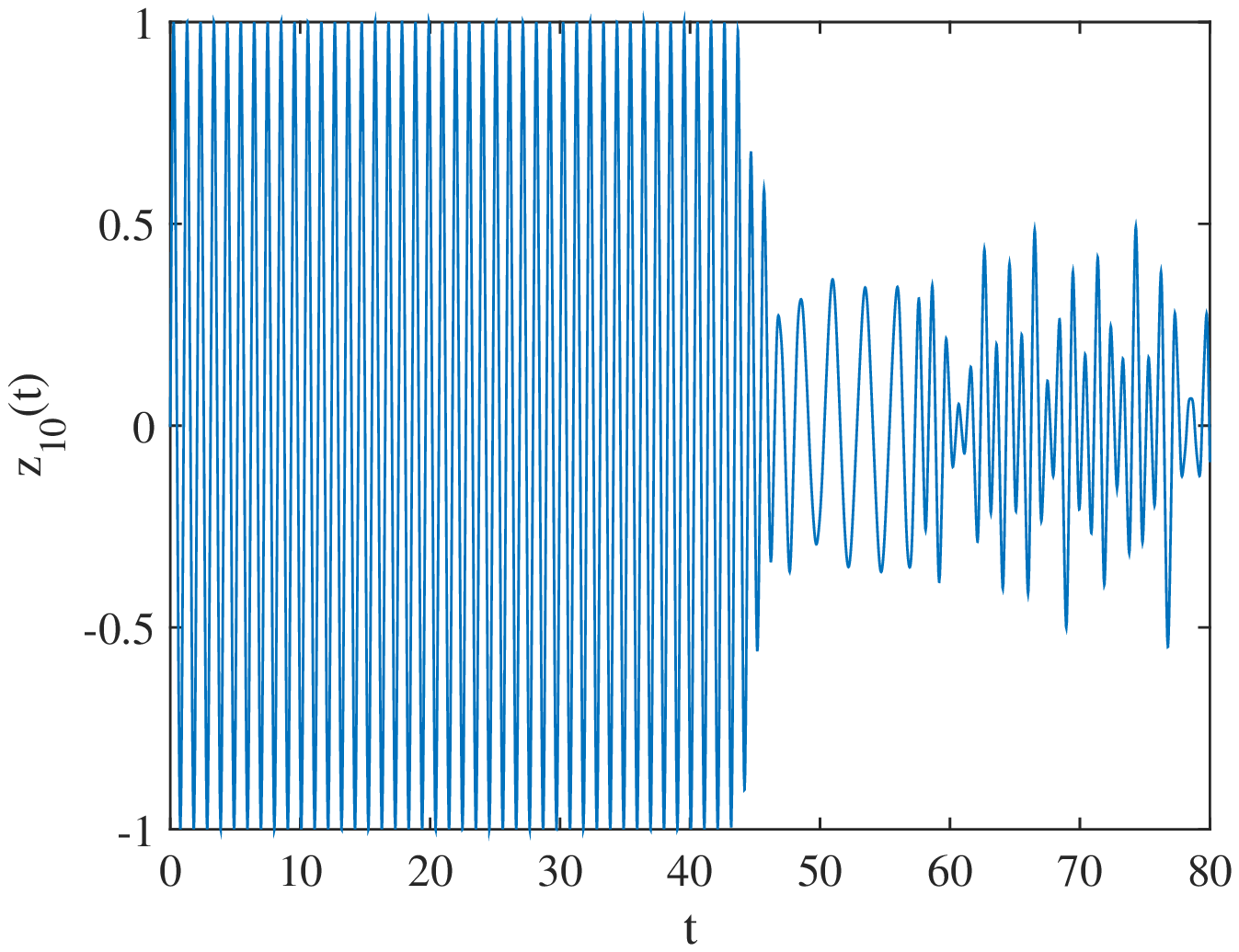}}
\end{center}
\caption{\small Central-site CB destruction for $N=19,\mu=1,\alpha=100,x'(1)|_{t=0}=-\frac{y'(N)|_{t=0}}{2}=10^{-4}$. The plots show  propagation maps of (top to bottom) $x_n(t)$, $y_n(t)$, $z_n(t)$, and also $z_{10}(t)$}
\label{Fig9}
\end{figure}

Figure \ref{Fig9} shows the destruction of a central CB of large amplitude ($\alpha=100$) once it is hit with a small zero-momentum perturbation introduced at the left boundary (by setting $x'(1)=-\frac{y'(1)}{2\mu}=10^{-4},\mu=1$).

It is thus demonstrated (in Figs. \ref{Fig8} and \ref{Fig9}) that a stable large-amplitude CB exists in the considered chain for $M=m$, and that for other mass ratios, such as $M=2m$, for example, the CB can be destroyed by phonons.

\subsubsection{Compact breather stability -- the moderate-amplitude regime}

\begin{figure}[H]
\begin{center}
\includegraphics[scale = 0.6]{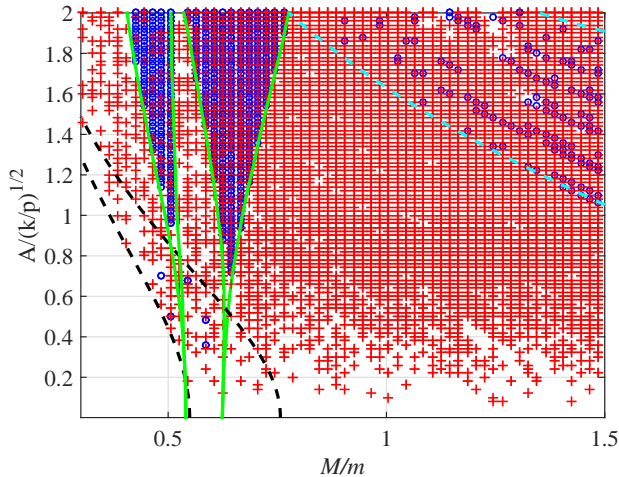}
\end{center}
\caption{\small Central-site CB stability map for $N=19$ for moderate amplitudes. Pitchfork instability points are denoted by (blue online) circles. Neimark-Sacker instability points are denoted by (red online) `+' symbols. The dashed curves (cyan online) bound the zone of resonance between symmetric-mode propagation frequencies and the main CB frequency. Solid (green online) curves represent the boundaries of the two main pitchfork instability tongues as calculated for the $N=3$ case. The dashed black curves at the lower left corner represent the boundaries of resonance of propagation frequencies (`optical' phonons) with the \emph{second} CB-frequency.}
\label{Fig11}
\end{figure}

Observation of Fig. \ref{Fig7} in a range of moderate amplitudes reveals an interesting picture. It appears that one can have reasonable prediction of pitchfork instability. Figure \ref{Fig11} shows that the boundaries of the two main pitchfork instability tongues, as calculated for the $N=3$ case, accurately represent the analogous regions for the $N\gg 1$ ($N=19$) case, implying that the associated instability mechanism is local. 

Moreover, the pitchfork-instability regions lying outside the two aforementioned tongues seem to be related to resonance of the CB frequencies with linear phonons. Those regions seem to form quasi-fractal sets bounded by amplitude--mass-ratio curves corresponding to CB frequencies falling within the propagation (optical) range (as calculated in the spectral analysis section). Interestingly enough, these instability zones, which are related to the CB resonance with phonons, seem to be manifested through pitchfork instability. In addition, there seem to emerge Hopf-bifurcation related instability zones that cannot be predicted by linear-spectrum CB analysis. The strong dependence of the geometry of the latter on the number of elements in a chain (as can be learned from comparison of Figs. \ref{Fig6} and \ref{Fig11}), suggests that these zones can be associated with resonance between the CB and \emph{weakly-nonlinear} propagating waves.

The dashed curves in Fig. \ref{Fig11} are obtained by comparing $1$- and $2$-multiples of the frequency given in Eq. (\ref{eq22}), using Eqs. (\ref{eq4}) and (\ref{eq18}), with the $q=\lbrace 0,\pi\rbrace$ limits of Eq. \ref{eq3}, taking the `+' sign for the $\mu \le 1$ case, as follows:

\begin{equation}
\label{eq37}
\begin{split}
\mu^{-}(\alpha)=\frac{4}{\pi^2 n^2}\frac{\left[K\left(\frac{\alpha/2}{\alpha+2}\right)\right]^2}{1+\alpha/2}, \\ \mu^{+}(\alpha)=\left\lbrace\frac{\pi^2 n^2}{4}\frac{1+\alpha/2}{\left[K\left(\frac{\alpha/2}{\alpha+2}\right)\right]^2}-1\right\rbrace^{-1} \ ; \ n=1,2
\end{split}
\end{equation}

\subsubsection{Compact breather stability -- the large-amplitude regime}

\begin{figure*}[ht]
\begin{center}
{\includegraphics[scale = 0.65,trim={0cm 4.75cm 14cm 4cm},clip]{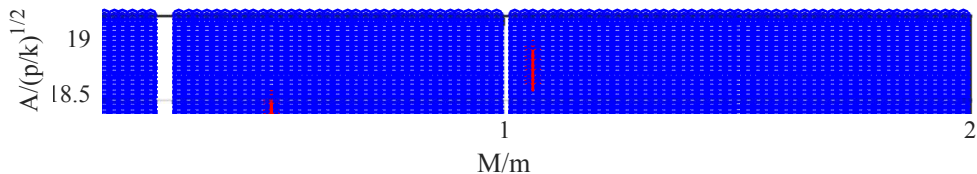}
\includegraphics[scale = 0.65,trim={1cm 8.75cm 13cm 0cm},clip]{N19Map_a_001_2_LT20CHp.eps}
\includegraphics[scale = 1,trim={2cm 9cm 0cm 0cm},clip]{N19Map_a_001_2_LT20CHp.eps}\\

\includegraphics[scale = 0.65,trim={4.85cm 0.6cm 9cm 10.0cm},clip]{N19Map_a_001_2_LT20CHp.eps}
\includegraphics[scale = 0.65,trim={5cm 0.6cm 9cm 10.0cm},clip]{N19Map_a_001_2_LT20CHp.eps}
\includegraphics[scale = 0.65,trim={5cm 0.6cm 9cm 10.0cm},clip]{N19Map_a_001_2_LT20CHp.eps}
\includegraphics[scale = 0.65,trim={5cm 0.6cm 9cm 10.0cm},clip]{N19Map_a_001_2_LT20CHp.eps}
\includegraphics[scale = 0.65,trim={5cm 0.6cm 9cm 10.0cm},clip]{N19Map_a_001_2_LT20CHp.eps}
\includegraphics[scale = 0.65,trim={5cm 0.6cm 9cm 10.0cm},clip]{N19Map_a_001_2_LT20CHp.eps}
\includegraphics[scale = 0.65,trim={5cm 0.6cm 9cm 10.0cm},clip]{N19Map_a_001_2_LT20CHp.eps}
\includegraphics[scale = 0.65,trim={5cm 0.6cm 7cm 10.0cm},clip]{N19Map_a_001_2_LT20CHp.eps}
\includegraphics[scale = 0.65,trim={4.7cm 0.6cm 9cm 10.0cm},clip]{N19Map_a_001_2_LT20CHp.eps}
\includegraphics[scale = 0.65,trim={5cm 0.6cm 9cm 10.0cm},clip]{N19Map_a_001_2_LT20CHp.eps}
\includegraphics[scale = 0.65,trim={5cm 0.6cm 9cm 10.0cm},clip]{N19Map_a_001_2_LT20CHp.eps}
\includegraphics[scale = 0.65,trim={5cm 0.6cm 9cm 10.0cm},clip]{N19Map_a_001_2_LT20CHp.eps}
\includegraphics[scale = 0.65,trim={5cm 0.6cm 9cm 10.0cm},clip]{N19Map_a_001_2_LT20CHp.eps}
\includegraphics[scale = 0.65,trim={5cm 0.6cm 9cm 10.0cm},clip]{N19Map_a_001_2_LT20CHp.eps}
\includegraphics[scale = 0.65,trim={5cm 0.6cm 9cm 10.0cm},clip]{N19Map_a_001_2_LT20CHp.eps}
\includegraphics[scale = 0.65,trim={12cm 0.6cm 0cm 10.0cm},clip]{N19Map_a_001_2_LT20CHp.eps}\\

\includegraphics[scale = 0.65,trim={2cm 0cm 0cm 10.5cm},clip]{N19Map_a_001_2_LT20CHp.eps}}
\end{center}
\caption{\small Central-site CB stability map for $N=19$ for large amplitudes. Pitchfork instability points are denoted by (blue online) dots. Neimark-Sacker instability points are denoted by (red online) `+' symbols (found only around $M/m=0.5$ and slightly above $M/m=1$)}
\label{Fig12}
\end{figure*}

The stability map for large amplitudes and mass ratios in the range $0.13<M/m<2$ (which is the range examined throughout this work) is shown in Fig. \ref{Fig12}. Aside from intractable Neimark-Sacker bifurcation-related instability points (shown in red online), there are two noteworthy features in the presented map, standing-out on an otherwise instability-indicating background. One feature is the finite stability gap situated around $M/m\approx 0.275$. This gap is related to nothing mysterious. It is simply the large-amplitude limit of a stable region between two instability tongues starting from the zero-amplitude axis and expanding upwards. These tongues, situated on the left and on the right of the aforementioned finite gap, are the continuation of the two principal-instability tongues shown in Fig. \ref{Fig11}, with the green (online) boundaries being estimates corresponding to the three-elements chain. 

Indeed, here, in the large-amplitude limit, too, just as in the small-to-moderate amplitude limit, analysis of the three-element system shows the same pitchfork bifurcation-related stability bounds that yield the finite gap of stability situated around $M/m\approx 0.275$. The difference between the $N=3$ case and the $N=19$ case is less than one percent. Therefore, in regard with this first (aforementioned) large-amplitude seemingly-tractable feature, there is no increase in pattern-complexity due to increase in the system size. In addition to this quantitative observation, it can be noted that even for a smaller system (that of a single representative element) it was already shown -- by rigorous asymptotic analysis of a rigorously derived Hill equation --  that a sequence of instability-tongues can be observed. For this sequence, one finds that the boundaries of the emerging instability tongues have vertical asymptotes, the quantitative description of which is provided. 

This is qualitatively similar to the reality that can be observed in Fig. \ref{Fig12}, with its vertical finite-width stripe corresponding to stability. 

Therefore, it may be concluded that the first aforementioned phenomenon, the finite gap between the two pitchfork bifurcation-related instability tongues, appears to be tractable. 


This result is expected (or, at least, reasonable), given that the two instability tongues surrounding the aforementioned gap are the two principal instability tongues, related to the two local modes with which the CB mode can locally exchange stability (in the considered three-on-site-elements chain). In principle, resonance-related instability-tongues are usually characterized by linearly-distinct boundaries. In the natural, $\alpha$--$\mu$ parametrization, these boundaries are indeed linearly distinct. In the amplitude--mass-ratio plane, however, one, obviously, observes a square-root-like initial increase in the tongue boundaries, as appears in Fig. \ref{Fig11}, hiding the linear distinction (between the boundaries).

The second, pitchfork bifurcation-related (and thus more ordered) distinct feature in the stability map in Fig. \ref{Fig12} is an infinitesimal-width stability stripe, situated, as it appears, at $M=m$. This is a noteworthy phenomenon that has the potential of being tractable. First of all, the stability at $M=m$ can be observed already in the analysis of the $N=3$ case. Moreover, the same argumentation as before holds for the boundaries of the principal local-instability-related tongues and their vertical asymptotes (this time, however, for the right tongue and its right boundary, rather than for the left tongue and its left boundary, as for case of the finite-width stability stripe).

However, here, further explanation would be of value, since a zero-width stability-gap is an unusual phenomenon. Usually, collision between boundaries of instability tongues is related to degeneracies in the system. In the considered case, a degeneracy clearly occurs for $M=m$, when two of the three nonlinear normal modes in the system become interchangeable. Such a degeneracy does not necessarily have to lead to collision of boundaries, let alone to an asymptotic one. In fact, it only occurs here for large amplitudes, which suggests that it is not only the mass symmetry that is at play. Rather, one should suspect that the phenomenon is related to combination between mass-symmetry and high energy, or strong nonlinearity. Strong nonlinearity in the interaction between the masses brings to mind the phenomenon of impact, and indeed one immediately notices that for $M=m$, impact-interaction between the masses has a special feature of conserving periodicity of antisymmetric local vibration. One would therefore find it reasonable to consider a limit-case analogy, where nearest-neighbor interaction in the chain is modeled by linear springs connected in parallel to strongly-nonlinear springs. However, those strongly-nonlinear springs would not be smooth ones, modeled by a power-law dependence on displacement, but rather ones of the vibro-impact kind.

\section{The vibro-impact limit}
\label{Sec5}
\subsection{Problem formulation}

The (dimensionless) equation of motion for the antisymmetric compact periodic mode in the smooth system would be:
\begin{equation}
\label{eq5.1}
\hat{z}''+\hat{z}+\frac{\alpha}{2}\hat{z}^3=0
\end{equation}

For a smooth description of a linear system augmented by a vibro-impact potential (instead of the quartic potential), the equation of motion would have the form \cite{Perchikov2015}
\begin{equation}
\label{eq5.2}
\hat{z}''+\hat{z}+\frac{\alpha}{2}\hat{z}^{\alpha-1}=0
\end{equation}
where $\alpha$ should be large.

One observes that the two equations become identical for $\alpha=4$, which is just the upper limit of the vertical axis in Fig. \ref{Fig11}. Thus, the general form of the smooth version of the vibro-impact system is equivalent to that of the cubic system for $\alpha=4$. However, this choice is only representative of actual impact if one agrees that 4 is a large enough number with respect to unity, which is a certain stretch. Nevertheless, an interesting result can be obtained if such stretched assumption is made.

One would therefore opt to explain the ``large''-amplitude limit of the stability map in Fig. \ref{Fig7} using the, perhaps, more tractable vibro-impact version of the considered system (having in mind specifically the $M=m$ case).

We thus assume a modified system, in which the interactions are delivered by coupling consisting of linear springs attached in parallel to vibro-impact `sleeves', allowing maximum relative (normalized) displacement of unity between the \emph{connected} masses.

\begin{figure*}[ht]
\begin{center}
{\includegraphics[scale = 0.4,trim={0 0cm 0 0cm},clip]{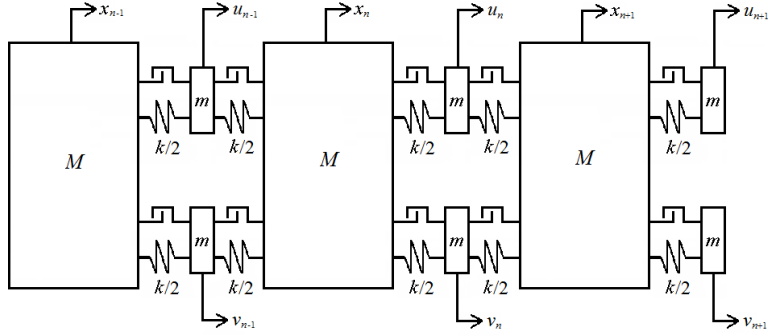}}
\end{center}
\caption{\small Sketch of the 1D horizontal chain with linear links and vibro-impact `sleeves' between masses $m$ and $M$, where the maximum relative free displacement between mass $m$ and any of its neighboring $M$ masses is $\Delta_{\alpha}$}
\label{Fig13}
\end{figure*}

The dispersion bands for this system are the same as for the first system addressed in the beginning of this paper. However, the stability of a CB solution becomes an interesting problem here, with reasonable inference on the stability of the original system, for the limit of large (but finite) amplitudes.

Figure \ref{Fig12} shows that there seems to be a unique isolated (`punctured') point of stability in the large-amplitude limit at $M=m$ (within the examined mass-ratios range). What one can do relatively easily is to examine the $M=m$ case, resorting to the system shown in Fig. \ref{Fig13}.

What should be done first is examination of the dynamics of a CB solution located, say, on the central site of the chain. The crucial instance is when the vibro-impact potential comes into action. If for $M=m$ the system is `impact-integrable', borrowing from a concept discussed in \cite{NPOVG2017}, then one can examine the linear stability of the CB using Floquet theory, employing analytical construction of the monodromy matrix. One should start by assuming a central-site CB, having the mass associated with displacement $u_{n_0}$ labeled as, say, element `1'; the mass associated with displacement $v_{n_0}$ labeled as, say, element `2'; the mass associated with displacement $x_{n_0}$ labeled as, say, element `3'; and the mass associated with displacement $x_{n_0+1}$ labeled as, element `4'. 

It could then be assumed, without loss of generality, that the initial condition is $u'_{n_0}(t=0)=-V_0,v_{n_0}(t=0)=V_0-\left|\mathcal{O}(\epsilon)\right|, x_{n_0}(t=0)=\left|\mathcal{O}(\epsilon)\right|, x_{n_0+1}(t=0)=\left|\mathcal{O}(\epsilon)\right|$. In line with this initial condition, one can assume, due to symmetry, that after elements `1' and `2' pass the distance $\Delta_{\alpha}$ each, a sequence of impacts begins, with the first impact occurring between elements `1' and `3'. Since the masses of elements `1' and `3' are identical (and even if they are only nearly identical), the result of the impact is that element `1' obtains infinitesimal velocity, and element `3' obtains approximately the velocity $-V_0$ (that is $V_0$ `leftwards'). 

The next impact has to involve element `2'. There are two options. One is impact between `2' and `3', and the other is impact between `2' and `4'. In the former case, after impact between `2' and `3', which are two elements of equal (or approximately equal) mass, moving with (approximately) opposite velocities, the result of the impact is the reversion of the directions of the two elements. In other words, after the second impact, the one between `2' and `3', element `2' obtains approximately the velocity $-V_0$. 

The third impact can then occur only between element `3', positioned to the left of element `1' and now moving towards element `1', and element `1', which has infinitesimal velocity. This impact, again, makes the elements `1' and `3' swap their velocities (even if only approximately).

Consequently, after this third impact, element `3' has infinitesimal velocity, while element `1' has approximately the velocity $V_0$ (that is, $V_0$ `rightwards'). Thus, after three impacts, the velocities of the two elements of the CB are reversed and the neighboring elements remain approximately at rest. This possibility can be labeled as sequence `1-3,2-3,3-1'.

The second possibility is that the second impact occurs between elements `2' and `4'. After this impact, the velocity of element `2' becomes infinitesimal, and the velocity of element `4' becomes approximately $V_0$, due to the approximately equal masses of elements `2' and '4'. The third impact in this version can be either between `4' and `1' or between `3' and `2'. There is no significance to the order. It can be assumed that the impact between `2' and `3' occurs first. This impact makes `3' and `2' approximately swap their velocities. Consequently, element `2' starts moving with velocity $-V_0$, and element `3' resorts back to infinitesimal velocity. Finally, element `4', which after the impact with element `2' moves with velocity $V_0$, now impacts element `1', which has infinitesimal velocity. Consequently, element `1' obtains velocity $V_0$ (`rightwards') and element `4' resorts back to infinitesimal velocity. In this option, as before, the sequence of four impacts reverses the velocities of the two elements of the CB. This possibility can be labeled `1-3,2-4,3-2,4-1' (or, alternatively, `1-3,2-4,4-1,3-2', which is equivalent).

This qualitative analysis shows that for $M/m=1+\mathcal{O}(\epsilon)$, the CB with the vibro-impact nonlinearity is indeed `impact-integrable'. Consequently, the monodromy matrix can be constructed using the saltation matrix. This saltation matrix would have to represent both scenarios, one in which there are three consecutive impact within a short instance, and the other one, in which there are \emph{four} consecutive impacts hidden in the instance of the reversal of the CB velocities. The `impact-integrability' is encompassed in the fact that the result is the same whether the two elements of the CB experience impact strictly simultaneously or, rather, consecutively (in each of the two possible scenarios discussed above), with infinitesimal time lags.

The next step is to construct the two saltation matrices representing the sequences of impacts. According to the result obtained in \cite{NPOVG2017}, under conditions of `impact-integrability', the total saltation matrix representing the entire sequence of consecutive impacts can be constructed as the product of saltation matrices representing the individual impacts, in the order of their occurrence. The expressions for the saltation matrices for the two scenarios, taking advantage of the derivations in \cite{NPOVG2017} (computed for the assumption of perturbations carefully taken to the zero limit), are given in \ref{AppendixB} (along with some additional details pertinent to the application of Floquet theory to vibro-impact systems).

\subsection{Stability-analysis results}

Numerical spectral analysis of the monodromy matrix (analytically-constructed as detailed in \ref{AppendixB}) for the two possible impact scenarios shows interesting results. It appears that both for the three-masses impacts and the four-masses impacts, there is a critical (normalized) frequency corresponding to (subcritical) pitchfork bifurcation. For lower frequencies there are two real eigenvalues, one of which is larger than unity. For higher frequencies all eigenvalues lie on the unit circle. The value of this (first, as shown below) critical normalized frequency is exactly the square root of three. This value is obtained by numerical analysis, in which consecutive digits in the decimal representation of this irrational number were obtained by carefully approaching the bifurcation. This number corresponds precisely to the maximum frequency of the propagation spectrum shown in the linear analysis in the beginning of the present paper. 

This correspondence is logical but not trivial. Indeed, one may logically anticipate that for CB frequencies below the largest propagation frequency, there may be resonance between the antisymmetric compact mode and a small propagating perturbation. This is reasonable. The nontrivial part is that we see here an example of a situation where the pitchfork bifurcation is related not to purely-local instability, but to something else. Rather, the bifurcation is related to instability corresponding to resonance between a local mode and an `anti-local' mode. This `anti-local' mode is the planar-wave propagating perturbation. Quantitatively, this result is summarized in the equation below, where use is made of Eq. \ref{eq3}).
\begin{equation}
\label{Vib1}
\begin{split}
\hat{\omega}_{\text{cr}}^{(\text{PF})}=\lim_{\mu\to1/2}{}\underset{q}{\text{max}}\lbrace{\hat{\omega}_{\text{prop}}(\mu,q)}\rbrace=\\=\lim_{\mu\to1/2}\sqrt{\frac{1+\mu}{\mu}}=\sqrt{\frac{1/2+1}{1/2}}=\sqrt{3}
\end{split}
\end{equation}

In the scenario of impacts involving only three masses, there are no additional bifurcations beyond the aforementioned (fully tractable) one. In contrast to that, in the case of four consecutive instantaneous impacts between four masses, in addition to the discussed bifurcation, which is still present, there appears to be one additional bifurcation, at a higher frequency value. This second bifurcation is also subcritical, representing stability of the periodic antisymmetric solution for high-enough frequency, but it is of the Neimark-Sacker type. Here, for subcritical frequencies there are two eigenvalues that are complex conjugate with absolute values larger than unity and two complex conjugate eigenvalues with absolute values smaller than unity (reciprocal to those of the first pair). For supercritical frequencies there is transition to four complex conjugate eigenvalues on the unit circle. The corresponding (second) critical normalized-frequency value is numerically identified to be
\begin{equation}
\label{Vib2}
\hat{\omega}_{\text{cr}}^{(\text{NS})}=2.8405
\end{equation}

\begin{figure}[ht]
\begin{center}
{\includegraphics[scale = 0.5,trim={0 0cm 0 1cm},clip]{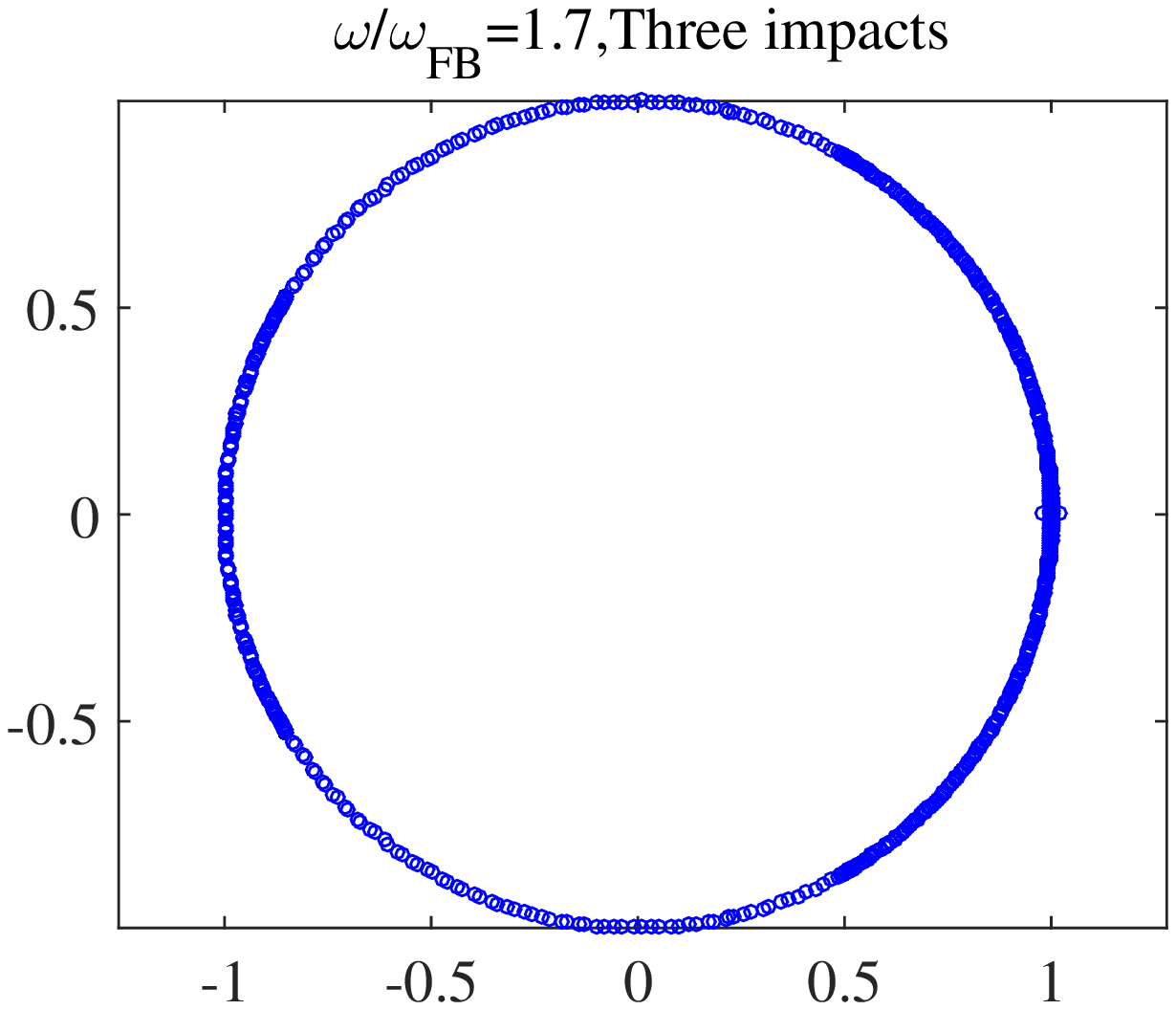} \\
\bigskip
\bigskip
\includegraphics[scale = 0.5,trim={0 0cm 0 1cm},clip]{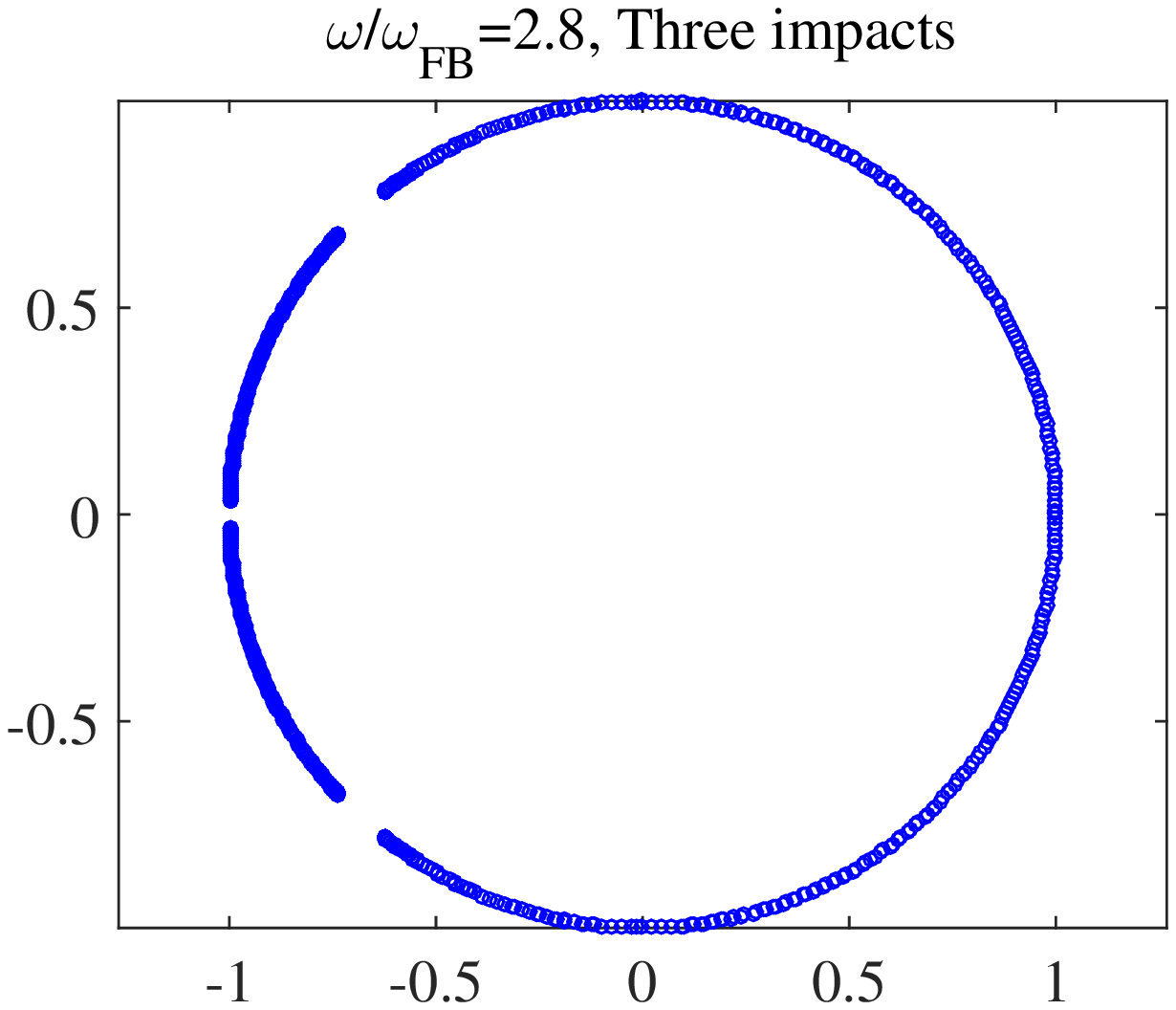}}
\end{center}
\begin{picture}(0,0)
  \put(118,182){Re $\lambda$}
  \put(12,267){Im $\lambda$}
  \put(118,12){Re $\lambda$}
  \put(12,100){Im $\lambda$}
\end{picture}
\caption{\small Monodromy matrix eigenvalues $\lambda$ for $\omega/\omega_{\text{FB}}=1.7$ (top) and $\omega/\omega_{\text{FB}}=2.8$ (bottom) for the three-impacts impact-integrable case for the nonsmooth system}
\label{Fig14}
\end{figure}

\begin{figure}[ht]
\begin{center}
{\includegraphics[scale = 0.5,trim={0 0cm 0 1cm},clip]{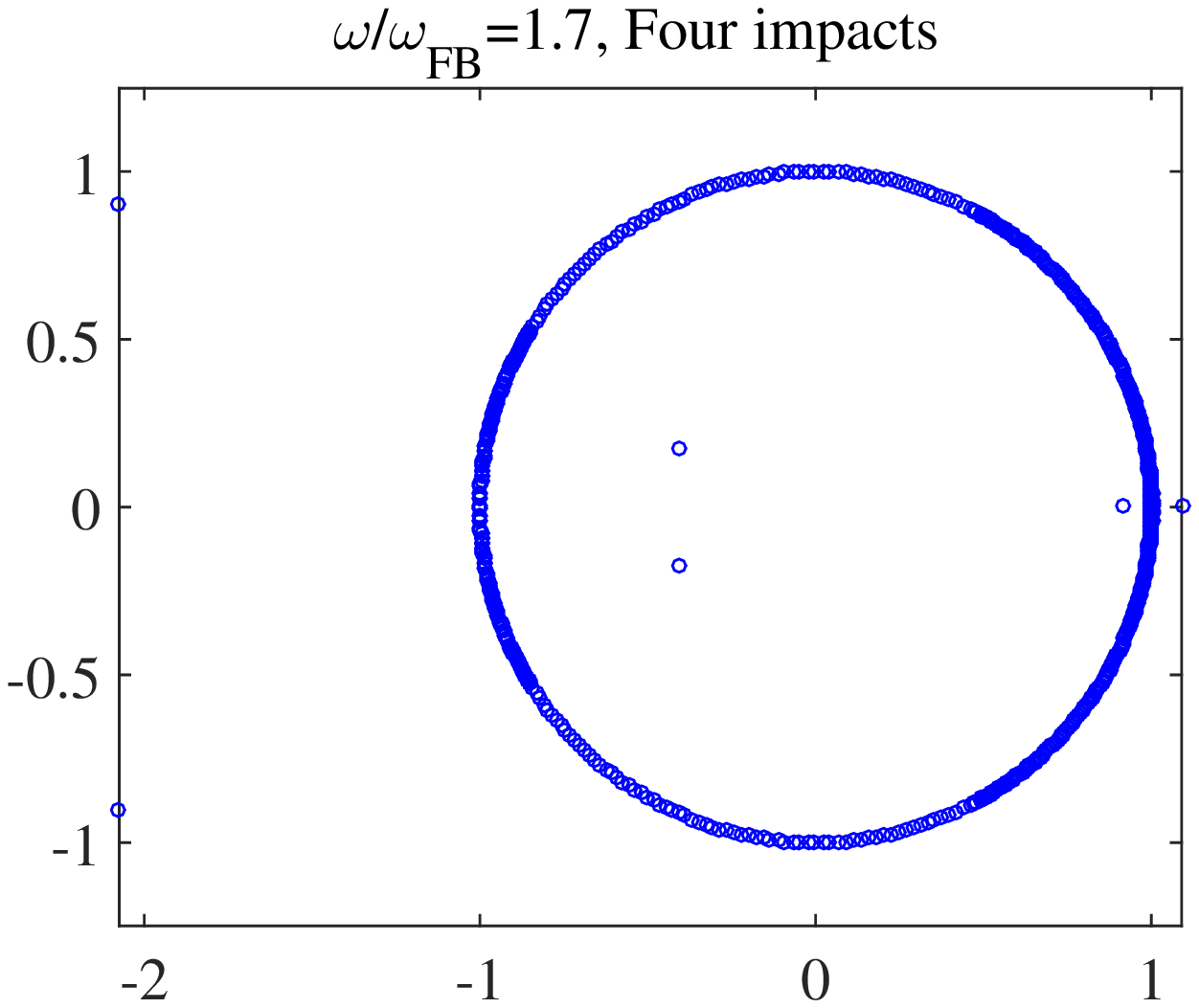} \\
\bigskip
\bigskip
\includegraphics[scale = 0.5,trim={0 0cm 0 1cm},clip]{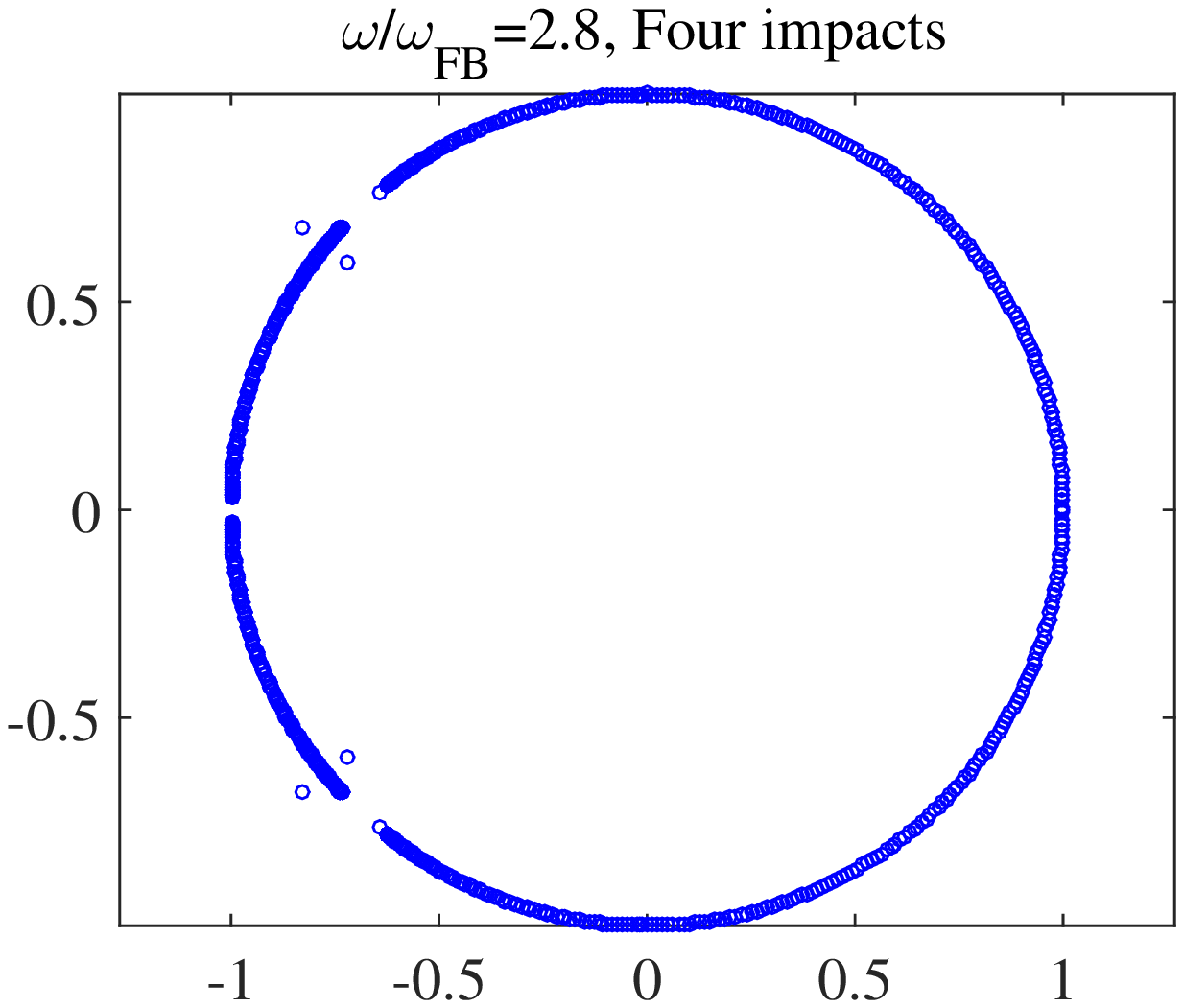}}
\end{center}
\begin{picture}(0,0)
  \put(118,182){Re $\lambda$}
  \put(12,267){Im $\lambda$}
  \put(118,12){Re $\lambda$}
  \put(12,100){Im $\lambda$}
\end{picture}
\caption{\small Monodromy matrix eigenvalues $\lambda$ for $\omega/\omega_{\text{FB}}=1.7$ (top) and $\omega/\omega_{\text{FB}}=2.8$ (bottom) for the four-impacts impact-integrable case for the nonsmooth system}
\label{Fig15}
\end{figure}

Unlike in the former critical-value case, this numerical value is harder to tract analytically by relation to a clear and simple instability mechanism. As before, we see here, again, that instabilities associated with pitchfork bifurcations are tractable, be them local or nonlocal-modes-related, whereas instabilities associated with the Neimark-Sacker bifurcations, are intractable (or, at least, less tractable) by direct observation. Illustration of the two bifurcations is given in Figs. \ref{Fig14}-\ref{Fig15} (for the case of $N=19$, to comply with the cubic-nonlinearity results).

One observes the two real eigenvalues just outside the unit circle in the top plot in Fig. \ref{Fig14}.

One observes two real eigenvalues just outside the unit circle in the top plot in Fig. \ref{Fig15}, corresponding to the pitchfork bifurcation in the four-impacts case. Those eigenvalues move to the unit circle in the plot on the right. In addition, in both plots in Fig. \ref{Fig15} there are two pairs of complex conjugate eigenvalues outside the unit circle, illustrating the Neimark-Sacker-bifurcation-related instability. Those eigenvalues vanish for normalized frequencies higher than the second critical (normalized) frequency as given in Eq. (\ref{Vib2}).

\subsection{Relation to the smooth system -- energy equivalence}

As shown above, pitchfork-bifurcation-related instability in the vibro-impact system, associated with resonance of the compact nonlinear mode with the propagation spectrum, can be avoided for (normalized) frequencies above the value of $\sqrt{3}$. The same is true for the system with the cubic nonlinearity. Substituting $\mu=1/2$ into the upper estimate in Eq. (\ref{eq37}), we obtain the value of $\alpha$ corresponding to which the principal frequency of the CB for $M=m$ in the cubic-nonlinearity case becomes equal to the upper limit of the (normalized) propagation frequencies spectrum, namely, $\sqrt{3}$. Solving this equation for $\alpha$, one obtains the value $\alpha_c=5.49$. This value is about 30 to 35 percents larger than the value 4, associated with the equivalence between the cubic system and a smooth system representing a vibro-impact potential under the assumption that 4 is a large-enough number. For our purposes, an error of 30 to 35 percents is tolerable, being about half an order of discrepancy.

Next, we can write the energy of the system (as it moves in the compact mode) for the cubic nonlinearity, as follows:
\begin{equation}
\label{eqEnc}
E_{\text{cu}}=\frac{1}{2}kA^2\left(1+\frac{\alpha}{4}\right)
\end{equation}

On the other hand, for the vibro-impact limit, the antisymmetric mode in compact periodic dynamics admits displacement which is sinusoidal in time until first impact. The associated time between impacts is related to inverse normalized frequency. This allows one to express the energy of the system as follows (assuming that the maximum displacement before impact is equal to 1 in units normalized the same way as for the cubic system, that is by the amplitude, $A$):
\begin{equation}
\label{eqEnvi}
E_{\text{v-i}}=\frac{1}{2}kA^2\frac{1}{\sin^2{\left(\frac{\pi}{2}\hat\omega^{-1}\right)}}
\end{equation}

Requiring energy equivalence between the cubic and the vibro-impact systems and substituting the critical (squared normalized amplitude) value $\alpha=5.49$, one obtains the critical normalized frequency above which resonance of the compact mode with the propagation spectrum is avoided -- if drawing from the (somewhat stretched) equivalence between the cubic and the vibro-impact nonlinearity types:
\begin{equation}
\label{eqwc}
\hat\omega_c^{\text{v-i/cu}}=2.22
\end{equation}

This is the value above which resonance with the propagation spectrum can be avoided both for the cubic and for the vibro-impact system, if one requires their equivalence. In contrast, if the two cases are treated independently, the corresponding value of the normalized frequency is $\sqrt{3}\approx 1.732$. This means that the self-consistency of the assumption $\alpha=4 \gg 1$, which leads to the 30 to 35 percents error in terms of $\alpha_c$, also leads to about 30 percents of error (0.2836, to be precise) in terms of $\hat\omega_c$. This error is not negligible, but it is also not very large, only half an order, meaning that the calculations are more correct than not, and that some of the essence of the equivalence is indeed captured. Clearly, this error could have been smaller had we solved a system with a \emph{quintic} force-term added to the linear interaction, instead of the cubic term. In that case, instead of the elliptic Jacobi function, a different function would have emerged, with a different frequency-amplitude relation, and the critical amplitude for avoiding resonance with the propagation spectrum would have had a different value. This value, used in the energy equivalence relation, would have yielded back, most probably, a value of the critical frequency much closer to $\sqrt{3}$. But the essence of the matter is qualitatively captured already in the analyzed case. The crucial point in drawing the equivalence with the vibro-impact system is the possibility to explain the isolated stability stripe around $M=m$ in the parameter plane. This isolated stability stripe has been shown here to emerge from considerations of impact-integrability for the implicitly-related nonsmooth limit.

In order to conclude the analysis, it would be valuable to look whether the smooth system has additional zero-width vertical stripes of stability in the parameter plane, for values of $M/m$ other than 1, at least in the considered range $M/m\in[0.13,2]$. Close observation of the numerical results of the application of Floquet theory gives a negative answer. However, in principle, a zero-width stripe could have been numerically hidden `between' the grid points. Thus, additional qualitative analysis might be welcome. Such analysis is described in the following subsection.

A final noteworthy remark can be given in relation to the equivalence between the smooth and the vibro-impact systems. We observe that the $\alpha \gg 1,M=m$ region is exclusively stable within its neighborhood. The $\alpha \gg 1$ case corresponds to the \emph{high-energy} limit in the smooth system, as can be learned from Eq. (\ref{eqEnc}). Without having formal equivalence in the equations of motion for $\alpha \gg 1$, it may still be interesting to address the question of  stability for the $M=m$ case in the high-energy limit for the vibro-impact system. From Eq. (\ref{eqEnvi}) we see that the only way to obtain high energy in the vibro-impact system is taking the $\hat\omega\gg1$ limit, for which the \emph{energy} scales as $\hat\omega^2$. Thus, the limit in which the vibro-impact system is energetically-equivalent to a high-energy smooth system for $M=m$ (for the case where the compact antisymmetric periodic solution is stable), is the \emph{high-frequency} limit. As previously discussed, above the derived critical frequency, the vibro-impact system for $M=m$ does indeed have its compact periodic antisymmetric solution linearly stable. In this sense, the large-amplitude-limit stable stripe at $M=m$ in the smooth system, associated with the compact periodic antisymmetric mode, can be considered qualitatively explained -- with the help of the nonsmooth-limit analysis.

\subsection{Absence of additional isolated (hidden) stability stripes in the parameter plane}

The existence of additional zero-width stability stripes in the parameter plane for the case of smooth nonlinearity is addressed here by referring to the related vibro-impact system. For the latter system, the parameter plane relevant for the stability diagram is represented by the coordinates $\mu$ and $\hat\omega$. In order for stable regions, as would emerge from application of Floquet theory, to exist in this plane, first, there need to exist $\mu$-values guaranteeing impact-integrability. It was already shown that impact-integrability is guaranteed for $M/m=1$. The question is, are there additional $M/m$ values in the range $[0.13,2]$, for which impact integrability is guaranteed (for a representative element of two masses $m$ positioned between two masses $M$). 

To answer this, we, as before, denote the momenta of the `upper' mass $m$, the `lower' mass $m$, the `left' mass $M$ and the `right' mass $M$ by $p_1,p_2,p_3$ and $p_4$, respectively. Next, it is assumed that one can express $p_4$ in terms of the other three momenta using conservation of total linear momentum for the four masses. Then, expressing $p_3$ (and thus also $p_4$) in terms of $p_1$ and $p_2$ using conservation of energy, we can obtain a two-dimensional mapping for the state vector $\textbf{p}=\lbrace p_1,p_2\rbrace^{\top}$. This mapping can represent all the possible impacts occurring between the four masses. To this end, the mapping would have to be defined by cases, with a different set of $\mu$-dependent coefficients for each of the four possible impacts (1-3,3-2,2-4,4-1). Clearly, this mapping would be nonlinear, due to the nonlinearity of the dependence of the Hamiltonian on the momenta. 

Now, even the one-dimensional nonlinear mapping can yield chaotic trajectories, let alone a two-dimensional one. Therefore, whether the exact-inversion mapping applied to the couplet $\lbrace -1,1\rbrace^{\top}$ exists for a specific value of $\mu$, would depend on the number of impacts occurring, and their order. This is not an easy problem to solve, and had we wanted to find a positive answer, we would have been forced to go the hard way of direct enumeration of the possible sequences. For each value of $\mu$, starting from $\textbf{p}=\lbrace p_1,p_2\rbrace^{\top}=\lbrace -1,1\rbrace^{\top}$, we would have been forced to check all possible combinations of the feasible impacts, examining the final value of the vector $\textbf{p}$ in each case and looking for the results that yield $\lbrace 1,-1\rbrace^{\top}$. 

Luckily, we do not have to follow this difficult route. As numerical analysis of the smooth system shows, there appear to be \emph{no} additional impact-integrable $\mu$-values in the examined range. This means that if the vibro-impact limit is any indicator, the instantaneous impact-dynamics of the four masses for the compact periodic antisymmetric mode is \emph{not} impact-integrable. This  means that not \emph{all} possible trajectories in the phase-space of this system are periodic. This implies that at least some trajectories are non-periodic. This, in turn, means that it would be sufficient to find a non-periodic trajectory in a sub-space of the phase-space of the four-mass system, for $\mu$ in the examined range, to prove the absence of additional isolated zero-width stable stripes in the parameter plane. This sub-space could be associated with specific initial conditions manifested as a specific infinitesimal perturbation of the antisymmetric compact mode.

A reasonable choice would be perturbation of the initial conditions that would lead to a sequence of impacts involving not all four but only \emph{three} masses. Then, instead of four different feasible impacts (on the order of occurrence of which one has to wonder), there would be only two `types' of impact, occurring one after the other.

This is a reasonable scenario. If we assume that the initial leftward velocity of particle 1 is slightly larger than the rightward velocity of particle 2, then the first impact would be between particles 1 and 3. Then, not only for the case of $M=m$, as mentioned earlier, but, in fact, for any mass ratio, the second impact would be between particles 3 and 2. 

The reason for this is that after the first impact, the distances between the impacting parts of particles 2 and 4 and those of particles 2 and 3 are the same. However, the speed of approach of particles 3 and 2 is larger than that of particles 2 and 4. The latter fact owes to the negative velocity that particle 3 would have after the first impact (for \emph{any} mass ratio). 

After the impact between particles 3 and 2, the speed of particle 2 would decrease. If the decrease is sufficient, impact with particle 4 could be avoided. In addition, particle 4 has to have initial negative infinitesimal displacement, large enough for impact between particles 1 and 4 to be avoided. In the same time, that displacement should be small enough for impact between particles 2 and 4 to be avoided. 

In the following, explicit kinetic and kinematic calculations are performed for the case of perturbation in the initial conditions which guarantees the occurrence of impacts between particles 1, 2 and 3 alone (with no participation of particle 4 in the impacts-sequence).

\subsubsection{Kinetic considerations}

The following perturbed antisymmetric initial conditions just before the first impact are assumed for the 4 particles:
\begin{equation}
\label{Kin1}
\begin{split}
u_{n_0}^{(0)}=-\Delta, \ \dot u_{n_0}^{(0)}=-v_0[1+\mathcal{O}(\epsilon)], \ v_{n_0}^{(0)}=\Delta(1-\epsilon), \\ \dot v_{n_0}^{(0)}=v_0,  x_{n_0}^{(0)}=0,  \dot x_{n_0}^{(0)}=0,  x_{n_0+1}^{(0)}=-\zeta,  \dot x_{n_0+1}^{(0)}=0
\end{split}
\end{equation}

Since the sequence of impacts is quasi-instantaneous, the coordinates do not change during the process, and the kinetics is fully described by the momenta.

Assuming ideal impacts, one can use conservation of linear momentum and energy for any two impacts (since in the perturbed antisymmetric case there are only impacts involving two particles, simultaneously), to obtain the transformation mapping for any two-dimensional velocity vector before and after impact:
\begin{equation}
\label{Kin2}
\begin{split}
v_{1,2}'=\frac{1-\hat{\mu}}{1+\hat{\mu}}v_{1,2}+\frac{2\hat{\mu}}{1+\hat{\mu}}v_3, \\ v_3'=\frac{2}{1+\hat{\mu}}v_{1,2}-\frac{1-\hat{\mu}}{1+\hat{\mu}}v_3, \  \hat\mu\triangleq \frac{M}{m}
\end{split}
\end{equation}
here the notation is changed and all the velocities are denoted by $v$, with the subscript referring to the particle index (unless it is zero, in which case the value refers to the nominal speed of a particle of mass $m$ of the CB just before impact). The subscript in the expressions below refers to the serial number of the latest impact that had already occurred.

The relations given above are true for impacts between particle 3 (having mass $M$) and either particle 1 or 2 (each having mass $m$). Particle 4 (with mass $M$) is assumed not to interact by impact, and this assumption is validated in the following subsection, dedicated to the kinematics.

Having the initial conditions (just before the first impact) and the transformation relations given above, we can `integrate' to obtain the velocities after impact. Doing this we bear in mind that the impact dynamics is quasi-instantaneous, in the sense that the corresponding displacements are infinitesimal. Consequently, linear elastic coupling (the springs) has no effect on the dynamics. As aforementioned, the first impact is between particles 1 and 3. Therefore, the velocities of those particles after impact become (neglecting $\epsilon$ with respect to 1 in the kinetic calculations here and onward):
\begin{equation}
\label{Kin3}
v_{1}^{(1)}=-\frac{1-\hat{\mu}}{1+\hat{\mu}}v_0, \ v_{2}^{(1)}=v_0, \ v_{3}^{(1)}=-\frac{2}{1+\hat{\mu}}v_0
\end{equation}

For $0<\zeta \ll 1,\hat\mu>1$, particle 1 changes the direction of its velocity before experiencing impact with particle 4. Consequently, particle 1 starts moving rightwards, and at the considered quasi-instance it will not experience impact (through the `sleeve', as suggested in the sketch in Fig. \ref{Fig13}) with particle 4. Therefore, we have shown that one can avoid impact between particles 1 and 4 by proper choice of initial perturbations, for the case of $\hat\mu>1$. After the first impact, for $\hat\mu>1$, particle 1 is still moving leftwards. The second impact involves particles 2 and 3. The velocities after this second impact would be as follows:
\begin{equation}
\label{Kin4}
\begin{split}
v_{1}^{(2)}=-\frac{1-\hat{\mu}}{1+\hat{\mu}}v_0, \\ v_{2}^{(2)}=\frac{1-4\hat\mu-\hat\mu^2}{(1+\hat{\mu})^2}v_0, \\ v_{3}^{(2)}=\frac{4}{(1+\hat{\mu})^2}v_0
\end{split}
\end{equation}

As far as the kinetics of particle 2 is concerned, it is evident that impact between particles 2 and 4 can be avoided at the considered quasi-instance if the velocity of particle 2 changes its direction. Negative velocity can be obtained for particle 2 after its first impact with particle 3 under the condition $\hat\mu=M/m>\sqrt{5}-2\approx 0.236$. Regarding the position of particle 2 at the time of this impact, and whether it satisfies the assumption of no impact with particle 4 -- the matter is addressed in the following subsection. The remaining question in the range $M/m\in[0.236,2]$ is that of the velocity of particle 1. It is clear from Eq. (\ref{Kin4}) that particle 3 will move toward particle 1 after the impact with particle 2, and thus the third impact will involve particles 1 and 3, and produce velocities as follows:
\begin{equation}
\label{Kin5}
\begin{split}
v_{1}^{(3)}=\frac{-1+9\hat{\mu}+\hat\mu^2-\hat\mu^3}{(1+\hat{\mu})^3}v_0, \\ v_{2}^{(3)}=\frac{1-4\hat\mu-\hat\mu^2}{(1+\hat{\mu})^2}v_0, \\ v_{3}^{(3)}=-2\frac{(1-\hat\mu)(3+\hat\mu)}{(1+\hat{\mu})^3}v_0
\end{split}
\end{equation}

Solution of the cubic equation in the numerator in the first expression given above shows that direction-reversal of the velocity of the first particle after its second impact with particle 3 is obtained for $\hat\mu=M/m\in[0.10992,3.4940]$. This range covers the range we are interested in, namely, $\hat\mu=M/m\in[0.13,2]$. Therefore, in the relevant parameter range, at most two impacts between particles 1 and 3 are sufficient for avoiding impact between particles 1 and 4. We still need to check the range $\hat\mu=M/m\in[0.13,0.236]$ in regard with possible impacts involving particles 2 and 4, from the perspective of the velocities (rather than the positions). In the range $\hat\mu=M/m\in[0.13,0.236]$, a fourth (overall) impact is required for the reversal of the direction of the velocity of particle 2. After the fourth impact, one has the following velocities:
\begin{equation}
\label{Kin6}
\begin{split}
v_{1}^{(4)}=\frac{-1+9\hat{\mu}+\hat\mu^2-\hat\mu^3}{(1+\hat{\mu})^3}v_0, \\ v_{2}^{(4)}=\frac{(1-\hat\mu)(1-15\hat\mu-9\hat\mu^2-\hat\mu^3)}{(1+\hat{\mu})^4}v_0, \\ v_{3}^{(4)}=8\frac{1-2\hat\mu-\hat\mu^2}{(1+\hat{\mu})^4}v_0
\end{split}
\end{equation}

For $\hat\mu<1$ (which is the range still remaining in question after three impacts in total), the sign of the velocity of particle 2 after its second impact with particle 3 is determined by the sign of the cubic function $1-15\hat\mu-9\hat\mu^2-\hat\mu^3$. This cubic function is negative for $\hat\mu >0.064178$, which covers the range $\hat\mu=M/m\in[0.13,0.236]$ in question. Therefore, we see that after at most two impacts of each of particles 1 and 2 with particle 3, the directions of the velocities of particles 1 and 2 are reversed, such that at the considered quasi-instance, impact with particle 4 can be avoided. Consequent impacts of particles 1 and 2 with particle 3, which ``chases'' them, can occur, but they will only affect the magnitudes of the velocities of particles 1 and 2, and not their signs. One observes from Eq. (\ref{Kin5}) that for $\hat\mu=1$, exactly three impact (in total) reverse the initial velocities of particles 1 and 2 and bring the velocity of particle 3 back to zero.

It has thus been shown that in the parameter range analyzed in the smooth case, small perturbation of initial antisymmetric compact conditions exists, for which particle 4 does not participate in the quasi-instantaneous impact-dynamics -- at least from the kinetics perspective (the velocities involved). The next subsection examines the consistency of the aforementioned assumption of no impacts involving particle 4 from the perspective of the kinematics -- the positions of the particles.

\subsubsection{Kinematic considerations}
We assume, with no loss of generality, and remaining accurate up to a second order correction in $\epsilon$, that one can start with a relaxed system at rest and supply velocities $v_1(0)$ and $-v_1(0)+\mathcal{O}(\epsilon)$ to particles 1 and 2, respectively, at positions $q_1(0)=q_2(0)=0$. Consequently, after certain time $t_1$, particle 1 will reach the position $-\Delta$, and particle 2 will reach the position $\Delta(1-\epsilon)$. The velocities of the particles just before the first impact would be $-v_0$ (of non-specified value) and $v_0-\mathcal{O}(\epsilon)$, for particles 1 and 2, respectively. In this framework, the first impact occurs between particles 1 and 3. The requirement for avoiding impact between particles 1 and 4 up to this instance is simply $-q_4(0)=\zeta>0$. The second impact occurs between particles 2 and 3. The time from the first impact to the second impact is the time it takes the right boundary of the sleeve of particle 3 to cover its distance from particle 2, moving in the velocity of approach between them:
\begin{equation}
\label{Kin7}
\delta{t}_1=\frac{\Delta\epsilon}{v_2^{(1)}-v_3^{(1)}}=\frac{1+\hat\mu}{3+\hat\mu}\frac{\Delta}{v_0}\epsilon+\mathcal{O}(\epsilon^2)
\end{equation}

During this time, the position of particle 1 changes to
\begin{equation}
\label{Kin8}
q_1^{(2)}=-\Delta-v_1^{(1)}\delta{t}_1=-\Delta-\frac{1-\hat\mu}{3+\hat\mu}\Delta\epsilon+\mathcal{O}(\epsilon^2)
\end{equation}

This means that up to this instance (or always if $\hat\mu>1$), avoiding impact with particle 4 becomes possible by setting: $\zeta=\frac{1-\hat\mu}{3+\hat\mu}\Delta\epsilon+\Delta\epsilon^{3/2}$. At the time of the first impact with particle 3, the position of particle 2 is
\begin{equation}
\label{Kin9}
q_2^{(2)}=\Delta-\Delta\epsilon+\frac{1+\hat\mu}{3+\hat\mu}\Delta\epsilon+\mathcal{O}(\epsilon^2)
\end{equation}

Taking into account the introduced displacement of particle 4 needed to guarantee avoidance of impact with particle 1, the requirement for no impact between particles 2 and 4, at this instance, is guaranteed by definition ($\hat\mu>0$).

For the case of $\hat{\mu}\in[0.236,2]$, particle 2 is no longer a concern from this instance on, in the sense of possible impact with particle 4. However, in the range $\hat\mu\in[0.236,1]$, particle 1 is of concern. The position of particle 1 at the instance of its second impact with particle 3 needs to be calculated. To this end, the time interval between the first impact of particles 2 and 3 and the second impact of particles 1 and 3 is required. This time interval can be computed from the initial distance (at the time of the first impact of particles 2 and 3) of particles 1 and 3 and their relative velocity (when the left end of the sleeve of particle 3 is removed from particle 2 by $-2\Delta$):
\begin{equation}
\label{Kin10}
\delta{t}_2=\frac{q_1^{(2)}-q_3^{(2)}}{v_3^{(2)}-v_1^{(2)}}=\frac{(1+\hat\mu)^3}{(3+\hat\mu)(5-\hat\mu^2)}\frac{\Delta}{v_0}\epsilon+\mathcal{O}(\epsilon^2)
\end{equation}

From the first impact of particles 2 and 3 till the second impact of particles 1 and 3, the time interval given above allows particle 1 to shift further to the left (for $\hat\mu<1$), to the following position:
\begin{equation}
\label{Kin11}
\begin{split}
q_1^{(3)}=q_1^{(2)}+v_1^{(2)}\delta{t}_2=-\Delta-\frac{1-\hat\mu}{3+\hat\mu}\Delta\epsilon \\ -\frac{1-\hat\mu}{3+\hat\mu}\frac{(1+\hat\mu)^2}{5-\hat\mu^2}\Delta\epsilon+\mathcal{O}(\epsilon^2)
\end{split}
\end{equation}

For self-consistency of the assumption of no impact between particles 1 and 4 for $\hat\mu<1$, the absolute value of the displacement of particle 4 just before the first (overall) impact, has to satisfy
\begin{equation}
\label{Kin12}
\zeta=\frac{2-2\hat\mu}{5-\hat\mu^2}\Delta\epsilon+\Delta\epsilon^{3/2}
\end{equation}

The reason for the presence of the second, fractional-order, term is that there has to be a positive addition of order higher than unity to maintain the tightest possible bound and allow the maximum freedom for particle 2. In the same time, it is necessary to have the additional term stronger than (possible) second-order corrections, emerging from expansion of rational functions originating from linear corrections to the initial conditions for the velocities. The chosen power is simply a symmetric compromise for the exponent, and the coefficient of unity is a symmetric positive choice of a coefficient of order unity with no magnitude-significance.

The last validation that has to be made now concerns the position of particle 2 at the time of its second impact with particle 3. This position can be calculated according to formulas similar to the ones used in previous calculations. First, one needs to calculate the position of particle 2 at the instance of the third (overall) impact:
\begin{equation}
\label{Kin13}
\begin{split}
q_2^{(3)}=q_2^{(2)}+v_2^{(2)}\delta{t}_2=\Delta-\Delta\epsilon+\frac{1+\hat\mu}{3+\hat\mu}\Delta\epsilon+\\+\frac{1+\hat\mu}{3+\hat\mu}\frac{1-4\hat\mu-\hat\mu^2}{5-\hat\mu^2}\Delta\epsilon+\mathcal{O}(\epsilon^2)
\end{split}
\end{equation}

Next, the time interval between the third and the fourth impact should be derived, using the same strategy as before:
\begin{equation}
\label{Kin14}
\begin{split}
\delta{t}_3=\frac{q_1^{(3)}+2\Delta-q_2^{(3)}}{v_2^{(3)}-v_3^{(3)}}=\\ \frac{1+\hat\mu}{3+\hat\mu}\frac{3+4\hat\mu+\hat\mu^2}{5-\hat\mu^2}\frac{(1+\hat\mu)^3}{7-7\hat\mu-7\hat\mu^2-\hat\mu^3}\frac{\Delta\epsilon}{v_0}+\mathcal{O}(\epsilon^2)
\end{split}
\end{equation}

A positive value for the time interval is guaranteed for $7-7\hat\mu-7\hat\mu^2-\hat\mu^3>0$, which holds for $\hat\mu\in[0,0.60388]$. This span covers the range relevant for the analysis concerning the \emph{second} impact between particles 2 and 3 necessary for reversing the direction of particle 2 (the range $\hat\mu\in[0.13,0.236]$).

Finally, using the obtained, third, time interval, one can derive the position of particle 2 at the fourth (overall) impact, as follows:
\begin{equation}
\label{Kin15}
\begin{split}
q_2^{(4)}=q_2^{(3)}+v_2^{(3)}\delta{t}_3= \\ \Delta-\Delta\epsilon+\frac{1+\hat\mu}{3+\hat\mu}\Delta\epsilon+\frac{1+\hat\mu}{3+\hat\mu}\frac{1-4\hat\mu-\hat\mu^2}{5-\hat\mu^2}\Delta\epsilon+\\+\frac{1+\hat\mu}{3+\hat\mu}\frac{1-4\hat\mu-\hat\mu^2}{5-\hat\mu^2}\frac{3+7\hat\mu+5\hat\mu^2+\hat\mu^3}{7-7\hat\mu-7\hat\mu^2-\hat\mu^3}\Delta\epsilon+\mathcal{O}(\epsilon^2)
\end{split}
\end{equation}

The self-consistency requirement for the absence of impact between particles 2 and 4 can then be expressed by the inequality $q_2^{(4)}+\zeta<\Delta$. This leads to the following requirement:
\begin{equation}
\label{Kin16}
\begin{split}
\frac{1+\hat\mu}{3+\hat\mu}+\frac{1+\hat\mu}{3+\hat\mu}\frac{1-4\hat\mu-\hat\mu^2}{5-\hat\mu^2}+\\ \frac{1+\hat\mu}{3+\hat\mu}\frac{1-4\hat\mu-\hat\mu^2}{5-\hat\mu^2}\frac{3+7\hat\mu+5\hat\mu^2+\hat\mu^3}{7-7\hat\mu-7\hat\mu^2-\hat\mu^3}+\frac{2-2\hat\mu}{5-\hat\mu^2}<1
\end{split}
\end{equation}

This requirement holds in conjunction with the assumption $\hat\mu\in[0.13,0.236]$, relevant for the analysis of a possibility of a \emph{second} impact between particles 2 and 3, necessary for reversing the direction of particle 2. In addition, one recalls the aforementioned requirement $7-7\hat\mu-7\hat\mu^2-\hat\mu^3>0$. Finally, the limits of the investigated parameter range yield the inequality $5-\hat\mu^2>0$. All these inequalities, when superimposed, lead to the following equivalent self-consistency condition necessary for avoiding impact between particles 2 and 4:
\begin{equation}
\label{Kin17}
\hat\mu<1
\end{equation}

Since the condition in Eq. (\ref{Kin17}) is satisfied in the examined range $\hat\mu\in[0.13,0.236]$ with a \emph{finite margin}, clearly the $\epsilon^{3/2}$ correction in $\zeta$ is insufficient for enabling contact between particles 2 and 4. Therefore, it is established that there exists an infinitesimal perturbation of compact antisymmetric initial conditions for the vibro-impact-limit system, for which the dynamics involves only particles 1, 2, and 3. This dynamics consists in possibly an infinite sequence of consecutive impacts of particle 3 with particles 1 and 2, eventually leading (or not) to exact reversal of the velocities of the particles assigned to them just before the first impact. If the aforementioned sequence consists of a \emph{finite} number of steps, then the 3-element subsystem can be considered impact-integrable. Otherwise, the subsystem can be considered not impact-integrable. Now, if a slightly-perturbed periodic solution is non-integrable, even for a specific (feasible) perturbation, then the exact periodic solution in question is, clearly, unstable. 

What is now left to do in order to answer the stability question for the compact periodic solution definitively, for the assumed parameter range, is to find the parameter values corresponding to impact-integrability of the three-body problem (for the examined system). For those mass ratios, a perturbation would have to be found that would lead to four-body dynamics. For this four-body dynamics, it would have to be shown that there is either no impact-integrability, or no frequency guaranteeing the stability of the compact periodic solution (as should be learned by saltation matrix analysis).

\subsubsection{Integration of the three-masses sub-system}

It is clear that perturbed compact antisymmetric initial conditions involving three of the masses cannot lead to quasi-instantaneous velocities-reversal with a single impact. The smallest number of sufficient impacts would be 2. From Eq. (\ref{Kin4}), the triplet $\lbrace{v_1^{(2)},v_2^{(2)},v_3^{(2)}}\rbrace$ becomes the triplet $\lbrace{v_0,-v_0,0}\rbrace$ for $\hat\mu_2\to\infty$. The next natural number of impacts to yield velocities-reversal is 3, which, as can be found from Eq. (\ref{Kin5}), is obtained for $\hat\mu_3=1$.

In order to find the values of $\hat\mu$ corresponding to velocities-reversal for 4,5,6 etc. impacts, the three-body problem needs to be further integrated, continuing the sequence presented in Eqs. (\ref{Kin3})-(\ref{Kin6}).

First, one observes that the triplet $\lbrace{v_1^{(4)},v_2^{(4)},v_3^{(4)}}\rbrace$ becomes the triplet $\lbrace{v_0,-v_0,0}\rbrace$ for $\hat\mu_4=\sqrt{2}-1\approx 0.4142$. This value lies within the examined range, $\hat\mu\in[0.13,2]$.

Next, applying Eq. (\ref{Kin2}) to Eq. (\ref{Kin6}) for the fifth impact, which would occur between particles 1 and 3, the following velocities are obtained (\emph{after} the fifth impact):
\begin{equation}
\label{Kin18}
\begin{split}
v_{1}^{(5)}=\frac{-1+25\hat{\mu}-30\hat\mu^2-26\hat\mu^3-\hat\mu^4+\hat\mu^5}{(1+\hat{\mu})^5}v_0, \\ v_{2}^{(5)}=\frac{(1-\hat\mu)(1-15\hat\mu-9\hat\mu^2-\hat\mu^3)}{(1+\hat{\mu})^4}v_0, \\ v_{3}^{(5)}=-2\frac{5-20\hat\mu-6\hat\mu^2+4\hat\mu^3+\hat\mu^4}{(1+\hat{\mu})^5}v_0
\end{split}
\end{equation}

The triplet $\lbrace{v_1^{(5)},v_2^{(5)},v_3^{(5)}}\rbrace$ becomes the triplet $\lbrace{v_0,-v_0,0}\rbrace$ for $\hat\mu_5=\sqrt{5}-2\approx 0.2361$. This value also lies within the examined range, $\hat\mu\in[0.13,2]$.

Furthermore, applying Eq. (\ref{Kin2}) to Eq. (\ref{Kin18}) for the sixth impact, which would occur between particles 3 and 2, the following velocities are obtained (\emph{after} the sixth impact):
\begin{equation}
\label{Kin19}
\begin{split}
v_{1}^{(6)}=\frac{-1+25\hat{\mu}-30\hat\mu^2-26\hat\mu^3-\hat\mu^4+\hat\mu^5}{(1+\hat{\mu})^5}v_0, \\ v_{2}^{(6)}=\frac{1-36\hat\mu+85\hat\mu^2+48\hat\mu^3}{(1+\hat{\mu})^6}v_0\\-\frac{21\hat\mu^4+12\hat\mu^5+\hat\mu^6}{(1+\hat{\mu})^6}v_0, \\ v_{3}^{(6)}=4\frac{(1-\hat{\mu})(3-17\hat\mu-15\hat\mu^2-3\hat\mu^3)}{(1+\hat{\mu})^6}v_0
\end{split}
\end{equation}

The triplet $\lbrace{v_1^{(6)},v_2^{(6)},v_3^{(6)}}\rbrace$ becomes the triplet $\lbrace{v_0,-v_0,0}\rbrace$ for $\hat\mu_6=\frac{2}{\sqrt{3}}-1\approx 0.1547$. This value, again, also lies within the examined range, $\hat\mu\in[0.13,2]$.

In order to see whether velocities-reversal can be obtained for exactly 7 impacts, we need to apply Eq. (\ref{Kin2}) to Eq. (\ref{Kin19}) for the seventh impact, occurring between particles 1 and 3. The following velocities are obtained (after the seventh impact):
\begin{equation}
\label{Kin20}
\begin{split}
v_{1}^{(7)}=\frac{-1+49\hat{\mu}-189\hat\mu^2-35\hat\mu^3}{(1+\hat{\mu})^7}v_0+\\+\frac{125\hat\mu^4+51\hat\mu^5+\hat\mu^6-\hat\mu^7}{(1+\hat{\mu})^7}v_0, \\ v_{2}^{(7)}=\frac{1-36\hat\mu+85\hat\mu^2+48\hat\mu^3}{(1+\hat{\mu})^6}v_0\\-\frac{21\hat\mu^4+12\hat\mu^5+\hat\mu^6}{(1+\hat{\mu})^6}v_0, \\ v_{3}^{(7)}=-2\frac{7-70\hat\mu+49\hat\mu^2+76\hat\mu^3}{(1+\hat{\mu})^7}v_0\\ -2\frac{9\hat\mu^4-6\hat\mu^5-\hat\mu^6}{(1+\hat{\mu})^7}v_0
\end{split}
\end{equation}

It turns out that the triplet $\lbrace{v_1^{(7)},v_2^{(7)},v_3^{(7)}}\rbrace$ becomes the triplet $\lbrace{v_0,-v_0,0}\rbrace$ for \emph{no} real positive value of $\hat\mu_7$.

Finally, in order to see whether velocities-reversal can be obtained for exactly 8 impacts, we apply Eq. (\ref{Kin2}) to Eq. (\ref{Kin20}) for the eighth impact, which would occur between particles 3 and 2. The following velocities are obtained (\emph{after} the eighth impact):
\begin{equation}
\label{Kin21}
\begin{split}
v_{1}^{(8)}=\frac{-1+49\hat{\mu}-189\hat\mu^2-35\hat\mu^3}{(1+\hat{\mu})^7}v_0+\\+\frac{125\hat\mu^4+51\hat\mu^5+\hat\mu^6-\hat\mu^7}{(1+\hat{\mu})^7}v_0, \\ v_{2}^{(8)}=\frac{1-64\hat\mu+364\hat\mu^2-112\hat\mu^3-410\hat\mu^4}{(1+\hat{\mu})^8}v_0+\\+\frac{-96\hat\mu^5+44\hat\mu^6+16\hat\mu^7+\hat\mu^8}{(1+\hat{\mu})^8}v_0, \\ v_{3}^{(8)}=16\frac{1-14\hat\mu+21\hat\mu^2+20\hat\mu^3}{(1+\hat{\mu})^8}v_0\\ -16\frac{5\hat\mu^4+6\hat\mu^5+\hat\mu^6}{(1+\hat{\mu})^8}v_0
\end{split}
\end{equation}

The triplet $\lbrace{v_1^{(8)},v_2^{(8)},v_3^{(8)}}\rbrace$ becomes the triplet $\lbrace{v_0,-v_0,0}\rbrace$ for $\hat{\mu}_8=\sqrt{4-2\sqrt{2}}-1\approx 0.08$. This value is already outside the examined range of $\hat\mu\in[0.13,2]$.

To conclude, we have obtained a decreasing sequence of six values $\hat\mu_j$, where $j$ is the number of impacts required for exact velocities-reversal and $\hat\mu_j$ is the corresponding mass ratio:
\begin{equation}
\label{Kin22}
\begin{split}
\hat\mu_2\to\infty, \ \hat{\mu}_3=1, \ \hat{\mu}_4=\sqrt{2}-1\approx 0.4142, \\ \hat\mu_5=\sqrt{5}-2\approx 0.2361, \ \hat\mu_6=\frac{2}{\sqrt{3}}-1\approx 0.1547, \\ \hat\mu_8=\sqrt{4-2\sqrt{2}}-1\approx 0.08
\end{split}
\end{equation}

It would be reasonable to conclude by informal induction that $\hat\mu_j$ is (at least weakly) monotonically decreasing with $j$ when sufficiently small.
We can also address this matter more rigorously \emph{asymptotically}. Performing first-order expansion of Eq. (\ref{Kin2}) for particle 1 for $\hat\mu\ll1$, one gets:
\begin{equation}
\label{Kin23}
v_1^{(k+2)}=v_1^{(k)}+2\hat\mu\left[v_3^{(k+1)}-v_1^{(k)} \right]
\end{equation}
where $k$ denotes the impact number. Formally, one can `integrate' this difference equation starting from the initial condition, which to first order is $v_1^{(0)}=-v_0$, up to the velocity after the last impact between particles 1 and 3, which for impact-integrability should be $v_1^{(j)}=v_0$. This leads to the following relation:
\begin{equation}
\label{Kin24}
\begin{split}
v_0=\hat\mu\sum_{k=1,3,5}^{\lfloor (j-1)/2 \rfloor}{\left[v_3^{(k+1)}-v_1^{(k)} \right]}=\\=\hat\mu\sum_{k=1,3,5}^{\lfloor (j-1)/2 \rfloor}{v_3^{(k+1)}} -\hat\mu\sum_{k=1,3,5}^{\lfloor (j-1)/2 \rfloor}{v_1^{(k)}}
\end{split}
\end{equation}

The sequence $v_1^{(k)}$ goes from $-v_0$ to $v_0$, passing through zero. Moreover, we look for values of $\hat\mu$ corresponding to impact-integrability. Therefore, all the assumed sequences are finite and periodic, and thus time-reversal symmetry applies, and hence $v_1^{(k)}$ is also symmetric on a period or \emph{antisymmetric} on half a period (which is what we consider in the velocity-reversal problem). Consequently, the sequence $v_1^{(k)}$ is antisymmetric with respect to the ``middle'' impact. Therefore, the last sum in Eq. (\ref{Kin24}) amounts to \emph{zero} for large-enough values of $j$, for which the sequence $v_1^{(k)}$ is smooth enough for the sum to approximate an (odd-function) integral.

Regarding, $v_3^{(k)}$, we can say the following. For $\hat\mu\to 0$ one observes from Eqs. (\ref{Kin3})-(\ref{Kin21}) that for small, even, values of $k$, one has $v_3^{(k)}=4(k/2)v_0$. As we consider periodic impact-sequences, one can assume that for large-enough values of $j$, for $k$ close-enough to $j$, symmetric behavior would emerge, namely: $v_3^{(k)}=4[(j-k)/2]v_0$. For non-zero values of $\hat\mu$, $v_3^{(k)}$ should initially increase with each \emph{even} impact. Then, when $v_1^{(k)}$ and $v_2^{(k)}$ would reach their intermediate values close to zero, all the energy would be stored in $v_3^{(k)}$, and thus, in the middle of the sequence, one should expect a maximum value of $v_3^{(k)}$, after which the energy will again `flow' to $v_1^{(k)}$ and $v_2^{(k)}$, and $v_3^{(k)}$ will return to its (initial) smallest value. This implies that there is a maximum value of $v_3^{(k)}$ and that one can estimate it from energy considerations, namely, assuming that particle 3 contains all (or almost all) the energy:

\begin{equation}
\label{Kin25}
v_3^{(\text{max})}\sim \frac{v_0}{\sqrt{\hat\mu/2}}
\end{equation}
(this would only hold for a large number of impacts, each changing the velocities $v_1$ and $v_2$ only slightly).

Substituting this result, along with the understanding regarding the sum over $v_1^{(k)}$ into Eq. (\ref{Kin24}), one obtains the following relation:
\begin{equation}
\label{Kin26}
\begin{split}
v_0\sim\hat\mu \frac{v_0}{\sqrt{\hat\mu/2}}\sum_{k=1,3,5}^{\lfloor (j-1)/2 \rfloor}\frac{v_3^{(k+1)}}{v_3^{(\text{max})}}\sim \\ \sim v_0\sqrt{2\hat\mu} \left\lfloor \frac{j-1}{2} \right\rfloor\rho, \ \rho \triangleq \frac{\langle v_3^{(|k\text{ even})}\rangle_{\text{avg}}}{v_3^{(\text{max})}}
\end{split}
\end{equation}

In order to estimate the last quotient in Eq. (\ref{Kin26}), one has to know the distribution $v_3^{(k+1)}$ for odd increasing values of $k$ (the `current' impact number). For a large number of impacts needed for completing half a period, a smooth approximation for this distribution would be a concave function, linear at the edges (initial and final values of $k$), and symmetric around the middle. In principle, for such a distribution, the value of the aforementioned quotient should be between 1/2 (for a triangular function) and 1 (for a nearly uniform distribution -- except the sharp change at the edges of the range). Any convexity in the distribution would imply increase in the forcing, which is unreasonable, since the force is exerted by particles 1 and 2, whose momenta decrease from the beginning of the range to its middle. Thus, the feasible range would be $\rho\in[1/2,1]$.

By resorting to smooth-limit analysis of the dynamic equations for the 3-particles impact problem for $\hat\mu\ll 1$, insight into the problem is given in \ref{AppendixC}. It is shown there that a good estimate for the velocity distribution of particle 3 at even-valued impact numbers is the \emph{sinusoidal} distribution on the positive half period. Consequently, the corresponding aforementioned quotient can be taken as $\rho\sim 2/\pi$.

Using this result and assuming large values of $j$ (many impacts for velocity reversal -- the assumption of large $j$ is crucial for the calculation of the estimate for $\rho$), one obtains the following asymptotic relation:
\begin{equation}
\label{Kin27}
\begin{split}
\hat\mu \underset{j\gg1}\sim \frac{\pi^2}{2j^2}
\end{split}
\end{equation}
which implies inversely quadratic decay of $\hat\mu$ with $j$.

Extrapolating \emph{back} to 8 impacts, the highest number of impacts for which we have exact integration results, one obtains from Eq. (\ref{Kin27}) the following estimate:
\begin{equation}
\label{Kin28}
\begin{split}
\hat\mu_8\sim \frac{\pi^2}{128} \approx 0.08
\end{split}
\end{equation}
which is fairly close to the exact value given in Eq. (\ref{Kin22}). This gives validation to the relevance of the asymptotic analysis presented herein.

Hence, one can conclude that, indeed, for $j>8$, one has $\hat\mu_j<\hat\mu_8$, and thus the only `impact-integrable' values of $\hat\mu$ falling into the chosen range of $[0.13,2]$ are the 4 values listed in Eq. (\ref{Kin22}).

It has thus been shown that there are only 4 (isolated) points (mass-ratio values) that should be suspected as stable in the discrete-limit system. If the analogy with the smooth (linear-cubic) system is worthwhile, as argued above, then there are only 4 points `suspected' to correspond to `isolated' stability of the compact antisymmetric solution in the smooth system for large amplitudes. One of those points, namely, the one for which $M=m$, was already shown to be stable.

Now, instead of scanning the entire parameter domain, one can assume a value of the amplitude at the upper bound of the examined range, namely $\alpha=400$, take $N=19$ representative elements, and perform numerical Floquet analysis only for the cases of $\hat\mu\triangleq M/m=\hat\mu_3,\hat\mu_4,\hat\mu_5,\hat\mu_6$. The case of $M/m=\hat\mu_3=1$ is already known to correspond to stability (for high-enough frequencies), but its results are shown in the figures below for clarity anyway, along with the results for the other three mass-ratio values.

\begin{figure}[ht]
\begin{center}
{\includegraphics[scale = 0.53,trim={0 0cm 0 0cm},clip]{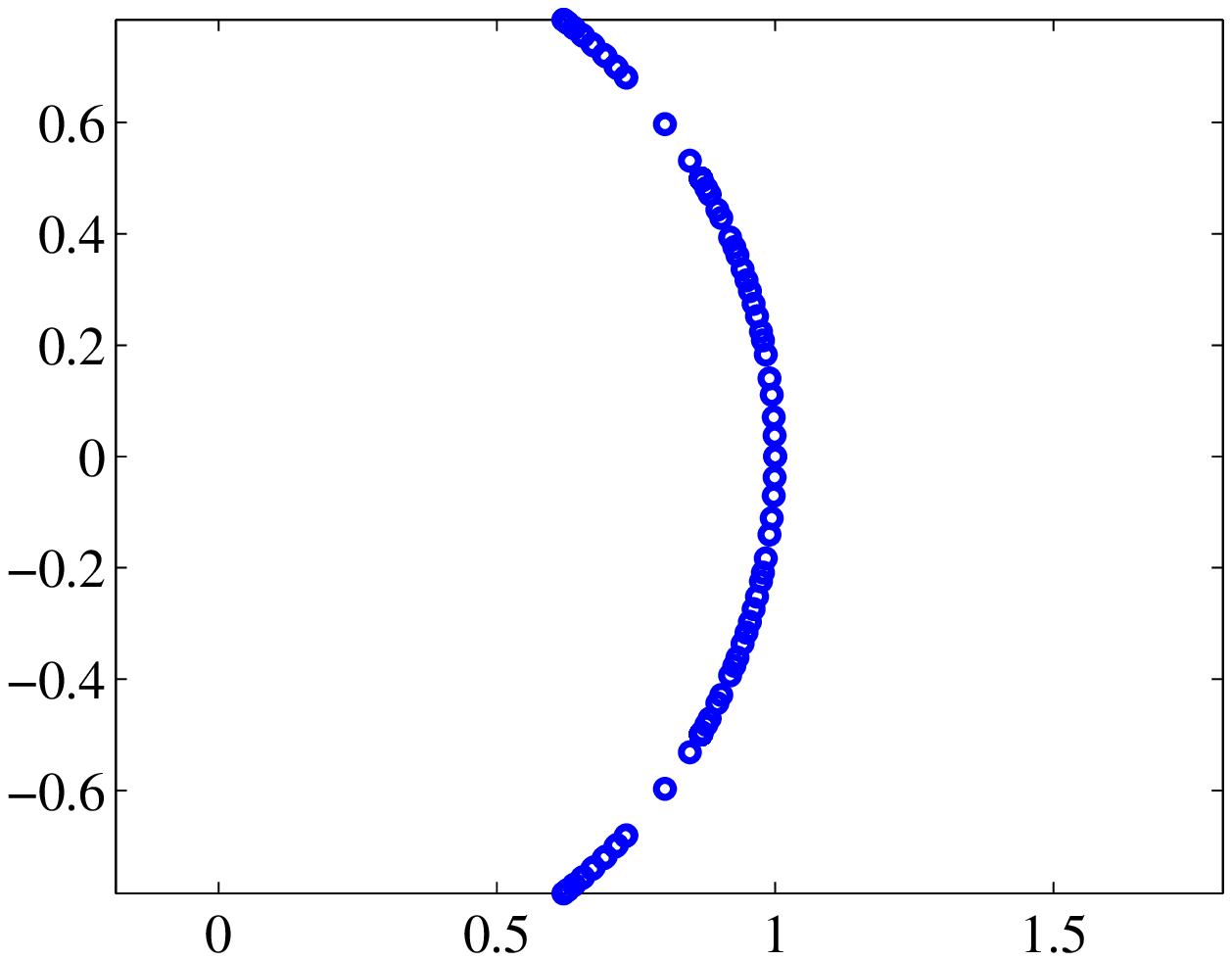} \\
\bigskip
\bigskip
\includegraphics[scale = 0.53,trim={0 0cm 0 0cm},clip]{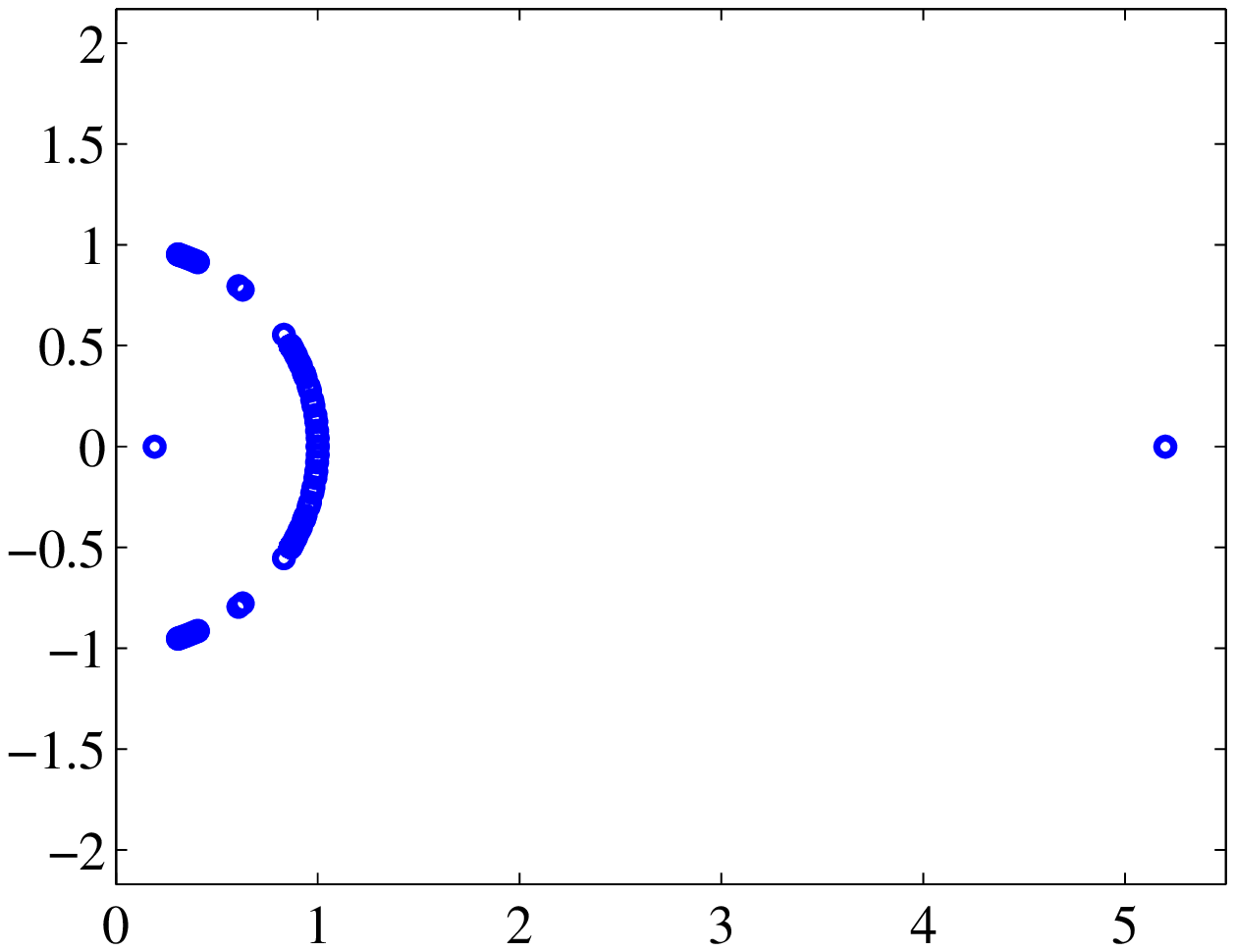}}
\end{center}
\begin{picture}(0,0)
  \put(118,212){Re $\lambda$}
  \put(7,300){Im $\lambda$}
  \put(118,19){Re $\lambda$}
  \put(7,106){Im $\lambda$}
\end{picture}
\caption{\small Monodromy matrix eigenvalues $\lambda$ for the compact antisymmetric periodic mode for $N=19,\alpha=400$ for $M/m=1$ (top) and $M/m=\sqrt{2}-1$ (bottom) for the \emph{smooth} system}
\label{Fig16}
\end{figure}

\begin{figure}[ht]
\begin{center}
{\includegraphics[scale = 0.53,trim={0 0cm 0 0cm},clip]{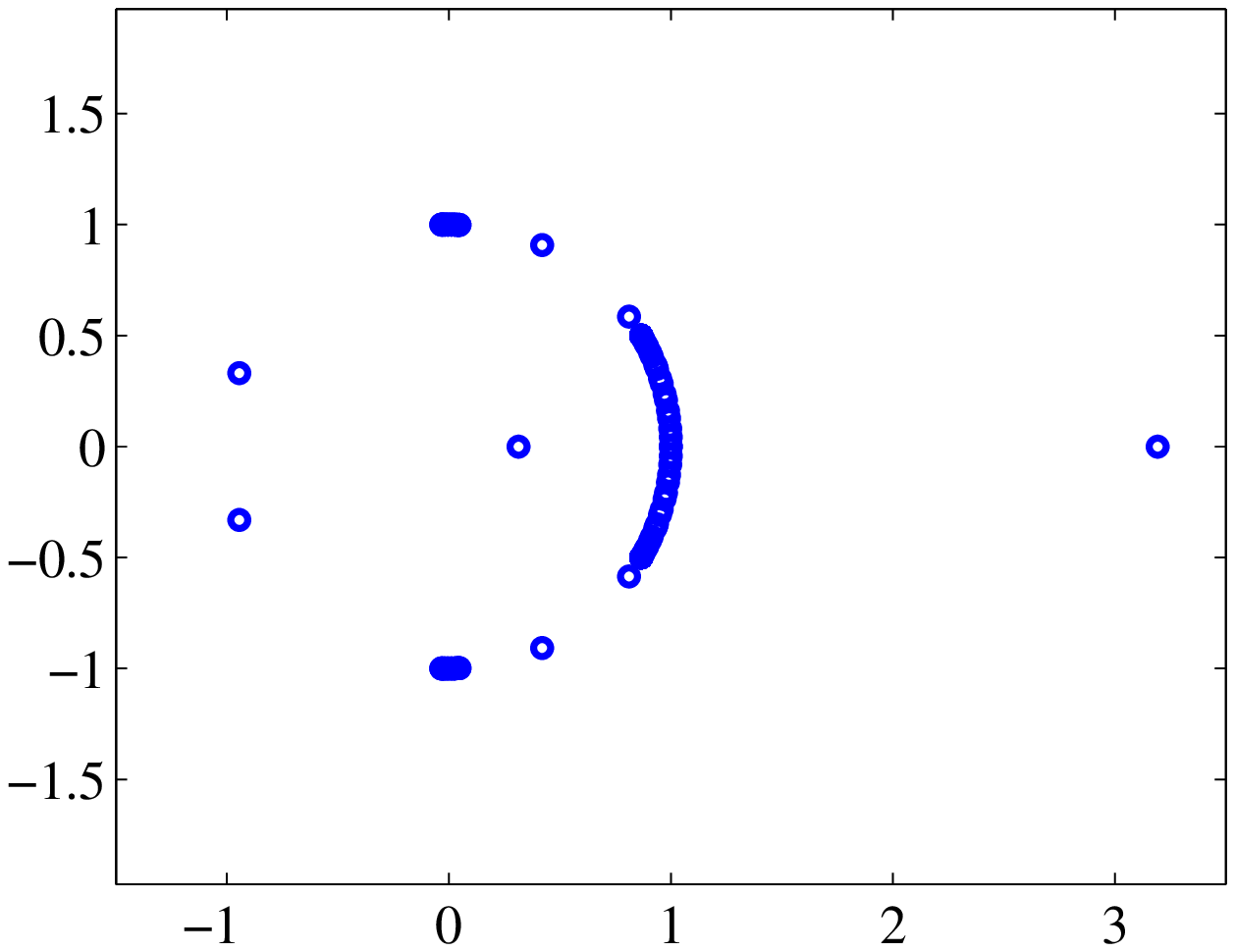} \\
\includegraphics[scale = 0.53,trim={0 0cm 0 0cm},clip]{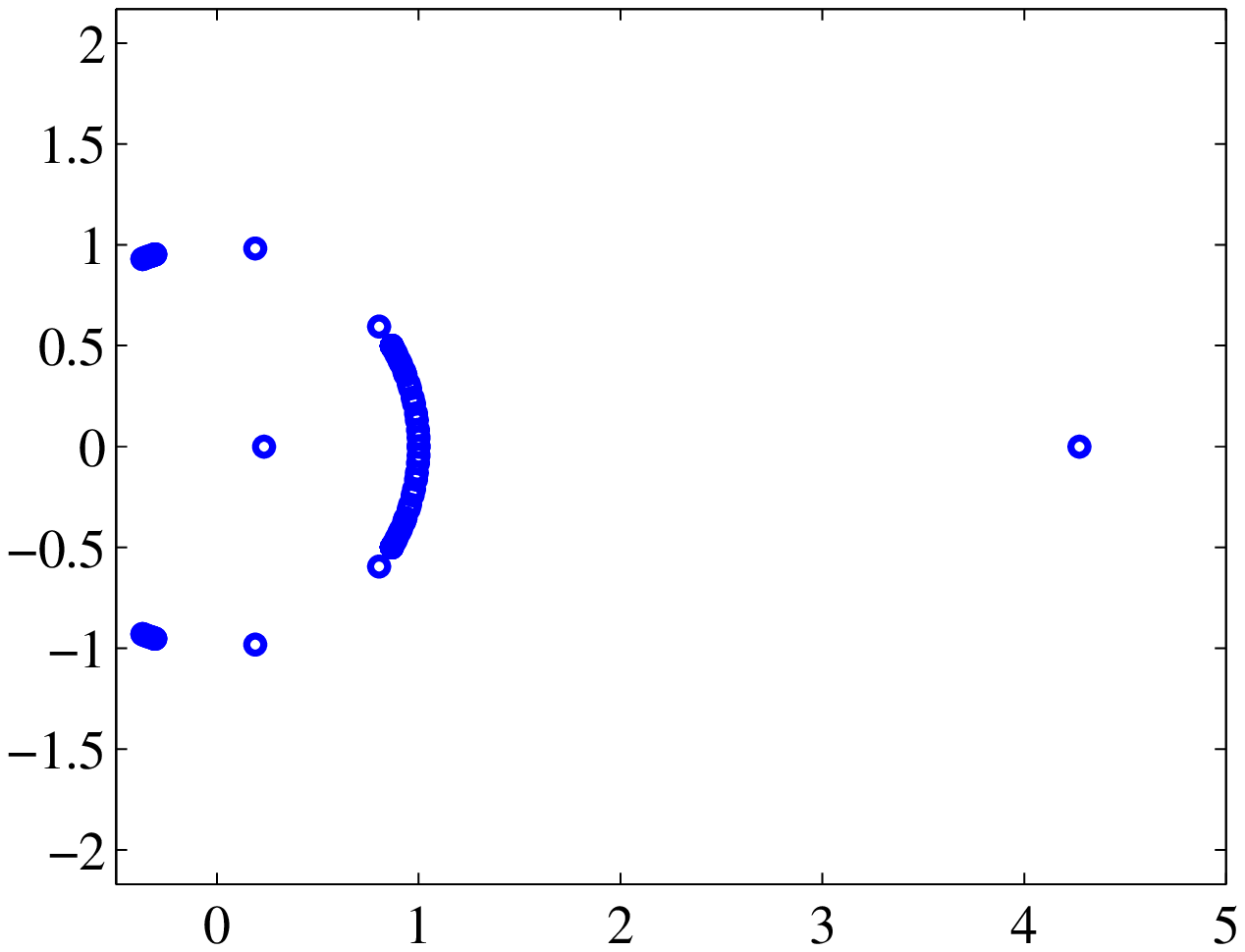}}
\bigskip
\bigskip
\end{center}
\begin{picture}(0,0)
  \put(118,212){Re $\lambda$}
  \put(7,300){Im $\lambda$}
  \put(118,40){Re $\lambda$}
  \put(7,130){Im $\lambda$}
\end{picture}
\caption{\small Monodromy matrix eigenvalues $\lambda$ for the compact antisymmetric periodic mode for $N=19,\alpha=400$ for $M/m=\sqrt{5}-2$ (top) and $\frac{M}{m}=\frac{2}{\sqrt{3}}-1$ (bottom) -- \emph{smooth} system}
\label{Fig17}
\end{figure}

Clearly, it could have been advantageous to examine four-particle dynamics for the four relevant mass-ratio values, however, this task may be rather cumbersome. It is sufficient, though, that the discrete-system analogy distilled a small number of isolated mass-ratio values to examine in the smooth large-amplitude case. When there are specific values to check the stability for in the parameter plane, the tendency of the stable domain to be quasi-fractal no longer poses an obstacle for analysis, which renders the problem tractable. The eigenvalues of the monodromy matrices of the compact periodic solutions for $\alpha=400,N=19$ and the aforementioned mass ratios (obtained by numerical implementation of Floquet theory for the smooth system) are presented in Figs. \ref{Fig16} and \ref{Fig17}.

One notes that of the four mass ratios, stability of the compact antisymmetric solution is observed for an isolated value of the mass ratio only for $M/m=1$, as suggested by cruder analysis shown already in Fig. \ref{Fig12}.

To conclude, it may be asserted that it was possible to explain the stability pattern of the compact antisymmetric periodic dynamic solution in a translationally-invariant nonlinear one-dimensional lattice, at least in what concerns stability in the context of pitchfork bifurcations. It is not surprising that pitchfork bifurcations appear to be tractable in the analyzed system, in contrast to Neimark-Sacker bifurcations. The reason for the tractability can be the fact that pitchfork bifurcations are normally associated with a symmetry group of higher symmetry, and higher-symmetry systems (modes, in this case) are naturally more tractable (this is, of course, the reason for the power of symmetry considerations when obtaining solutions in physics in general). Indeed, the two instability cases that were explained here were related to resonance between the compact antisymmetric nonlinear normal mode and either phonons (least-localized modes), or the (complementary) \emph{symmetric} nonlinear normal mode, which is (exponentially) localized. In both cases, the corresponding bifurcation was of the pitchfork type. Both the strong (exponential) and the vanishing (planar-wave) localization are examples of extreme levels of localization, and as such, they can be reasonably-well argued to be associated with high (spatial) symmetry.

Thus, the proposed translationally-invariant nonlinear one-dimensional mechanical lattice shown here to possess stable compact modes can be considered well-enough studied, at least for a first look into such a system. Higher-order effects can be addressed in future work.

The following section suggests a possible application for the studied phenomena.

\section{Application -- a moving acoustic sensing devise}
\label{Sec6}

A natural application of the studied system could be in the field of acoustic sensor engineering. It may be necessary to detect a small acoustic perturbation to ambient conditions, such as, for example, emanation from working equipment. It may be important to detect the spatial location of the perturbation with high accuracy. To this end, one might look for a sensor with the effect of amplified resolution.

It may be advantageous to construct an arrangement of chains as the one shown in Fig. \ref{Fig1}, in the form of a \emph{parallel array} (of such chains). The idea is to excite compact antisymmetric periodic modes localized at different positions along the chain. For example, positioning the CB site at locations shifted a given number of representative elements in the direction of the chains, such that this shift is increased by a fixed amount from one chain to the next one. This design may have advantages. If the magnitude of a single shift along the chain, when passing from one chain to the next one, is an order of magnitude larger than the spatial distance between two adjacent chains, then there is already an order of magnitude gain in the amplification factor of the sensor.

The concept of the device is that when a small (localized) acoustic perturbation reaches a specific chain in the array, it can destroy the CB-mode excited (in advance) in that chain. The underlying assumption is that the parameter (the mass ratio, for example) was chosen such that the CB is marginally-stable (so that it would require a finite, even if small, perturbation to cause instability).

\begin{figure}[H]
\begin{center}
\includegraphics[scale = 0.5,trim={0 0cm 0 0cm},clip]{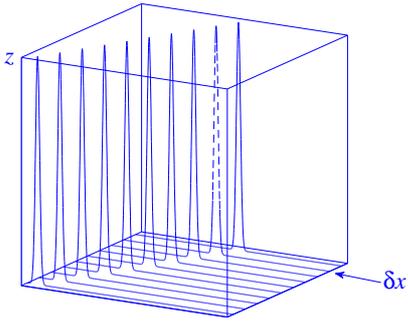}
\end{center}
\caption{\small Schematic sketch of the suggestion for the sensor -- the spikes represent antisymmetric ($z$-mode) discrete CBs located on shifted localization-sites in an array of translationally-invariant chains, with a schematic spatially-resolved external-mode incident acoustic perturbation}
\label{Fig18}
\end{figure}

By observing the location of the destroyed CB, it would be possible to know the position of the external perturbation in question (see Fig. \ref{Fig18}). The high accuracy of the sensing would be based on the large distance between pre-existing CBs in the chain direction, as compared to the distance between the chains (which is the actual spatial resolution).

As for the mechanism of `reading' the device, clearly it should not be problematic to determine the CB of which position (along the chain) was destroyed. Due to the relatively large intervals between the CB sites, a low-resolution sensor can be used to `read' the proposed device, and in conjunction with the calibration (associating the CB positions with the chain positions in the array -- which could be an affine relation), an overall high-resolution sensor could be obtained.

Possible challenges in the engineering of the sensor could be the assembling of the chains in a casing, such that interaction between the chains is avoided. A second crucial issue is damping. The solution to the second challenge may be the addition of small background periodic excitation.

Advanced-stage analysis of the device may require studying an array of the chains as suggested, with the addition of weak strongly-nonlinear coupling between the chains and introduction of small forcing and damping. A stability chart for the overall device with the mentioned additions would have to be constructed.

Two points are importing to note. The first is that the translational invariance of the lattice is crucial for detecting moving acoustic perturbations, especially if a moving (tracking) device is sought. The second point is that in order to avoid false positives, a threshold excitation should be designed for, meaning that the system cannot simply be tuned to be unstable. Rather marginal stability should be sought, assuming that \emph{nonlinear} instability is likely to be associated with marginal \emph{linear} stability. To this end, it is crucial for the stability bounds to be tractable. Hence the large emphasis put on the matter in the present study, which proved to be reasonably successful, at least for high-symmetry instabilities.

\section{Conclusions}
\label{Sec7}

The present paper proposes a system consistent of a one-dimensional nonlinear non-integrable classical lattice describing the exact dynamics of the fundamental degrees of freedom (rather than averaged quantities) in the mechanical setting. The system is comprised of `on-site` complex blocks containing several degrees of freedom each, characterized by the existence of symmetry between the degrees of freedom within each block.

The aforementioned symmetry is one that allows spatial decoupling of a single block from the rest of the lattice, corresponding to antisymmetric (momentumless) mode of motion. In linear analysis (for the case of nonlinearity having a weakly-nonlinear limit), the system shows three dispersion curves, an acoustic branch, an optical branch, and a flat band, which has a tangent point with one of the two other bands.

Nonlinear analysis was performed by assuming a $\beta$-FPU type interaction potential. The analysis of a single representative cell, assuming conservation of momentum and energy, can be performed by rigorous implementation of Hill's method. Standard instability tongues emerge. A qualitatively similar picture of instability tongues is obtained for a chain of three blocks, by numerical integration in the framework of the Floquet theory, with exact detection of the two boundaries for each tongue. A very large number of overlapping tongues is observed. Two of the tongues, related to the pitchfork bifurcation, are reproduced also for the case of a chain comprised of 19 blocks, where Floquet theory is, again, implemented using numerical integration. This time, however, the instability-tongues' boundaries are not detected exactly. Rather, the parameter plane (mass-ratio--amplitude) is divided by a dense grid and each node is determined to be either stable or unstable. A very complicated instability pattern emerges. Apart from the two principal pitchfork-bifurcation-related tongues evident already for the three-blocks chain, an additional, comb-like, array of pitchfork-instability-related tongues is observed. This comb-like array of instability tongues falls into a region the boundaries of which can be theoretically derived. In fact, those boundaries appear to be related to resonance between the compact antisymmetric nonlinear normal mode (associated with a flat dispersion band) and the linear propagation spectrum. This renders the stability map tractable for low amplitudes, at least for the case of the high-symmetry pitchfork-bifurcation-related instability.

For the case of large amplitudes, almost the entire parameter plane appears to become unstable, with two exceptions. One is a narrow but finite stability stripe, related to a gap between the two principal pitchfork-bifurcation-related instability tongues emerging already for the chain with three elements. The second exception corresponds to stability for an isolated value of the mass ratio.

A major part of the paper is dedicated to the effort to explain the aforementioned emerging isolated stability-preserving mass-ratio value. To this end, the smoothly nonlinear large-amplitude regime is associated with a nonsmooth system with conservative impact interaction (and linear dynamics between impacts). We use a construction of logical arguments, energy considerations, and direct integration of a dynamic mapping equation. In addition, we employ the saltation matrix realization of the Floquet theory for piecewise linear systems, applying asymptotic considerations of various sorts. The overall effort turns out to be relatively successful and the isolated-parameter-value-related stability is explained.

Finally, based on the property of translational invariance of the system, in which internal symmetry and the associated flat-band produce a perfectly-localized periodic mode, and relying on the more-or less tractable stability picture of the system, possible practical implementation is suggested. This implementation consists of an acoustic motion sensor (where translational invariance is important), for which high spatial resolution of detecting acoustic perturbations can be obtained as follows. Spatially-close local perturbations are related to well-distant `standing waves' in an array of chains. In such a device, the perfect compactness of the standing waves (before they are destructed by the perturbations, while those are being detected), can allow `cleaner' detection (since even after destruction, the compactness holds for some time). Furthermore, the analysis capabilities, as presented in this work, allow better design and fine-tuning of the aforementioned (still hypothetical) device (for example, by tuning the parameters to marginal stability for then relying on possible associated nonlinear instability, to obtain an excitation threshold).

The present work lies within the emerging body of literature of the recent years related to such concepts as applications of flat bands, dynamic localization, meta-materials and resonance-based sensors.

The main contribution of this work is the implementation of a perfectly-compact-breather solution in a translationally-invariant nonlinear mechanical system, and the use of the vibro-impact system analogy for study of otherwise hardly-tractable stability characteristics.

\section*{Acknowledgments}

The financial help of the Israel Science Foundation, Grant No. 1696/17, is gratefully acknowledged.



\renewcommand{\theequation}{{A.}\arabic{equation}}
\setcounter{equation}{0}
\setcounter{section}{0}

\gdef\thesection{Appendix \Alph{section}}
\section{Details of the Floquet-theory analysis of the CB for the smooth system}
\label{AppendixA}

The matrix ${\textbf{B}}(\tau)$ appearing in Eq. (\ref{eq31}) can be presented in a block-form as:
\begin{equation}
\label{eq32}
{\textbf{B}}(\tau)=\begin{bmatrix}
    -\textbf{B}^{(xx)} & \textbf{B}^{(xy)} & \textbf{0}_{N-1,N}\\
    \textbf{B}^{(yx)} & -\textbf{B}^{(yy)} & \textbf{0}_N\\
    \textbf{0}_{N,N-1} & \textbf{0}_N & -\textbf{B}^{(zz)}
  \end{bmatrix}
\end{equation}
where $\textbf{0}_{p,q}$ is a rectangular zeros-matrix of $p$ rows and $q$ columns. The components of the four defining rectangular matrices (with occasional commas between subscripts added for clarity) are given by
\begin{equation}
\label{eq33}
\begin{split}
B^{(xx)}_{ij}=\beta^{(xx)}_i\delta_{ij}, \
\beta^{(xx)}_{i}\triangleq -\left.\frac{\partial{\hat{x}_{i+1}''}}{\partial{\hat{x}_{i+1}}}\right\rvert_{\hat{\textbf{x}}=\hat{\textbf{y}}=0}=\\ \frac{1}{\mu}\left[1+\frac{3}{4}\alpha\hat{z}^2(\tau)(\delta_{N_0,i+1}+\delta_{N_0,i})\right],   1 \le i \le N-1, \\
B^{(xy)}_{ij}=\left.\frac{\partial{\hat{x}_{i+1}''}}{\partial{\hat{y}_{j}}}\right\rvert_{\hat{\textbf{x}}=\hat{\textbf{y}}=0}=\frac{1}{4\mu}\left[\delta_{i+1,j}+\delta_{ij}+\right. \\ \left.+\frac{3}{2}\alpha\hat{z}^2(\tau)(\delta_{N_0,i+1}\delta_{i+1,j}+\delta_{N_0,i}\delta_{ij})\right], \ 1 \le j \le N ; \\
B^{(yx)}_{ij}=\left.\frac{\partial{\hat{y}_{i}''}}{\partial{\hat{x}_{j+1}}}\right\rvert_{\hat{\textbf{x}}=\hat{\textbf{y}}=0}=\left[1+\frac{3}{2}\alpha\hat{z}^2(\tau)\delta_{N_0,i}\right]\times\\ \times(\delta_{i,j+1}+\delta_{i,j}), \ 2 \le i \le N-1, 1 \le j \le N-1; \\
B^{(yx)}_{N,j}=\left.\frac{\partial{\hat{y}_{N}''}}{\partial{\hat{x}_{j+1}}}\right\rvert_{\hat{\textbf{x}}=\hat{\textbf{y}}=0}=\\=\left[1+\frac{3}{2}\alpha\hat{z}^2(\tau)\delta_{N_0,N}\right]\delta_{N,j+1}; \\
B^{(yy)}_{ij}=-\left.\frac{\partial{\hat{y}_{i}''}}{\partial{\hat{y}_{j}}}\right\rvert_{\hat{\textbf{x}}=\hat{\textbf{y}}=0}=\left[1+\frac{3}{2}\alpha\hat{z}^2(\tau)\delta_{N_0i}\right]\delta_{ij}, \\ 2 \le i \le N-1, 1 \le j \le N;\\
B^{(yy)}_{N,j}=-\left.\frac{\partial{\hat{y}_{N}''}}{\partial{\hat{y}_{j}}}\right\rvert_{\hat{\textbf{x}}=\hat{\textbf{y}}=0}=\\=\left[\frac{1}{2}+\frac{3}{4}\alpha\hat{z}^2(\tau)\delta_{N_0,N}\right]\delta_{N,j},  1 \le j \le N
\end{split}
\end{equation}
\begin{equation}
\label{eq33b}
\begin{split}
B^{(zz)}_{ij}=-\left.\frac{\partial{\hat{z}_i''}}{\partial{\hat{z}_{j}}}\right\rvert_{\hat{\textbf{x}}=\hat{\textbf{y}}=0}=\left[1+\frac{3}{2}\alpha\hat{z}^2(\tau)\delta_{N_0i}\right]\delta_{ij}, \\ 1 \le i \le N-1, 1 \le j \le N;
B^{(zz)}_{N,j}=-\left.\frac{\partial{\hat{z}_{N}''}}{\partial{\hat{z}_{j}}}\right\rvert_{\hat{\textbf{x}}=\hat{\textbf{y}}=0}\\=\left[\frac{1}{2}+\frac{3}{4}\alpha\hat{z}^2(\tau)\delta_{N_0,N}\right]\delta_{N,j},  1 \le j \le N;
\end{split}
\end{equation}
The remaining expressions for the components of $B^{(yx)}_{1,j}$ and $B^{(yy)}_{1,j}$ are given by
\begin{equation}
\label{eq34}
\begin{split}
B^{(yx)}_{12}=\left.\frac{\partial{\hat{y}_{1}''}}{\partial{\hat{x}_{2}}}\right\rvert_{\hat{\textbf{x}}=\hat{\textbf{y}}=0}=0, \
B^{(yx)}_{1,j}=\left.\frac{\partial{\hat{y}_{1}''}}{\partial{\hat{x}_{j}}}\right\rvert_{\hat{\textbf{x}}=\hat{\textbf{y}}=0}=\\-1-\frac{3}{2}\alpha\hat{z}^2(\tau)\delta_{N_0,1} \ \forall \  j>2, \
B^{(yy)}_{11}=-\left.\frac{\partial{\hat{y}_{1}''}}{\partial{\hat{y}_{1}}}\right\rvert_{\hat{\textbf{x}},\hat{\textbf{y}}=0} \\= \frac{1+2\mu}{2\mu}\left[1+\frac{3}{2}\alpha\hat{z}^2(\tau)\delta_{N_0,1}\right], 
B^{(yy)}_{1,j}=\\-\left.\frac{\partial{\hat{y}_{1}''}}{\partial{\hat{y}_{j}}}\right\rvert_{\hat{\textbf{x}}=\hat{\textbf{y}}=0}=\frac{1}{2\mu}\left[1+\frac{3}{2}\alpha\hat{z}^2(\tau)\delta_{N_0,1}\right] \ \forall \ j  \ge 2
\end{split}
\end{equation}
here $\delta_{ab}$ or $\delta_{a,b}$ is Kronecker's delta, and $\hat{z}(\tau)$ is the (periodic) CB solution at the CB localization-site, which is identical to the solution given in Eqs. (\ref{eq18}) and (\ref{eq20}).

\bigskip

\gdef\thesection{\Alph{section}}
\subsection{Periodic boundary conditions}

\bigskip

For periodic boundary conditions (and the assumption of zero total momentum and displacement, as before), Eq. (\ref{eq29}) should be replaced by the following expressions:

\begin{equation}
\label{eq35}
\begin{split}
\hat{y}_N''= -\frac{1+2\mu}{2\mu}\hat{y}_N-\sum\limits_{n=2}^{N-1}\hat{x}_n-\frac{1}{2\mu}\sum\limits_{n=1}^{N-1}\hat{y}_n+\\ +\frac{\alpha}{4}(2\hat{x}_N-\hat{y}_N)[(2\hat{x}_N-\hat{y}_N)^2+3\hat{z}_N^2]\\-\frac{\alpha}{4}\left(2\sum\limits_{n=2}^{N}\hat{x}_n+\frac{1}{\mu}\sum\limits_{n=1}^{N-1}\hat{y}_n+\frac{1+\mu}{\mu}\hat{y}_N\right)\times \\ \times\left[\left(2\sum\limits_{n=2}^{N}\hat{x}_n+\frac{1}{\mu}\sum\limits_{n=1}^{N-1}\hat{y}_n+\frac{1+\mu}{\mu}\hat{y}_N\right)^2+3\hat{z}_N^2\right] \\
\hat{z}_N''= -\hat{z}_N-\frac{\alpha}{4}\left\lbrace \vphantom{ \left. \left(2\sum\limits_{n=2}^{N}\hat{x}_n+\frac{1}{\mu}\sum\limits_{n=1}^{N-1}\hat{y}_n+\frac{1+\mu}{\mu}\hat{y}_N\right)^2\right]\hat{z}_N+2\hat{z}_N^3} 3\left[(2\hat{x}_N-\hat{y}_N)^2+\right. \right.\\ \left. \left. \left(2\sum\limits_{n=2}^{N}\hat{x}_n+\frac{1}{\mu}\sum\limits_{n=1}^{N-1}\hat{y}_n+\frac{1+\mu}{\mu}\hat{y}_N\right)^2\right]\hat{z}_N+2\hat{z}_N^3\right\rbrace
\end{split}
\end{equation}

Then, the fourth and sixth rows in Eqs. (\ref{eq33}) and the second row in Eq. (\ref{eq33b}) should be replaced by the following expressions (much simplified, since in a ring, the CB site can be chosen arbitrarily and there is no point in setting $N_0=N$):
\begin{equation}
\label{eq36}
\begin{split}
B^{(yx)}_{N,j}=\left.\frac{\partial{\hat{y}_{N}''}}{\partial{\hat{x}_{j+1}}}\right\rvert_{\hat{\textbf{x}}=\hat{\textbf{y}}=0}=-1, 1 \le j \le N-2; \\ B^{(yx)}_{N,N-1}=\left.\frac{\partial{\hat{y}_{N}''}}{\partial{\hat{x}_{N}}}\right\rvert_{\hat{\textbf{x}}=\hat{\textbf{y}}=0}=0\\
B^{(yy)}_{N,j}=-\left.\frac{\partial{\hat{y}_{N}''}}{\partial{\hat{y}_{j}}}\right\rvert_{\hat{\textbf{x}}=\hat{\textbf{y}}=0}=\frac{1}{2\mu}, 1 \le j \le N-1; \\ B^{(yy)}_{N,N}=-\left.\frac{\partial{\hat{y}_{N}''}}{\partial{\hat{y}_{N}}}\right\rvert_{\hat{\textbf{x}}=\hat{\textbf{y}}=0}=\frac{1+2\mu}{2\mu}\\
B^{(zz)}_{N,j}=-\left.\frac{\partial{\hat{z}_{N}''}}{\partial{\hat{z}_{j}}}\right\rvert_{\hat{\textbf{x}}=\hat{\textbf{y}}=0}=\delta_{N,j}, 1 \le j \le N
\end{split}
\end{equation}

\renewcommand{\theequation}{{B.}\arabic{equation}}
\setcounter{equation}{0}
\gdef\thesection{Appendix \Alph{section}}
\section{Details of the Floquet-theory analysis of the CB -- nonsmooth system}
\label{AppendixB}

The matrices required for analytical construction of the monodromy are as follows:
\begin{equation}
\label{FN2a}
\begin{split}
\textbf{C}_{j,l} \triangleq\begin{bmatrix}
\textbf{I}_{\hat{N}}&\textbf{0}_{\hat{N}}\\
\textbf{0}_{\hat{N}}&\textbf{I}_{\hat{N}}+\Delta\hat{\textbf{C}}^{(j,l)}
\end{bmatrix},
\textbf{A}_* \triangleq\begin{bmatrix}
\textbf{0}_{\hat{N}}&\textbf{I}_{\hat{N}}\\
\hat{\textbf{A}}&\textbf{0}_{\hat{N}}\\
\end{bmatrix},
\\
\hat{\textbf{A}}\triangleq\begin{bmatrix}
    -\frac{1}{\mu}\textbf{I}_{N-1} & \frac{1}{4\mu}\textbf{G}_{N-1,N} & \frac{1}{4\mu}\textbf{G}_{N-1,N}\\
    \frac{1}{2}(\textbf{G}_{N-1,N})^{\top} & -\textbf{I}_N & \textbf{0}_N\\
    \frac{1}{2}(\textbf{G}_{N-1,N})^{\top} & \textbf{0}_N & -\textbf{I}_N
  \end{bmatrix}
\end{split}
\end{equation}
where $\textbf{0}_{N}$ is a matrix of zeros of size $N$, $\textbf{I}_{N}$ is the identity matrix of rank $N, \ \hat{N}\triangleq3N-1$ and $G_{ij}\triangleq \delta_{ij}+\delta_{i,j-1}$.
\begin{equation}
\begin{split}
\label{FN2b}
\Delta\hat{C}^{(1,3)}_{ik}=\Delta\hat{C}^{(3,1)}_{ik}\triangleq -\delta_{ik}(\delta_{i,n_0-1}+\delta_{i,3n_0-2})+\\ \delta_{i,n_0-1}\delta_{k,3n_0-2}+\delta_{n_0-1,k}\delta_{i,3n_0-2} \\+
\Delta\hat{C}^{(2,3)}_{ik}=\Delta\hat{C}^{(3,2)}_{ik}\triangleq -\delta_{ik}(\delta_{i,n_0-1}+\delta_{i,5n_0-3})+\\ \delta_{i,n_0-1}\delta_{k,5n_0-3}+\delta_{n_0-1,k}\delta_{i,5n_0-3}  \\+
\Delta\hat{C}^{(1,4)}_{ik}\triangleq -\delta_{ik}(\delta_{i,n_0}+\delta_{i,3n_0-2})+\\+\delta_{i,n_0}\delta_{k,3n_0-2}+\delta_{n_0,k}\delta_{i,3n_0-2} \\
\Delta\hat{C}^{(2,4)}_{ik}\triangleq -\delta_{ik}(\delta_{i,n_0}+\delta_{i,5n_0-3})+\\+\delta_{i,n_0}\delta_{k,5n_0-3}+\delta_{n_0,k}\delta_{i,5n_0-3}
\end{split}
\end{equation}
\begin{equation}
\label{FS2}
\textbf{S}_{j,l}=\textbf{C}_{j,l}+(\textbf{A}_*\textbf{C}_{j,l}-\textbf{C}_{j,l}\textbf{A}_*)\textbf{D}_{j,l}
\end{equation}
\begin{equation}
\label{FS3}
\begin{split}
\textbf{D}_{1,3} \triangleq \frac{\textbf{L}\textbf{x}^0[\textbf{n}^{(1,3)}]^\top}{[\textbf{n}^{(1,3)}]^\top\textbf{A}_*\textbf{L}\textbf{x}^0} , \\ \textbf{D}_{2,3} \triangleq \frac{\textbf{C}_{1,3}\textbf{L}\textbf{x}^0[\textbf{n}^{(2,3)}]^\top}{[\textbf{n}^{(2,3)}]^\top\textbf{A}_*\textbf{C}_{1,3}\textbf{L}\textbf{x}^0}, \\ \textbf{D}_{3,1} \triangleq \frac{\textbf{C}_{2,3}\textbf{C}_{1,3}\textbf{L}\textbf{x}^0[\textbf{n}^{(1,3)}]^\top}{[\textbf{n}^{(1,3)}]^\top\textbf{A}_*\textbf{C}_{2,3}\textbf{C}_{1,3}\textbf{L}\textbf{x}^0}, \\
\textbf{D}_{2,4} \triangleq \frac{\textbf{C}_{1,3}\textbf{L}\textbf{x}^0[\textbf{n}^{(2,4)}]^\top}{[\textbf{n}^{(2,4)}]^\top\textbf{A}_*\textbf{C}_{1,3}\textbf{L}\textbf{x}^0}, \\ \textbf{D}_{3,2} \triangleq \frac{\textbf{C}_{2,4}\textbf{C}_{1,3}\textbf{L}\textbf{x}^0[\textbf{n}^{(2,3)}]^\top}{[\textbf{n}^{(2,3)}]^\top\textbf{A}_*\textbf{C}_{2,4}\textbf{C}_{1,3}\textbf{L}\textbf{x}^0}, \\ \textbf{D}_{4,1} \triangleq \frac{\textbf{C}_{2,3}\textbf{C}_{2,4}\textbf{C}_{1,3}\textbf{L}\textbf{x}^0[\textbf{n}^{(1,4)}]^\top}{[\textbf{n}^{(1,4)}]^\top\textbf{A}_*\textbf{C}_{2,3}\textbf{C}_{2,4}\textbf{C}_{1,3}\textbf{L}\textbf{x}^0}
\end{split}
\end{equation}
where $n_i^{(1,3)}=\delta_{i,3n_0-2} -\delta_{i,3n_0-1},n_i^{(2,3)}=\delta_{i,3n_0-2} -\delta_{i,3n_0}, n_i^{(1,4)}=\delta_{i,3n_0-1} -\delta_{i,3n_0+1}$, and $n_i^{(2,4)}=\delta_{i,3n_0} -\delta_{i,3n_0+1}$, for $i=1,2,..,3N$ and $\delta_{a,b}$ is Kronecker's delta.
\begin{equation}
\label{FS1}
\begin{split}
\textbf{M}^{(p)}=[\textbf{L}\textbf{S}^{(p)}\textbf{L}]^2 \ , \ \textbf{L}=\exp{\left(\frac{\pi}{2\hat\omega}\textbf{A}_*\right)} \ , \\ \textbf{S}^{(3)}=\textbf{S}_{3,1}\textbf{S}_{2,3} \textbf{S}_{1,3} \ , \ \textbf{S}^{(4)}=\textbf{S}_{4,1}\textbf{S}_{3,2} \textbf{S}_{2,4} \textbf{S}_{1,3}
\end{split}
\end{equation}
where $\hat\omega=\omega/\sqrt{k/m}$ is the normalized frequency of the periodic solution ($\omega$ being the frequency), which should be above the flat-band frequency ($\hat\omega_{\text{FB}}=1$). The initial condition is $\textbf{x}^0=(\textbf{q}_{\hat{N}},\textbf{p}_{\hat{N}})^{\top},q_i=0,p_i=\delta_{i,3n_0-1}-\delta_{i,3n_0-2}$. The (absolute) value of the initial velocity of the elements of the CB appears explicitly only in Eq. (\ref{FS3}), where it contracts, and thus one can understand $\textbf{x}^0$ as a normalized quantity. The initial velocity still influences the value of the \emph{period} of the CB, represented by $\hat{\omega}$. Equation (\ref{FN2a}) assumes that one (say, the leftmost) mass $M$ is either fixed or its displacement and velocity are known from total momentum conservation (`free-free' boundary conditions). Therefore, the number of independent displacements is assumed to be $3N-1$. Also, the second boundary needs a condition, say that of a free boundary. Unlike for the case of moderate amplitudes, where boundary conditions affect the emergence of Hopf instabilities, it appears that for large amplitudes the Hopf instabilities nearly vanish and only pitchfork instabilities remain, for which the specific choice of boundary conditions is less important, for a long chain. 

Moreover, the choice of boundary conditions does not affect the emergence of a stable CB for large amplitudes for the ratio $M/m=1$ (the value for which there is reason to construct the monodromy matrix in the nonsmooth case). Therefore, Eq. (\ref{FN2a}) is corrected to represent the simplest set of boundary conditions -- `fixed-free'. It is assumed that $x_1$ is known and needs not an equation. Thus, the quantity $\textbf{G}_{N-1,N}$ is understood to be having its first row corresponding to the second of Eqs. (\ref{eq1}). The same goes for the equations defining $\textbf{C}_{j,l},\textbf{D}_{j,l}$ and $\textbf{S}_{j,l}$. Accordingly, the matrix $\textbf{G}_{N,N-1}$ has its first column corresponding to the variable $x_2$. This covers the `left' (fixed) boundary condition. Furthermore, if one assumes that the `right' boundary condition is free, then the matrix $\textbf{I}_N$ in the diagonal blocks of $\hat{\textbf{A}}$ is defined as $I_{ij}=\delta_{ij} \ \forall \ i<N, \ I_{NN}=1/2$. For the specific case addressed, the substitution $\mu=1/2$ should be made for consistency (rendering the matrix $\hat{\textbf{A}}$ symmetric for an infinite chain or periodic boundary conditions). An odd value of $N$ should be taken, preferably with $n_0=(N+1)/2$, for maximum distance of the CB from the boundaries.

\renewcommand{\theequation}{{C.}\arabic{equation}}
\setcounter{equation}{0}
\gdef\thesection{Appendix \Alph{section}}
\section{Analysis of the three-body dynamics in the smooth limit for $\hat\mu\ll1$}
\label{AppendixC}

Rearranging Eq. (\ref{Kin23}), we get the following two-impacts finite-difference equation:
\begin{equation}
\label{A1}
v_1^{(k+2)}-v_1^{(k)}=2\hat\mu\left[v_3^{(k+1)}-v_1^{(k)} \right]
\end{equation}

The expression for $v_3^{(k+1)}$ can be obtained from Eq. (\ref{Kin2}), by taking the $\hat\mu\ll1$ limit:
\begin{equation}
\label{A2}
v_3^{(k+1)}=2v_2^{(k)}-v_3^{(k)}+2\hat\mu\left[v_3^{(k)}-v_2^{(k)} \right]
\end{equation}

Substituting Eq. (\ref{A2}) into Eq. (\ref{A1}), one gets, to first order, the following explicit two-impacts difference equation:
\begin{equation}
\label{A3}
v_1^{(k+2)}-v_1^{(k)}=2\hat\mu\left[2v_2^{(k)}-v_3^{(k)} -v_1^{(k)} \right]
\end{equation}

Next, in order to obtain $v_3^{(k+2)}$, one resorts again to Eq. (\ref{Kin2}) in the $\hat\mu\ll1$ limit, to get:
\begin{equation}
\label{A4}
v_3^{(k+2)}=2v_1^{(k)}-v_3^{(k+1)}+2\hat\mu\left[v_3^{(k+1)}-v_1^{(k)} \right]
\end{equation}

Substituting Eq. (\ref{A2}) into Eq. (\ref{A4}) results, to first order in $\hat\mu$, in the following explicit two-impacts difference equation::
\begin{equation}
\label{A5}
\begin{split}
v_3^{(k+2)}-v_3^{(k)}=2v_1^{(k)}-2v_2^{(k)}+\\+2\hat\mu\left[3v_2^{(k)}-2v_3^{(k)}-v_1^{(k)} \right]
\end{split}
\end{equation}

Recalling the equation of conservation of linear momentum for the three-body system, namely,
\begin{equation}
\label{A6}
v_2^{(k)}=-v_1^{(k)}-\hat\mu v_3^{(k)}
\end{equation}
allows eliminating $v_2^{(k)}$, and writing Eqs. (\ref{A3}) and (\ref{A5}) as a closed system:
\begin{equation}
\label{A7}
\begin{split}
v_1^{(k+2)}-v_1^{(k)}=-2\hat\mu\left[3v_1^{(k)}+v_3^{(k)} \right] \\
v_3^{(k+2)}-v_3^{(k)}=4v_1^{(k)}-2\hat\mu\left[4v_1^{(k)}+v_3^{(k)}\right]
\end{split}
\end{equation}
(omitting terms nonlinear in $\hat\mu$).

In Eqs. (\ref{A7}), the updated variables refer to the values of the velocities after impacts involving particles 1 and 3. Hence, the change due to two impacts in the velocity of particle 3 is negative as long as the velocity of particle 1 is negative. For the positive points in the history of the velocity of particle 3, one would have to use the solution of the system in Eqs. (\ref{A7}), along with Eqs. (\ref{A2}) and (\ref{A6}) -- this is what the calculation in Eq. (\ref{Kin26}) requires.

The sequence $v_3^{(k)}$ is sign-changing (between consecutive entries in the sequence). However, if only odd values of $k$ are observed, then a negative sequence emerges, with no sign-changes within it.

Now, the changes between elements in this sequence are finite, as well as the differences between the instances. Therefore, the left-hand sides in Eqs. (\ref{A7}) can be understood as numerical derivatives with respect to the integer variable $\bar k$, where $k=2\bar k-1$, such that $v_{1,3}'\sim [v_{1,3}^{(\bar k+1)}-v_{1,3}^{(\bar k)}]/[(\bar k+1)-\bar k]$, where the prime implies differentiation with respect to increments (of magnitude 1) of $\bar k$. Then one gets the following relations:
\begin{equation}
\label{A8}
\begin{split}
v'_1\sim-2\hat\mu(3v_1+v_3), \
v'_3\sim 4v_1-2\hat\mu(4v_1+v_3)
\end{split}
\end{equation}
(the superscripts are omitted henceforth for clarity).

Defining $\tilde v_{1,3}^{(\bar k)}\sim v_{1,3}^{(k)}$, one can rewrite the above relations as two coupled first-order differential equations, which can be decoupled analytically into a single second-order linear equation and an auxiliary relation, as follows:
\begin{equation}
\label{A9}
\begin{split}
\tilde v''_3+8\hat\mu\tilde v'_3+8\hat\mu\tilde v_3=0, \
\tilde v_1 =\left(\frac{1}{4}+\frac{\hat\mu}{2}\right)\tilde v'_3+\frac{\hat\mu}{2}\tilde v_3
\end{split}
\end{equation}
(omitting terms nonlinear in $\hat\mu$).

The first (second-order) equation in Eqs. (\ref{A9}) can be readily solved, yielding the solution
\begin{equation}
\label{A10}
\begin{split}
\tilde v_3 ^{(\bar k)}=V_0 e^{-4\hat\mu \bar k}\sin{\left(\sqrt{8\hat\mu}\bar k+\phi\right)}
\end{split}
\end{equation}
(where for the sake of the well-posedness of the derivative, $\bar k$ can be hypothetically extended over the field of reals).

Consequently, the velocity associated with particle 1 would to leading order in $\hat\mu$ be
\begin{equation}
\label{A11}
\begin{split}
\tilde v_1 ^{(\bar k)}=\frac{\sqrt{8\hat\mu}}{4}V_0 e^{-4\hat\mu \bar k}\cos{\left(\sqrt{8\hat\mu}\bar k+\phi\right)}
\end{split}
\end{equation}

Therefore, using the definitions introduced above and employing Eqs. (\ref{A6}) and (\ref{A2}), one would have, to leading order in $\hat\mu$,
\begin{equation}
\label{A12}
\begin{split}
v_{1}^{(k \text{ even})}\sim\frac{\sqrt{8\hat\mu}}{4}V_0 e^{-4\hat\mu \bar k}\cos{\left[\sqrt{8\hat\mu}\bar k+\phi\right]} =\\=v_{1}^{(k \text{ odd})}, \ 
v_{2}^{(k \text{ odd})}\underset {\hat\mu\ll 1}\to - v_{1}^{(k \text{ odd})}\Rightarrow \ v_{3}^{(k \text{ even})}\\ \underset {\hat\mu\ll 1}\to -2v_{1}^{(k \text{ odd})}-v_{3}^{(k \text{ odd})}\underset {\hat\mu\ll 1}\to  -v_{3}^{(k \text{ odd})}\sim \\ \sim V_0 e^{-4\hat\mu \bar k}\sin{\left[\sqrt{8\hat\mu}\bar k+\phi-\pi\right]}
\end{split}
\end{equation}
which implies that, to leading order in $\hat\mu$, each next impact merely changes the sign of the velocity of particle 3 after a given impact (rather than also doubling its magnitude, as in the strict $\hat\mu= 0$ case -- a sign of singularity of the integrability, in the sense of velocity-reversal, with $\hat\mu$ as it approaches zero).

Now, we see that the sequence of velocities of particle 3 after an even number of impacts is governed by two numerical time scales, one associated with exponential decay on a typical time of $\tau_e\sim (4\hat \mu)^{-1}$, and the other associated with sinusoidal change with a period of $\tau_s\sim 2\pi(8\hat\mu)^{-1/2}$, which for $\hat\mu\ll1$ is a much smaller time scale (than $\tau_e$). Since we are interested in integrable (periodic) solutions, the relevant time scale of consideration would be (half) a period (namely, $\tau_s/2$). During this time, the exponential prefactor has the minimum value of $e^{-\pi\sqrt{2\hat\mu}}$, which for $\hat\mu\ll1$ expands as $1-\pi\sqrt{2\hat\mu}$, and leads to a two-terms expression for $v_{3}^{(k \text{ even})}$. In this two-terms expression, the second term is negligible with respect to the first term in the limit of $\hat\mu\ll1$. Therefore, to leading order in $\hat\mu$, one has:
\begin{equation}
\label{A13}
\begin{split}
v_{3}^{(k \text{ even})}\sim V_0\sin{\left[\sqrt{8\hat\mu}\bar k+\phi-\pi\right]}
\end{split}
\end{equation}

Next, enforcing the initial condition, $v_3^{(0)}=0$, leads to the determination of the phase as $\phi=\pi-\sqrt{2\hat\mu}$ and this, along with determining the amplitude from Eq. (\ref{Kin25}), yields the following expression (correct to first-order) for the sequence of velocities of particle 3 after an even number of impacts:
\begin{equation}
\label{A14}
\begin{split}
v_{3}^{(k)}\sim \frac{v_0}{\sqrt{\hat\mu/2}}\sin{\left[\sqrt{2\hat\mu}k\right]}, \ \ \forall \ \ k \ \text{even}
\end{split}
\end{equation}
(where one recalls that $k=2\bar k-1$, as defined earlier).

Clearly, the obtained velocity evolution is \emph{sinusoidal}, as `assumed' in the derivation of Eq. (\ref{Kin27}). Indeed, the sine function is concave on the half-period where it is positive; it is symmetric around the middle of this range; and it is linear in the vicinity of the edges of this range, starting and ending as zero -- all the requirements from the distribution as dictated by periodic dynamics in the $\hat\mu\ll1$ limit (as discussed in the derivation of Eq. (\ref{Kin27})).

Finally, clearly, the average of the function in Eq. (\ref{A14}) over half a period (where the function is positive), normalized by the maximum value of the function (which is the definition of $\rho$ from Eq. (\ref{Kin26})), amounts to $2/\pi$, which is indeed the value used for the derivation of Eq. (\ref{Kin27}).

It is noteworthy that another way to validate Eq. (\ref{Kin27}) is to require that the argument of the sine in Eq. (\ref{A14}) becomes exactly $\pi$ just when $k=j$ and $\hat\mu$ takes the value suggested by Eq. (\ref{Kin27}) (and one bears in mind that $v_3^{(k=j)}=0$ is a condition required for velocity reversal).

This is not trivial, since the functional form of $\hat\mu_j$ in Eq. (\ref{Kin27}) was obtained from energy conservation and kinematic considerations for the discrete system represented by the mapping. However, here, the same functional form of inverse square dependence is obtained due to the structure of the coefficients of a second-order linear differential equation derived by taking `the smooth' limit (converting the mapping into a flow). The fact that identical functional forms were obtained, as aforementioned, is validation of the self-consistency of the process of resorting to the smooth-system limit (for the $\hat\mu \ll1 $ case). This is noteworthy, since it is the smooth-limit-validity assumption that enables the calculation of $\rho$, which is required for obtaining Eq. (\ref{Kin27}). Therefore, the analysis presented in this appendix, although it may appear as (partially) based on cyclic logic, does in fact provide independent validation necessary for checking overall self-consistency.

As a last note, one observes that taking the strict limit $\hat\mu\to 0$ (rather than $\hat\mu\ll1$) in Eq. (\ref{A14}), leads to the result $v_3^{(k \text{ even})}=2k v_0$ for small values of $k$. A complementary-symmetric result is also obtained for $k\to \pi/\sqrt{2\hat\mu}$. This is not mere evidence of the aforementioned `linear' behavior at the range edges, but is in fact an exact and correct (although asymptotic) result, as obtained by direct integration of the original discrete mapping. This fact also contributes to the overall self-consistency of the analysis.


\begin{thebibliography}{6}
\expandafter\ifx\csname url\endcsname\relax
  \def\url#1{\texttt{#1}}\fi
\expandafter\ifx\csname urlprefix\endcsname\relax\def\urlprefix{URL }\fi
\expandafter\ifx\csname href\endcsname\relax
  \def\href#1#2{#2} \def\path#1{#1}\fi


\bibitem{Aubry}
S.~Aubry, Breathers in nonlinear lattices: Existence, linear stability and quantization, Physica D103 (1997) 201-250.

\bibitem{Rosenau}
P.~Rosenau and A.~Zilburg, On a strictly compact discrete breather in a
{K}lein-{G}ordon model, Physics Letters A 379~(43-44) (2015) 2811--2816.

\bibitem{Flach1}
D.~Leykam, S.~Flach, O.~Bahat-Treidel, and A.~Desyatnikov, Flat band states: disorder and nonlinearity, Physical Review B 88 (2013) 224203.

\bibitem{Flach2}
W.~Maimaiti, A.~Andreanov, H.~C. Park, O.~Gendelman, and S.~Flach, Compact localized states and flatband generators in one dimension, Physical Review B 95 (2017) 115135.

\bibitem{Johansson2015b}
M.~Johansson, U.~Naether and R.~A. Vicencio, Compactification tuning for nonlinear localized modes in sawtooth lattices, Physical Review E 92 (2015) 032912.

\bibitem{ZegadloMalomed}
K.~Zegadlo, N.~Dror, N.~V. Hung, M.~Trippenbach, and B.~A. Malomed, Single and double linear and nonlinear flatband chains: spectra and modes, Physical Review E 96 (2017) 012204.

\bibitem{DanieliFlach}
C.~Danieli, A.~Maluckov, and S.~Flach, Compact descrete breathers on flat-band networks, Low Temperature Physics 44 (2018) 678-687

\bibitem{NPOVG2017}
N.~Perchikov and O.~V. Gendelman, Flat bands and compactons in mechanical lattices, Physical Review E 96 (2017) 052208.

\bibitem{Sergyeyev}
A.~Sergyeyev, S.~Skurativskyi, and V.~Vladimirov, Compacton solutions and(non)integrability of nonlinear evolutionary PDEs associated with a chain of prestressed granules, Nonlinear Analysis: Real World Applications 47 (2109) 68--84.

\bibitem{Maraver}
J.~Cuevas-Maraver, P.~G. Kevrekidis, B.~A. Malomed, and L.~Gao, Solitary waves in the Ablowitz--Ladik equation with power-law nonlinearity, Journal of Physics A: Mathamatical and Theoretical 52 (2019) 065202.

\bibitem{Aravena}
G.~C\'aceres-Aravena and R.~A. Vicencio, Perfect localization on flat band binary one-dimansional photonic lattices, arXiv:1903.00377.

\bibitem{James}
G.~James, Travelling breathers and solitary waves in strongly nonlinear lattices, Philosophical Transactions of the Royal Society A 376 (2018) 1--25.

\bibitem{Real}
B.~Real and R.~A. Vicencio, Controlled mobility of compact discrete solitons in nonlinear Lieb photonic lattices, Physical Review A 98 (2018) 053845.

\bibitem{Qin}
J.~Qin, Z.~Liang, B.~A. Malomed, and G.~Dong, Tail-free self-accelerating solitons and vortices, Physical Review A 99 (2019) 023610.

\bibitem{Kim}
S.~W. Kim and S.~Kim, Fano resonances in translationally-invariant nonlinear chains, Physical Review B 63 (2000) 212301.

\bibitem{Maimaiti}
W.~Maimaiti, S.~Flach, and A.~Andreanov, Universal $d=1$ flat band generator from compact localized states, Physical Review B 99 (2019) 125129.

\bibitem{Maraver2}
J.~Cuevas-Maraver, P.~G. Kevrekidis, D.~J. Frantzeskakis, N.~I. Karachalios, M.~Haragus, and G.~James, Floquet analysis of Kuznetsov-MA breathers: A path toward spectral stability of rogue waves, Physical Review E 96 (2017) 012202.

\bibitem{Maraver3}
J.~Cuevas-Maraver, P.~G. Kevrekidis, A.~Vainchtein, and H.~Xu, Unifying perspective: Solitary traveling waves as discrete breathers in Hamiltonian lattices and energy criteria for their stability, Physical Review E 96 (2017) 032214.

\bibitem{Doi}
Y.~Doi and K.~Yoshimura, Symmetric Potential Lattice and Smooth Propagation of Tail-Free Discrete Breathers, Physical Review Letters 117 (2016) 014101.

\bibitem{Suchkov}
S.~V. Suchkov, B.~A. Malomed, S.~V. Dmitriev, and Y.~S. Kivshar, Solitons in a chain of parity-time-invariant dimers, Physical Review E 84 (2011) 046609.

\bibitem{Perchikov2016}
N.~Perchikov, O.~V. Gendelman, Nonlinear dynamics of hidden modes in a system
with internal symmetry, JSV 377 (2016) 185--215.

\bibitem{Perchikov2015}
N.~Perchikov, O.~V. Gendelman, Dynamics and stability of a discrete breather in a harmonically excited chain with vibro-impact on-site potential, Physica D 292 (2015) 8--28.


\end{thebibliography}
\end{document}